\documentclass[acmsmall,screen]{acmart}
\AtBeginDocument{%
  }

\usepackage{xcolor}
\usepackage{tikz}
\usetikzlibrary{arrows.meta, positioning, fit, calc, shadows, shadings}
\usepackage{svg}
\usepackage{bbding}
\usepackage{stmaryrd}
\usepackage{microtype}
\usepackage{paralist}
\usepackage{csvsimple}
\usepackage{placeins}
\usepackage[inline]{enumitem} 
\usepackage{svg}
\usepackage{xurl}

\usepackage{booktabs}   

\RequirePackage[textsize=tiny,textwidth=1.2cm]{todonotes}
\RequirePackage{siunitx} 
\newif\ifblind
\blindtrue

\RequirePackage{symbolic-bisim} 
\usepackage{thmtools}
\usepackage[capitalise]{cleveref}

\newcommand{\pluseq}{\mathrel{+}=}

\newcommand{\yultext}[1]{\ma{\text{\yulinline{#1}}}}

\newcommand{\semcomp}{\mathbin{\oslash}}
\newcommand{\syncomp}[2]{#1\mathbin{\cup}#2}
\newcommand{\intcomp}{\mathbin{\curlywedge}}
\newcommand{\compatible}{\mathbin{\asymp}}
\newcommand{\gamesem}[1]{\llbracket #1 \rrbracket}

\newcommand\cmain{\yultext{Main}}
\newcommand\amain{\alpha_{main}}

\newcommand{\rR}{\ma{\mathcal{R}}}
\newcommand{\defEndsymbol}{$\blacksquare$}
\newcommand{\defEnd}{\hfill\defEndsymbol}

\newcommand{\conf}[1]{\ma{\langle #1 \rangle}} 

\newcommand{\deployl}{\ma{\mathsf{deploy}}}
\newcommand{\preturnl}[1]{%
  \ma{%
    \mathsf{p%
      \ifthenelse{\equal{#1}{i}}%
        {\mathnormal{i}}
        {#1}
      \text{-}ret}%
  }%
}

\newcommand{\calll}{\ma{\mathsf{call}}}

\newcommand{\createl}{\ma{\mathsf{create}}}
\newcommand{\ppcalll}{\ma{\mathsf{pp\text{-}call}}}
\newcommand{\pocalll}{\ma{\mathsf{po\text{-}call}}}
\newcommand{\oreturnl}{\ma{\mathsf{o\text{-}ret}}}
\newcommand{\eocall}{\ma{\mathsf{eo\text{-}call}}}
\newcommand{\ocall}{\ma{\mathsf{o\text{-}call}}}
\newcommand{\waitl}{\ma{\mathsf{wait}}}

\newcommand{\Uintset}{\ma{\mathsf{Uint}}} 
\newcommand{\Addrset}{\ma{\mathsf{Addr}}} 

\newcommand{\newCallMsg}{\ma{\mathtt{callMsg}}}
\newcommand{\newCreateMsg}{\ma{\mathtt{createMsg}}}

\newcommand{\Msgset}{\ma{\mathsf{Msg}}}
\newcommand{\Envset}{\ma{\mathsf{Env}}}
\newcommand{\Stateset}{\ma{\mathsf{State}}}
\newcommand{\MemSet}{\ma{\mathsf{Mem}}}
\newcommand{\StoreSet}{\ma{\mathsf{Store}}}
\newcommand{\Accset}{\ma{\mathsf{Acc}}}
\newcommand{\bytesset}{\ma{\mathsf{Bytes}}}

\newcommand{\AbiSet}{\ma{\mathsf{ABI}}}
\newcommand{\CALL}{\ma{\mathsf{call}}}
\newcommand{\CREATE}{\ma{\mathsf{create}}}

\newcommand{\evmOut}{\ma{\mathsf{out}}}

\newcommand{\msgCaller}{\ma{\mathsf{caller}}}
\newcommand{\msgTarget}{\ma{\mathsf{target}}}

\newcommand{\msgGas}{\ma{\mathsf{gas}}}
\newcommand{\msgVal}{\ma{\mathsf{value}}}
\newcommand{\msgData}{\ma{\mathsf{data}}}
\newcommand{\msgCAddr}{\ma{\mathsf{caddr}}}

\newcommand{\msgTransfer}{\ma{\mathsf{transfer}}}
\newcommand{\msgStatic}{\ma{\mathsf{static}}}

\newcommand{\stateSto}{\ma{\mathsf{storage}}}

\newcommand{\accNonce}{\ma{\mathsf{nonce}}}
\newcommand{\accBal}{\ma{\mathsf{bal}}}
\newcommand{\accCode}{\ma{\mathsf{code}}}
\newcommand{\accabi}{\ma{\mathtt{abi}}}
\newcommand{\abihash}{\ma{\mathtt{hash}}}
\newcommand{\abitype}{\ma{\mathtt{type}}}

\usepackage{listings-solidity} 
\definecolor{verylightgray}{rgb}{.97,.97,.97}

\usepackage{newtxtt}

\lstMakeShortInline[style=inlinecode]|
\lstdefinestyle{inlinecode}{
	language=Solidity,
	extendedchars=true,
	basicstyle=\small\ttfamily,
	showstringspaces=false,
	showspaces=false,
	showtabs=false,
    keywordstyle=\small\ttfamily,
    keywordstyle=[1]\small\ttfamily,
    keywordstyle=[2]\small\ttfamily,
    keywordstyle=[3]\small\ttfamily,
    identifierstyle=\small\ttfamily,
    commentstyle=\small\ttfamily,
    stringstyle=\small\ttfamily,
}
\newcommand{\solinline}{\lstinline[style=inlinecode]}
\usepackage{listings-evmtrace}
\usepackage{listings-yul}
\lstdefinestyle{yul-inline-small}{
	language        = Yul,
	basicstyle      = \ttfamily\small,     
	showspaces      = false,
	showstringspaces= false,
	keepspaces      = true,
}
\newcommand{\yulinline}{\lstinline[style=yul-inline-small]}
\lstdefinestyle{evm-inline-small}{
	language        = EVMTrace,
	basicstyle      = \ttfamily\small,     
	showspaces      = false,
	showstringspaces= false,
	keepspaces      = true,
}

\newcommand{\retop}{\yultext{return}}
\newcommand{\callop}{\yultext{call}}
\newcommand{\createop}{\yultext{create}}

\DeclareMathOperator{\rlp}{RLP}
\DeclareMathOperator{\keccak}{keccak256}

\newcommand{\Sblock}[1]{\ma{\textup{\texttt{\{}} #1 \textup{\texttt{\}}}} }  
\newcommand{\funk}{\ma{\mathsf{function}}}                  
\newcommand{\Sfundef}[4]{\ma{\funk\,#1\textup{\texttt{(}}#2\textup{\texttt{)}}#3#4}}
\newcommand{\Sfundefto}[4]{\Sfundef{#1}{#2}{{\to}#3}{#4}}
\newcommand{\assignk}{\ma{\textup{\texttt{:=}}}}                     
\newcommand{\Svardecl}[2]{\ma{\letk\,#1\assignk#2}}
\newcommand{\Sassign}[2]{\ma{#1\assignk#2}}
\newcommand{\Scond}[2]{\ma{\ifk #1 #2}}
\newcommand{\switchk}{\ma{\mathsf{switch}}}                 
\newcommand{\casek}{\ma{\mathsf{case}}}                     
\newcommand{\defaultk}{\ma{\mathsf{default}}}               
\newcommand{\Sswitch}[2]{\ma{\switchk\,#1#2}}
\newcommand{\fork}{\ma{\mathsf{for}}}                       
\newcommand{\Sfor}[4]{\ma{\fork#1#2#3#4}}
\newcommand{\breakk}{\ma{\mathsf{break}}}                   
\newcommand{\continuek}{\ma{\mathsf{continue}}}             
\newcommand{\leavek}{\ma{\mathsf{leave}}}                   
\newcommand{\Efuncall}[2]{\ma{#1\textup{\texttt{(}}#2\textup{\texttt{)}}}}
\newcommand{\opcode}[1][]{\ma{\mathop{\ifempty{#1}{op}{#1}}}}
\newcommand{\objectk}{\ma{\mathsf{object}}}                 
\newcommand{\codek}{\ma{\mathsf{code}}}                     
\newcommand{\yulobj}[3]{\ma{\objectk\,#1\Sblock{\codek\,#2\,#3}}}
\newcommand{\datak}{\ma{\mathsf{data}}}                     
\newcommand{\yuldat}[2]{\ma{\datak\,#1\,#2}}

\newcommand{\regk}{\ma{\mathsf{regular}}}

\newcommand{\nameN}{\ma{\mathcal{N}}}

\newcommand{\ssframe}[1]{\ma{\llparenthesis #1 \rrparenthesis}}
\newcommand{\sscont}[1]{\ma{\llbracket #1 \rrbracket_{\mathsf{cnt}}} }
\newcommand{\ssbreak}[1]{\ma{\llbracket #1 \rrbracket_{\mathsf{brk}}} }

\newcommand{\str}{\ma{\mathsf{str}}} 
\newcommand{\StrLit}{\ma{\mathsf{StrLit}}} 
\renewcommand{\hex}{\ma{\mathsf{hex}}} 
\newcommand{\HexLit}{\ma{\mathsf{HexLit}}} 
\newcommand{\Stmt}{\ma{\mathsf{Stmt}}} 
\newcommand{\ExpC}{\ma{\mathsf{ECxt}}} 
\newcommand{\StmtC}{\ma{\mathsf{SCxt}}} 

\newcommand{\ObjSet}{\ma{\mathsf{Obj}}} 
\newcommand{\DataSet}{\ma{\mathsf{Data}}} 

\newcommand{\yult}{\ma{\textsc{YulTracer}}} 

\usepackage{pifont}
\newcommand{\passed}{\ding{51}}
\newcommand{\rejected}{\ding{55}\,}

\usepackage{tabularx}
\usepackage{multirow}

\renewcommand{\texttt}[1]{{\small\ttfamily #1}}

\AtBeginDocument{
\theoremstyle{acmdefinition}
\newtheorem{remark}[theorem]{Remark}
}
\crefname{remark}{Remark}{Remarks}
\Crefname{remark}{Remark}{Remarks}
\begin{document}

\title{Open-World Assertion Checking for Smart Contracts via Game Semantics}

\author{Vasileios Koutavas}
\email{Email: Vasileios.Koutavas@tcd.ie}
\orcid{0000-0002-3970-2486}
\affiliation{%
	\institution{Trinity College Dublin}
	\country{Ireland}
}
\author{Yu-Yang Lin}
\email{linhouy@tcd.ie}
\orcid{0000-0001-5783-9454}
\affiliation{%
	\institution{Trinity College Dublin}
	\country{Ireland}
}

\author{Nikos Tzevelekos}
\email{nikos.tzevelekos@qmul.ac.uk}
\orcid{0000-0001-8509-8059}
\affiliation{%
	\institution{Queen Mary University of London}
	\country{UK}}


\begin{abstract}
We present a game semantics framework for open-world safety analysis of Ethereum smart contracts.
We model the interaction between a contract and its environment as a two-player game between the contract and the environment, and prove up to gas model approximations soundness:
every assertion violation found corresponds to a real execution; and completeness: every open-world execution is captured. 
To our knowledge, this provides the first formal open-world interaction semantics for Ethereum smart contracts with mathematical guarantees of soundness and completeness.

We implement this framework in \yult, an assertion reachability tool for real-world Solidity contracts, built on Yul, the intermediate language of the Solidity compiler.
\yult uses concrete execution and exhaustively explores game traces within user-specified bounds.
We evaluate it on reentrancy benchmarks, where \yult achieves 100\% recall and precision --- the only tool to do so from those we examined --- and on two large real-world exploits (the DAO and PredyPool), where it detects the known vulnerabilities and produces no false positives on fixed versions.
To our knowledge, \yult is the first tool to achieve this level of precision on real-world contracts without false positives.
We additionally demonstrate generality of the approach via the examination of access control benchmarks.
\end{abstract}

\maketitle

\section{Introduction}
\label{sec:intro}

Programmable blockchains extend distributed ledgers with \emph{smart contracts}:
programs invoked in transactions that can read and write on-chain state.
They underpin decentralised applications, notably in decentralised finance (DeFi).
On Ethereum, one of the largest programmable blockchains, smart contracts are
\emph{stateful programs} typically written in high-level languages such as Solidity, and compiled to Ethereum Virtual Machine (EVM) bytecode.
They manage large amounts of digital assets,%
\footnote{As of this writing, Ethereum contracts hold over USD\,59 billion in cryptocurrency~\cite{defillama}.}
and their code is public and executes in an \emph{open-world} environment, where it may interact with arbitrary, potentially adversarial code.
Hence, smart contracts have been repeatedly targeted, resulting in multi-billion-dollar losses~\cite{crypto-crimes-2024,eth-vulnerabilities,eth-bugs}, making them a critical area of security research.

The open-world setting, combined with the stateful nature of Ethereum smart contracts, poses a distinctive challenge to contract safety.
A contract’s public interface may be invoked by any entity, including unknown and adversarial contracts.
Moreover, when a contract performs an external call, it explicitly transfers control to untrusted code, which may execute arbitrary logic before returning.
Such interactions can steer execution through intermediate states that the contract’s developer did not anticipate, exposing windows in which key invariants may not hold.

A particularly subtle and dangerous form of interaction is \emph{reentrancy}.
In a reentrancy scenario, a contract calls untrusted code, which then calls back into the contract before the original invocation completes.
This allows the external entity to observe and manipulate the contract in intermediate states, potentially violating assumptions about control flow or state consistency.
Reentrancy has enabled some of the most severe smart contract exploits, including the 2016 DAO attack~\cite{dao-reentrancy-overview}, and remains a prevalent attack vector until today~\cite{pcaversaccio-reentrancy-attacks}.
More broadly, reentrancy exemplifies the challenges of reasoning about contracts that interact with an unconstrained environment.

Despite these challenges, a large ecosystem of tools aims to detect vulnerabilities and verify the safety of smart contracts.
Surveys catalogue an extensive body of work, with a recent one listing over ninety distinct tools~\cite{vidalVulnerabilityDetectionTechniques2024}.
These tools differ in how they balance automation against precision, particularly in their treatment of interactions between a contract and its environment.

Broadly, existing tools fall into two categories.
\emph{Automatic tools} analyse contracts for known vulnerability signatures without requiring user-provided specifications.
  They rely on techniques such as
  \emph{static analysis} (e.g.~\cite{ethainter,slither,symvalic}),
  \emph{symbolic execution} (e.g.~\cite{mythril,oyente,manticore}),
  \emph{abstract interpretation} (e.g.~\cite{madmax,securify}),
  or \emph{machine learning} (e.g.~\cite{zhuangSmartContractVulnerability2020}).
\emph{Specification-based tools} require developers or auditors to explicitly state correctness properties, often as assertions, which are expressive enough to capture any \emph{safety property}~\cite{lamport}.
  These tools are based on
  \emph{property-based fuzzing}
  (e.g.~\cite{contractfuzzer,echidna,foundry-forge}),
  \emph{symbolic testing}
  (e.g.~\cite{hevm,halmos}),
  and
  \emph{bounded} or \emph{SMT-based model checking}
  (e.g.~\cite{smartpulse,ethbmc,certora,solc-smtchecker}).

Automatic tools over-approximate the possible interactions between a contract and its environment, enforcing conservative disciplines such as the checks-effects-interactions pattern~\cite{code_patterns}.
While effective at flagging suspicious code and large-scale blockchain scans, this approach produces \emph{false positives}, particularly in contracts with intricate control flow, which require manual effort to triage.

Specification-based tools may also fail to accurately model all executions of a contract in an open-world setting.
Some over-approximate interactions, for example by treating all state as unknown when control is passed to an external contract~\cite{certora}, or relying on imprecise verification techniques~\cite{solc-smtchecker}.
Others under-approximate behaviour by limiting the number of transactions considered (e.g.~\cite{hevm} supports one),
or accounting only for limited reentrancy patterns~\cite{smartpulse}.
Testing-based approaches require users to supply concrete attacker contracts for sampling executions~\cite{contractfuzzer,echidna,foundry-forge} or symbolic exploration~\cite{halmos}.
Even tools that aim to exhaustively explore precise interactions (e.g.~\cite{ethbmc}) struggle to scale to real-world contracts due to the cost of symbolic execution.
Finally, to our knowledge, no tool targeting real-world Solidity contracts comes with a formal semantic foundation of open-world interactions and mathematical guarantees of soundness or completeness.



In this work, we present a formal, open-world theory of interaction between Ethereum smart contracts and their environment, and its implementation in \yult, an assertion reachability tool for real-world contracts.
Our approach offers a unique combination of benefits:
\begin{description}

  \item[Formal foundation:]
    We define a compositional interaction semantics for Ethereum smart contracts in the form of a 
    \emph{game semantics}~\cite{AJM,HO,Nickau} which concretely models all calls and returns between a contract and an arbitrary blockchain environment.
    This naturally captures the open-world setting of smart contracts, where any external entity may call into the contract or be called by it.
    To our knowledge, this provides the first formal open-world interaction semantics for Ethereum smart contracts.

  \item[Soundness:] We prove that the game semantics admits only actual contract executions within valid environments, up to gas model realisability (cf.~\cref{rem:opponent-gas}).
    Any assertion violation \yult discovers and does not depend on Opponent gas consumption, corresponds to a genuine, exploitable vulnerability, eliminating false positives in all cases we have examined.

  \item[(Bounded-)Completeness:]
    The game semantics is \emph{complete}: it encodes all possible interactions between contracts and environments, up to gas-dependent branching.
    We prove that if no assertion violations are derivable in the game semantics, then no attacker can trigger the assertions.
    Although this semantics induces infinite traces in general, \yult is \emph{bounded complete}: it exhaustively enumerates game semantics traces up to user-specified bounds (similar to bounded model checking~\cite{hwBMC,CBMC}).

\item[Real-world scalability:]
  \yult analyses real-world Solidity contracts emitted by the Solidity compiler (version 0.8.0 and above),\footnote{For our case studies we have transliterated older Solidity source code to Solidity 0.8 to compile with a modern compiler (see \cref{sec:upgrades}).}
  and detects known historic vulnerabilities in large-scale contracts with intricate reentrancy behaviour (\cref{sec:motivating}).
  When these vulnerabilities are fixed, \yult reports no spurious assertion violations.
  To our knowledge, \yult is the first tool that can precisely pinpoint such interaction-based vulnerabilities in large-scale contracts without false positives.

\end{description}


As Solidity is a feature-rich and rapidly evolving language, we build our theory and tool on Yul,
a stable, minimal intermediate language emitted by the Solidity compiler since version 0.8.0 (\cref{sec:lang}).
Yul has a standard operational semantics~\cite{yul:semantics} and is parametric in the underlying EVM semantics, which defines opcodes and contract calls and returns.
\yult users, however, never interact with Yul directly: contracts and assertions are written in Solidity, and the tool reports EVM call and return traces that can be straightforwardly mapped back to the source code.

Building on this language, we develop a game semantics for Yul smart contracts (\cref{sec:games}), modelling the interaction between a contract and its environment as a two-player game between a \emph{Proponent} (the contract) and an \emph{Opponent} (the environment).
We prove soundness: every play corresponds to a real contract execution, up to gas model realisability; and completeness: every open-world execution is captured by a play, up to gas-dependent control flow. The qualifications about gas are discussed in \cref{rem:opponent-gas}.

We implement this in \yult (\cref{sec:tool}).
To scale to large real-world contracts, \yult adopts \emph{concrete execution}, rather than SMT-based symbolic execution,
operating on concrete unsigned 256-bit words (U256), the native EVM data type.
This avoids frequent and expensive SMT-solver invocations required by symbolic approaches, which quickly become a scalability bottleneck.
The main challenge is choosing values supplied by the environment: \yult addresses this using \emph{environment knowledge domains}, finite sets of U256 values initialised with user-provided constants and extended at runtime with values produced by the contract, enabling exhaustive exploration within user-specified bounds.



We evaluate \yult in \cref{sec:evaluation} by comparing its performance on reentrancy benchmarks from the Gigahorse suite~\cite{grech-gigahorse-benchmarks}
and by using it to reproduce two real-world incidents: the DAO (2016; \cref{sec:dao:explain,sec:dao}), and PredyPool (2024; \cref{sec:predy:explain,sec:predy}).\footnote{%
Also on the Lendf.Me incident (2020; \cref{sec:lendf}).}
On the reentrancy benchmarks, \yult achieved 100\% recall and precision --- the only tool to do so --- while remaining performant.
On the real-world incidents, with modest domain-specific guidance, it detected the exploits without producing false positives when vulnerabilities were fixed.

Our case studies and benchmarks focus on reentrancy, as it is particularly difficult to detect without false positives due to complex cross-contract callbacks and intermediate states.
It cannot always be avoided, as it is often a desirable feature of smart contracts (e.g.~\cref{sec:predy:explain}), and it remains a prevalent class of exploits~\cite{pcaversaccio-reentrancy-attacks}.
\yult is not inherently limited to reentrancy, however, as it is an assertion-based analysis framework and can detect violations of arbitrary \emph{safety properties} when expressed as assertions~\cite{lamport}.
We demonstrate this on 16 access control benchmarks from the Gigahorse suite, discuss the scope and limits of assertion-based checking in \cref{sec:beyond}, and conclude in \cref{sec:conclusions}.


\section{Two Real-World Vulnerabilities}
\label{sec:motivating}

\subsection{The DAO}
\label{sec:dao:explain}

The DAO project~\cite{dao-victim}, deployed on the 30th of April 2016 on the Ethereum blockchain, was one of the first large-scale attempts to create a decentralised venture fund.
It pooled ETH and issued DAO tokens that conferred proportional voting rights.
Token holders could propose and vote on funding proposals; when approved, funds were disbursed on-chain to contractors according to the contract’s rules.
Governance, treasury management and disbursements were fully automated on the blockchain, with a ``split'' mechanism allowing investors to withdraw into a child DAO rather than support proposals they opposed.
By the 17th of June 2016, the DAO was infamously hacked~\cite{dao-reentrancy-overview} and drained of about 3.6 million ETH.

\begin{lstlisting}[
  float,
  caption={The vulnerable DAO functions.},
  label={lst:dao-vulnerable}]
contract DAO is DAOInterface, TokenCreation {
  ManagedAccount public rewardAccount; // defined in DAOInterface
  ...
  function splitDAO(uint _proposalID, address _newCurator) onlyTokenholders external override
        returns (bool _success) {
    Proposal storage p = proposals[_proposalID];
    ... // checks
    uint fundsToBeMoved =(*@\label{ln:dao:a}@*)
        (balances[msg.sender] * p.splitData[0].splitBalance) / p.splitData[0].totalSupply;
    assert(!(p.splitData[0].newDAO.createTokenProxy{value : fundsToBeMoved}(msg.sender) == false));(*@\label{ln:dao:b}@*)
    ...
    withdrawRewardFor(msg.sender);(*@\label{ln:dao:c}@*)
    totalSupply -= balances[msg.sender];
    balances[msg.sender] = 0;(*@\label{ln:dao:f}@*)
    paidOut[msg.sender] = 0;
    return true;
  }
  ...
  function withdrawRewardFor(address _account) internal override returns (bool _success) {
    ...
    assert(rewardAccount.payOut(_account, reward));(*@\label{ln:dao:d}@*)
    ...
} }
contract ManagedAccount is ManagedAccountInterface{
  ...
  function payOut(address _recipient, uint _amount) public override payable returns (bool) {
    ...
    (bool success, ) = _recipient.call{value: _amount}("");(*@\label{ln:dao:e}@*)
    ...
} }

\end{lstlisting}%

The vulnerable part of the code (upgraded to Solidity~0.8\footnote{See \cref{sec:upgrades} for the source code upgrades applied.})
is shown in \cref{lst:dao-vulnerable}.
The call to function |splitDAO| succeeds only if the caller (who is the sender of the call message, |msg.sender|) is a DAO token holder who voted in favour of the ``split proposal'' with id |_proposalID|,
the voting period has ended but the split-execution window is still open, and the caller has not voted for another proposal in the meantime.
When these (omitted) checks are satisfied, the funds to be moved to the child DAO are computed from the caller's balance and transferred (ll.~\ref{ln:dao:a}--\ref{ln:dao:b}).
The function then sends any accrued rewards to the caller directly via |withdrawRewardFor| (l.~\ref{ln:dao:c}),
which calls |ManagedAccount.payOut| (l.~\ref{ln:dao:d}) to transfer the rewards to the caller (l.~\ref{ln:dao:e}).
The caller's balance is updated at the end of |splitDAO| on line \ref{ln:dao:f}.

Transferring ETH via |call| (l.~\ref{ln:dao:e}) hands control to the recipient contract.
The DAO attacker leveraged this to re-enter |splitDAO| during these calls.
Because the caller’s balance is updated only after the external interactions,
the same funds are moved repeatedly to the attacker-controlled child DAO before the update at line \ref{ln:dao:f} takes effect; the process can iterate as many times as transaction and gas limits allow.

As we show in \cref{sec:dao}, \yult discovers this attack under suitable parameters.
Moreover, making |splitDAO| non-reentrant via a Solidity guard does \emph{not} fully mitigate risk: after control returns via line \ref{ln:dao:e},
the attacker can invoke the DAO's |transfer| function (not shown here) to move its balance to a cooperating contract, causing the balance to be drawn twice\,---\,once to the child DAO and once to the other contract.
Coordinated multi-transaction traces between the two attacker contracts can again drain the main DAO.\footnote{The actual DAO attacker combined the two attacks to maximise the speed of draining the DAO.}
\yult discovers this secondary attack with two Opponent addresses.
The contract is secured by additionally moving the call to |withdrawRewardFor| (l.~\ref{ln:dao:c}) \emph{after} the caller’s balance update at line \ref{ln:dao:f}. With this change, \yult finds no exploit within the tested bounds.

This exploit is hard to detect precisely: it spans multiple contracts (|DAO|, |ManagedAccount|) and requires a specific sequence of environment actions (creating a split proposal, voting, waiting for the voting window to close, then re-entrant calls to |splitDAO|).
Indeed, \textsc{Slither}~\cite{slither} reports effectively the same warnings on both the vulnerable and a corrected, safe version of the DAO, unable to distinguish the two.
Solidity compiler's \textsc{SMTChecker}~\cite{smtchecker} in BMC mode times out on the |Deployer| of the vulnerable version after 1 hour without producing a finding.


\subsection{PredyPool}
\label{sec:predy:explain}

\begin{lstlisting}[
  float,
  caption={The vulnerable PredyPool functions.},
  label={lst:predy-vulnerable}]
contract PredyPool is IPredyPool, IUniswapV3MintCallback, Initializable, ReentrancyGuardUpgradeable {
  ...
  function supply(uint256 pairId, bool isQuoteAsset, uint256 supplyAmount) external nonReentrant(*@\label{ln:predy:supply:s}@*)
        returns (uint256 finalSuppliedAmount) {
    return SupplyLogic.supply(globalData, pairId, supplyAmount, isQuoteAsset);
  }(*@\label{ln:predy:supply:e}@*)
  function trade(TradeParams memory tradeParams, bytes memory settlementData) external(*@\label{ln:predy:trade:s}@*)
        returns (TradeResult memory tradeResult) {
      ... // pre-trade checks
    return TradeLogic.trade(globalData, tradeParams, settlementData);(*@\label{ln:predy:tradelogic-trade-call}@*)
  }(*@\label{ln:predy:trade:e}@*)
  function take(bool isQuoteAsset, address to, uint256 amount) external onlyByLocker {(*@\label{ln:predy:take:s}@*)
    globalData.take(isQuoteAsset, to, amount);
  }(*@\label{ln:predy:take:e}@*)
  ...
}
library TradeLogic {
  ...
  function trade(GlobalDataLibrary.GlobalData storage globalData, 
                 IPredyPool.TradeParams memory tradeParams, bytes memory settlementData) external
      returns (IPredyPool.TradeResult memory tradeResult) {
    ... // performs trade and calculates tradeResult
    // The caller deposits or withdraws margin from the callback that is called below.
    callTradeAfterCallback(globalData, tradeParams, tradeResult);(*@\label{ln:predy:callTradeAfterCallback}@*)
    // check vault safety
    tradeResult.minMargin = PositionCalculator.checkSafe(...);(*@\label{ln:predy:checksafe}@*)
    ...
  }
  function callTradeAfterCallback(GlobalDataLibrary.GlobalData storage globalData,
        IPredyPool.TradeParams memory tradeParams, IPredyPool.TradeResult memory tradeResult) internal {
    globalData.initializeLock(tradeParams.pairId);(*@\label{ln:predy:initlock}@*)
    IHooks(msg.sender).predyTradeAfterCallback(tradeParams, tradeResult);(*@\label{ln:predy:predyTradeAfterCallback}@*)
    ...
  }
  ...
}
\end{lstlisting}

Predy Finance \cite{predy-finance, predy-victim} implemented a set of market contracts that shared a central lending and settlement vault, PredyPool.
Markets held the trading logic; PredyPool held ERC-20 tokens, lent to markets on behalf of traders, and settled profits and losses.
An excerpt of its code is shown in \cref{lst:predy-vulnerable}. 
It uses helper libraries and complex arguments that we do not describe here; relevant parts are discussed below.

PredyPool was organised by token pairs.
Registering a pair bound it to a specific Uniswap v3 pool~\cite{uniswap-protocol}.
Uniswap v3 pools act as decentralised exchanges between two ERC-20 tokens, holding token reserves, and automatically updating their relative exchange rate as swaps execute.
PredyPool used the Uniswap pools as oracles relying on its time-weighted average price (TWAP) for risk and price checks.
The pair owner was able to optionally whitelist which market contracts could perform trades for that pair.

Lenders provided liquidity tokens to PredyPool via |supply| (ll.~\ref{ln:predy:supply:s}--\ref{ln:predy:supply:e} in \cref{lst:predy-vulnerable})
and withdrew via |withdraw| (omitted).
Whitelisted market contracts invoked |trade| on a pair (ll.~\ref{ln:predy:trade:s}--\ref{ln:predy:trade:e}), which run pre-checks and called
the library function |TradeLogic.trade| (l.~\ref{ln:predy:tradelogic-trade-call}).
This function performed the trade operation and called the inner |callTradeAfterCallback| (l.~\ref{ln:predy:callTradeAfterCallback})
which in turn called back |predyTradeAfterCallback| (l.~\ref{ln:predy:predyTradeAfterCallback}) on the market contract.
During this callback, the market contract could borrow tokens from PredyPool's vault associated with the pair via |take|%
\footnote{Line~\ref{ln:predy:initlock} sets the caller to be the ``locker'' of the contract during the callback, making \lstinline{take}'s \lstinline{onlyByLocker} modifier succeed.}
(ll.~\ref{ln:predy:take:s}--\ref{ln:predy:take:e}) and return leftover funds via |supply|.
After the callback, PredyPool computed per-token balance differences and decided whether to accept or revert the trade (l.~\ref{ln:predy:checksafe}).

On the 14th of May 2024 an attacker exploited a vulnerability in PredyPool's code and extracted approximately USD\,\num{440000} worth of tokens (approx.~219,585 USDC and 83.9 WETH), resulting
in the suspension of PredyPool market services \cite{predyfinance-postmortem}.
The attacker registered a duplicate USDC/WETH pair and invoked trade on that pair.
Inside the trade callback they used |take| to borrow USDC and WETH from PredyPool's global vault which contained funds provided for a preexisting USDC/WETH pair.
Within the same callback, they used |supply| to return the same amounts back to PredyPool, this time credited to the attack pair.
Because the end-of-trade checks validated global per-token balances rather than per-pair, the transaction passed.
In a subsequent transaction, the attacker withdrew from their pair as if they had provided the liquidity.

Callbacks and reentrancy are an essential part of Predy's design.
In such a codebase spanning thousands of lines of code (tens of thousands including all of the \num{56} internal and external libraries), this vulnerability would likely be easy to miss,
even with support from existing tools.
In \cref{sec:predy} we show how \yult, with modest instrumentation similar to that of a test suite, produced a trace highlighting the exploit
without additional false positives.

Notably, |supply| is declared with a |nonReentrant| modifier (\cref{lst:predy-vulnerable}, l.~\ref{ln:predy:supply:s}) but |trade| is not, so
|supply| \emph{can} be called from within the trade callback, enabling the cross-pair movement of funds at the heart of the attack.
One mitigation is to make |trade| non-reentrant (in combination with a different method for traders to return leftover funds in the trade callback).
Indeed, with |nonReentrant| added to |trade|, \yult produced no exploit trace within our analysis bounds.

\section{Language}
\label{sec:lang}
\newcommand\memo{\mathcal{M}}
\newcommand\ioproj{{\iota o}}
\newcommand\lL{\mathcal{L}}
\newcommand\cC{\mathcal{C}}
\newcommand\dD{\mathcal{D}}
\newcommand\minuseq{\mathrel{-}=}

We choose Yul~\cite{yul,yul:semantics} as our term language for EVM contract code. 
Yul is an intermediate representation (IR) language between Solidity and EVM bytecode.
It is parameterised by a \emph{dialect} that defines available data types, primitive operations, and
any other supporting constructs. Core Yul, independent of the dialect, provides a small set of structured
control-flow constructs, avoiding low-level stack manipulation and jumps.
In this paper we consider an EVM-flavoured Yul syntax (\cref{fig:yul}) extended with an object syntax $\ObjSet$, a minimal EVM dialect, and an assertion instruction to specify safety properties. We introduce a formalisation of Yul objects because the EVM operates on contracts rather than on isolated code fragments. EVM-flavoured Yul accordingly organises executable Yul code into Yul objects, each of which represents a contract-level unit.

\begin{figure}[t]
\[\begin{array}{r@{\;\;}r@{\,}c@{\,}l}
    \textsc{\Stmt}\ni  & S & \mis & \Sblock{S^*}
                           \mor \Sfundefto{x}{\vec x}{\vec x}{\Sblock{S^*}}
                           \mor \Svardecl{\vec x}{M}
                           \mor \Sassign{\vec x}{M}
                           \mor M
                           \mor \Scond{M}{\Sblock{S^*}}\\
                     & & & \mor \Sswitch{M}{(\casek\,v\Sblock{S^*})^*\defaultk\Sblock{S^*}}
                           \mor \Sfor{\Sblock{S^*}}{M}{\Sblock{S^*}}{\Sblock{S^*}}\\
                     & & & \mor \breakk
                           \mor \continuek
                           \mor \leavek
                           \mor \regk
                           \\
    \textsc{\Exp}\ni & M & \mis & \Efuncall{x}{\vec M}
                           \mor \Efuncall{\opcode}{\vec M}
                           \mor x
                           \mor v
                           \mor <\!\vec v\!>
            \quad\quad \textsc{\Val}\ni v,w,c
            \qquad \textsc{\Var}\ni  x,y,z,f 
        \\
    \noalign{\vskip 2pt}
    \hline
    \noalign{\vskip 2pt}
    \textsc{\StmtC}\ni & E & \mis & \hole
                           \mor \Sblock{ E,\vec S}_{l}^{\nameN}
                           \mor \Svardecl{\vec x}{E}
                           \mor \Sassign{\vec x}{E}
                           \mor E_{\Exp}
                           \mor \sscont{E}
                           \mor \ssbreak{E}
                           \mor \Scond{E_{\Exp}}{\Sblock{S^*}}
                           \\
                     & & & \mor \Sswitch{E_{\Exp}}{(\casek\,v\Sblock{S^*})^*\defaultk\Sblock{S^*}} \\
    \textsc{\ExpC}\ni & E_{\Exp} & \mis & \hole
                           \mor \Efuncall{x}{\vec M,E_{\Exp},\vec v}
                           \mor \Efuncall{\opcode}{\vec M,E_{\Exp},\vec v}
                           \mor \ssframe{E}_{l}^{\vec x}\\
    \noalign{\vskip 2pt}
    \hline
    \noalign{\vskip 2pt}
    \textsc{\ObjSet}\ni & O & \mis & \yulobj{\str}{S}{D^*}\\
    \textsc{\DataSet}\ni & D & \mis & O \mor \yuldat{\str}{\hex}
        \qquad \textsc{\StrLit}\ni \str
        \qquad \textsc{\HexLit}\ni \hex
      \end{array}\]
    \caption{Yul syntax (top) and evaluation contexts (middle) from \cite{yul:semantics} extended with objects (bottom).}\label{fig:yul}
    \end{figure}
In \cite{yul:semantics}, Yul is instantiated with a dialect by defining the semantics of evaluating primitive instructions/opcodes $\opcode$, a type for values/machine-words, and a definition of the global state $G$. We shall provide here a minimal extension to express inter-contract control flow in the form of message calls, create and return opcodes, value types, and a definition of EVM components to instantiate $G$ with. This means intra-contract classes of opcodes such as for arithmetic, bitwise manipulation, comparison, etc, shall be omitted for brevity; we only care about inter-contract opcodes and their necessary supporting components.

The semantics in \cite{yul:semantics} is defined on configurations of the form $\conf{S\mid G ; L ; \nameN}$. These configurations comprise a global state $G$, a local state $L$, and a local function name space $\nameN$. We shall define a reduction semantics based on an extension of \cite{yul:semantics} to handle both objects and intra-contract EVM opcodes. Like in \cite{yul:semantics}, if a reduction occurs purely within Yul and does not involve $G$, we shall omit $G$ and write $\conf{S\mid L ; \nameN}$.

We next define a semantics for Yul objects.
First, we address a technicality. For the EVM dialect, Yul introduces three data opcodes: \yultext{datasize}, \yultext{dataoffset}, \yultext{datacopy}. These are built-in functions used to access data and objects nested within the current object being evaluated. For this reason, we can consider Yul terms to actually be pairs $(S, D)$ where $S$ is a statement and $D$ is the data component of objects; data functions operate on $D$. 
\[\begin{array}{@{}lll@{}}
\irule[Object][yul-object]{
  {}
}{
  \nbox{
    \conf{\yulobj{s}{S}{D} \mid G; L; \nameN}
    \trans{}
    \conf{ (S,D) \mid G; L; \nameN}
  }
}
\end{array}\]
However, since we omit intra-contract semantics herein, the $D$ component shall also be omitted henceforth unless required. Accordingly, when $D$ is omitted and $S = O.\codek$, the semantics of an object $O$ is understood to be simply the semantics of its code fragment $S$.

\subsection{An Inter-Contract EVM Dialect}
For the EVM dialect, we introduce byte sequences (\bytesset) as a base data type. We assume every sequence in \bytesset uniquely corresponds to a natural number and every object has a canonical \bytesset representation. We write $\bytesset$ for the set of all byte sequences and $\bytesset(N)$ for those of length $N$. Words in EVM-flavoured Yul\,---\,i.e.\ values that opcodes can operate on\,---\,are 256-bit unsigned integers ($\Uintset$) equivalent to $\bytesset(32)$. The EVM also introduces the notion of addresses ($\Addrset$), which are 160-bit unsigned integers equivalent to $\bytesset(20)$. We assume a transformation between $\Uintset$ and $\Addrset$ by truncating to their low bytes and left-padding with zeroes respectively.
\begin{align*}
bs &\in \bytesset\\
v,w &\in \Val = \Uintset = \bytesset(32)\\
  \alpha &\in \Addrset = \bytesset(20)\\
  \memo &\in \MemSet = \bytesset\times\bytesset(N)
\end{align*}
We also assume that all byte sequences support byte-wise indexing, so $bs(i)$ is the $i$th element of $bs$.
%
We introduce a memory component $\memo$ embedded with an I/O buffer of size $N$.
Given $\memo= (f,b)$ we shall write $\memo(i)$ for $f(i)$ and $\memo.\ioproj$ for the buffer $b$.
We shall be also viewing (the first component of) $\memo$ as a partial function $f\in\Uintset\rightharpoonup \bytesset(1)$ with contiguous domain $[0,\lvert f\rvert-1)$.
We shall be using interval notation $\memo[i,j)$, with $i\leq j$, for the array of bytes $[\memo(i),\dots,\memo(j-1)]$.
Memory update is written $\memo\left[[i,j)\mapsto bs\right]$ and is 0-padding when ${\mid bs\mid}< j-i$. The empty memory $(\varnothing,\varepsilon)$ is denoted by $\bot$.
It is also useful to introduce an operation for extending the current memory $\memo$. We let
$\mathtt{extend}(\memo,i)$ be an operation extending the memory $\memo$ to the interval $[0,j)$, that is, it returns $\memo'=\memo[k\mapsto 0\mid k\in[0,j)\setminus\dom\memo]$ along with a gas cost that is incurred by this extension.

We next introduce the remaining components needed for our presentation of the EVM dialect, focusing in particular on inter-contract behaviour. These are messages $m$ for inter-contract calls; the EVM environment, including time $T$ and state $\mathcal A$; and gas $g$. State is defined as an account map $\mathcal{A}$ from addresses  $\alpha$ to accounts $A$ (for internal accounts) or to the constant $\bullet$ (for external accounts).
Each account $A$ is a tuple containing, among other components, storage.
\[\begin{array}{r@{\;\;}r@{\;\;}c@{\;\;}l}
\textsc{\Msgset} \ni
  & m & \mis &
    \conf{%
      \msgCaller   \mathop{;} 
      \msgTarget   \mathop{;} 
      \msgGas      \mathop{;} 
      \msgVal      \mathop{;} 
      \msgData     \mathop{;} 
      \msgCAddr    \mathop{;} 
      \msgTransfer \mathop{;} 
      \msgStatic   \mathop{;} \ldots
    } \\
\textsc{\Envset} \ni
  & (T,\mathcal{A})& & 
    T \in \Uintset
    \qquad \textsc{\Stateset} \ni \mathcal{A} : \Addrset \rightharpoonup \Accset\uplus\{\bullet\}
    \\
\textsc{\Accset} \ni
  & A & \mis & 
    \conf{%
      \accNonce \mathop{;} 
      \accBal \mathop{;} 
      \accCode \mathop{;} 
      \stateSto \mathop
    }
    \qquad \StoreSet \ni \stateSto : \Uintset \to \Uintset
    \\
 &&&
    \accNonce \in \Uintset
    \qquad \accBal \in \Uintset
    \qquad \accCode \in \bytesset
\end{array}\]
Messages are records containing the caller and target addresses, supplied gas, transferred value, call data, code address, and flags for whether value transfer and state modification are allowed, written $m.\msgCaller$, $m.\msgTarget$, $m.\msgGas$, $m.\msgVal$, $m.\msgData$, $m.\msgCAddr$, $m.\msgTransfer$, and $m.\msgStatic$, respectively. For simplicity, we treat all message fields as elements of \bytesset, with conversions performed as needed. The precise choice of remaining fields is not essential here and may vary by implementation. Some fields are interpreted differently by different call variants\,---\,for example, \yultext{staticcall} constrains $m.\msgTransfer$ and $m.\msgStatic$, whereas \yultext{delegatecall} distinguishes $m.\msgTarget$ from $m.\msgCAddr$\,---\,and we defer these details unless necessary.
Time is measured in seconds. Any account whose \accCode\ component contains executable code is treated as a smart contract and carries a \stateSto, a word-indexed value map. We further assume that every deployed contract comes equipped with an ABI.
\[\begin{array}{r@{\;\;}r@{\;\;}c@{\;\;}l}
{\AbiSet \mathsf{fun}} \ni
  & f & \mis & 
    \conf{\abihash,\abitype}
    \qquad \abihash \in \bytesset(4)
    \qquad \AbiSet = \mathcal{P}_{\mathrm{fin}}({\AbiSet \mathsf{fun}})\\
  & \abitype & \mis & 
    \text{\solinline{bool}} \mor \text{\solinline{uint}} \mor \text{\solinline{int}} \mor \text{\solinline{address}} \mor \text{\solinline{bytes}} \mor \text{\solinline{string}} \mor \ldots
\end{array}\]
For our purposes, an ABI is a set of pairs consisting of a 4-byte selector and a type. The precise nature of the type is again implementation-dependent and not essential here. We assume that, for each callable selector, the ABI specifies an expected type such that a value of that type can be encoded in the message data supplied to the contract.

With all the above, we define EVM-Yul configurations as follows, by instantiating the global state $G$ as a tuple $\conf{\mathcal{A};T;g;K;\alpha;\memo}$:
\[
\conf{S \mid \conf{\mathcal{A};T;g;\alpha;\memo};L;\nameN}
\]
where the environment $(\mathcal{A}, T)$ consists of the state and the time; $g$ is the remaining gas; and $\alpha$, $\memo$, and $(S,L,\nameN)$ denote, respectively, the address of the account whose code is being executed, the memory of the current instance, and the Yul configuration of the current instance under evaluation. For brevity, we may henceforth write $S^L_{\nameN}$ for $(S,L,\nameN)$. 
\subsection{Inter-Contract EVM Instructions}
We consider a minimal extension of the semantics in \cite{yul:semantics} with an EVM dialect suitable for our game semantics. The Ethereum execution specification, up to the Cancun update, organises opcodes into 12 categories: \textit{arithmetic}, \textit{bitwise}, \textit{block}, \textit{comparison}, \textit{control-flow}, \textit{environment}, \textit{keccak}, \textit{log}, \textit{memory}, \textit{stack}, \textit{storage}, and \textit{system}, with an additional unofficial category of Yul-only builtins that are not part of the EVM instruction set (i.e.\ \yultext{datasize}, \yultext{dataoffset}, \yultext{setimmutable}, \yultext{loadimmutable}, and \yultext{linkersymbol}). Since inter-contract communication is mediated primarily by opcodes in the \textit{system} category, we focus on a representative fragment of the instruction set sufficient to capture such behaviour. To support the specification of safety properties, we additionally include a custom assertion opcode which is not part of the EVM instruction set. 
\begin{align*}
\opcode\ & \mis\  
        \yultext{return} \mor
        \yultext{call} \mor
        \yultext{create}\mor
        \yultext{ASSERT}
\end{align*}
Note that $\yultext{call}$ and $\yultext{create}$ are chosen as the standard representatives of related opcodes to keep the semantics concise. We could accommodate variants like $\yultext{delegatecall}$, $\yultext{staticcall}$ and $\yultext{create2}$ by slight modification of the game rules in \cref{sec:games}.

Apart from $\yultext{ASSERT}$, opcodes are not reduced by the internal Yul semantics of \cite{yul:semantics}. Hence evaluation may produce configurations that are stuck with respect to the operational semantics, namely those in which the next redex is an inter-contract opcode:
%
\[
\conf{E[op(\vec v)]\!\mid\! G ; L ; \nameN}\ \text{ and }\
op\in\{\yultext{return},\yultext{call},\yultext{create}\}.
\]
Such configurations are intentional: they are precisely the location where our game semantics supplies the next step.
Finally, we specify the relevant arguments for each opcode:
\begin{itemize}
\item $\yultext{return}(v_1,v_2)$, where $[v_1,v_1+v_2)$ is the memory block containing the return data;
\item $\yultext{call}(g,\alpha,v,v_i,v_s,v_o,v_z)$, where $g$ is the gas associated with the call, $v$ is the amount of Ether transferred with the call, $[v_i,v_i+v_s)$ is the memory block containing the call (input) data, and $[v_o,v_o+v_z)$ is the block reserved for the return (output) data;
\item $\yultext{create}(v,v_1,v_2)$, where $v$ is the balance to assign to the newly created contract, and $[v_1,v_1+v_2)$ the block in memory where this is going to be stored.
\end{itemize}


\section{Game Semantics}
\label{sec:games}
\newcommand\Pproj{_{\upharpoonright P}}
\newcommand\Oproj{_{\upharpoonright O}}

\begin{figure}[t]
\centering
\scalebox{0.5}{\begin{tikzpicture}[
  >=Stealth,
  arrstyle/.style={
    line width=1.6pt,
    draw={rgb,255:red,27;green,38;blue,49},
    shorten >=4pt, shorten <=4pt
  }
]

\definecolor{liborg}  {RGB}{243,156, 18}   
\definecolor{liborgdk}{RGB}{176, 98,  0}   
\definecolor{cligrn}  {RGB}{ 88,177, 88}   
\definecolor{cligrndk}{RGB}{ 28,115, 28}   
\definecolor{daered}  {RGB}{213, 78, 71}   
\definecolor{daereddk}{RGB}{148, 25, 18}   
\definecolor{onbg}    {RGB}{235,246,255}   
\definecolor{onborder}{RGB}{ 41,128,185}   
\definecolor{navycol} {RGB}{ 27, 38, 49}   


\fill[onbg, rounded corners=14pt]
    (3.6,-1.8) rectangle (11.4,1.8);
\draw[onborder, line width=2.2pt, rounded corners=14pt]
    (3.6,-1.8) rectangle (11.4,1.8);

\node[
  rectangle, rounded corners=8pt,
  minimum width=2.2cm, minimum height=2.2cm, inner sep=4pt,
  top color=liborg, bottom color=liborgdk,
  text=white, font=\bfseries, align=center,
  drop shadow={shadow xshift=2.5pt, shadow yshift=-2.5pt,
               opacity=0.45}
] (L) at (0,0) {\Huge $\mathcal L$\\[4pt]\small Library};

\node[
  rectangle, rounded corners=8pt,
  minimum width=2.2cm, minimum height=2.2cm, inner sep=4pt,
  top color=cligrn, bottom color=cligrndk,
  text=white, font=\bfseries, align=center,
  drop shadow={shadow xshift=2.5pt, shadow yshift=-2.5pt,
               opacity=0.45}
] (C) at (5.5,0) {\Huge $\mathcal C$\\[4pt]\small Client};

\node[
  rectangle, rounded corners=8pt,
  minimum width=2.2cm, minimum height=2.2cm, inner sep=4pt,
  top color=daered, bottom color=daereddk,
  text=white, font=\bfseries, align=center,
  drop shadow={shadow xshift=2.5pt, shadow yshift=-2.5pt,
               opacity=0.45}
] (D) at (9.5,0) {\Huge $\mathcal D$\\[4pt]\small Daemon};

\draw[arrstyle, <->] (L.east) -- (C.west);
\draw[arrstyle, ->]  (D.west) -- (C.east);

\node[font=\bfseries\large, text=darkgray, anchor=south west]
    at (1.75, 2.1) {On-Chain};

\node[font=\bfseries\large, text=darkgray, anchor=south west]
    at (8.5, 2.1) {Off-Chain};

\node[font=\bfseries\large, text=onborder, anchor=south west]
    at (-1.25, -2.1) {Proponent};

\node[font=\bfseries\large, text=onborder, anchor=south west]
    at (9.5, -2.5) {Opponent};

\draw[darkgray, line width=1pt, dash pattern=on 6pt off 4pt]
    (7.5, 2.6) -- (7.5,-2.5);
\end{tikzpicture}}
\caption{On-chain and off-chain components: the Library and Client contracts
         reside on-chain; the Daemon operates off-chain and interacts with
         the Client. Proponent represents the Library; Opponent consists of the on-chain Client and the off-chain Daemon. The Daemon is as an Externally Owned Account (EOA).}
\label{fig:onchain-offchain-v3}
\end{figure}


We follow an approach based on operational game semantics~\cite{Laird07,TzeGhica} to define a trace semantics for EVM inter-contract interactions, that is, behaviours that either initiate a new EVM execution frame or return to a previous one. Borrowing terminology from the client--library paradigm, we refer to a collection of known contracts as a \emph{library} ($\lL$), and to external contracts that use the library as a \emph{client} ($\cC$). We also include an off-chain entity, called a \emph{daemon} ($\dD$) who interacts with the client by issuing one-way calls with arbitrary time periods in between (see \Cref{fig:onchain-offchain-v3}). 

Our game semantics is given as an LTS over \emph{Proponent} configurations, modelling the library, and \emph{Opponent} configurations, modelling the client and daemon:
\begin{align*}
  \conf{\mathcal{A} ; T ; g ; K 
  \vdash \alpha ; \memo ; S^L_\nameN}_p
\qquad
  \conf{\mathcal{A} ; T ; g ; K 
  }_o
\end{align*}
where the annotations $p$ and $o$ are optional. We introduce the stack $K$ of frames $F$ built as follows, 
\begin{align*}
  F\ &::=\ (\CALL^{v_o}_{v_z},g,\alpha,\memo,E^L_\nameN)_p \mor (\CREATE,g,\alpha,\memo,E^L_\nameN)_p \mor \CALL_o
\end{align*}
again, with optional $o/p$ annotations.
Proponent stack frames are pushed on $K$ when Yul-EVM execution is stuck in a configuration
$
  \conf{E[\yultext{call}(\vec v)]\!\mid\! G ; L ; \nameN}$ or
 $ \conf{E[\yultext{create}(\vec v)]\!\mid\! G ; L ; \nameN}
$
and record the continuation point once the call/create is returned. Frames $\CALL_o$ are useful for general bookkeeping. 

On the Proponent side, a game configuration extends an EVM-Yul configuration with $K$, whereas 
on the Opponent side it consists only of the corresponding environment-level components together with $K$. We also exclude $\alpha$ in the latter case as there is no need to record in game configurations the current Opponent contract address.

Intuitively, the Proponent represents a known collection of contracts, whereas the Opponent represents arbitrary unknown clients with undefined code and off-chain capabilities that may interact with them. We call a library \emph{open} if execution may pass to external agents of unknown code, and \emph{closed} otherwise. Concretely, a library initially consists of a top-level Yul object, possibly containing nested Yul objects representing contracts to be deployed; evaluating this object may in turn deploy a collection of smart contracts. Thus, for a library $\lL$, initial game configurations are Proponent configurations of the form:\
$
  \conf{[\alpha_0\mapsto\conf{0;{v_0};\varepsilon;\varepsilon}]; T_0 ; g_0 ; \varepsilon 
    \vdash 
\alpha_0 ; \bot ; \lL^\varnothing_\varnothing}_p
$.
\begin{definition}[Trace Semantics]\label{def:trace:sem}
  Let $\lL$ be a library. Given an initial time $T_0$, gas $g_0$, {balance $v_0$} and deployment address $\alpha_0$, the semantics of $\lL$ is:
\[\gamesem{\lL} = \{ (t,\rho) \mid  
\conf{[\alpha_0\mapsto\conf{0;v_0;\varepsilon;\varepsilon}] ; T_0 ; g_0 
\vdash 
\alpha_0 ; \bot; \lL^\varnothing_\varnothing}_p
\trans{t} \rho\}\]
where configuration $\rho$ and trace $t$ are produced by a sequence of moves according to the game-semantic rules in this section, and where a \textbf{\em trace} is a sequence of move labels.
We say that $\gamesem{\lL}$ \textbf{\em fails} {(under $T_0,g_0,v_0,\alpha_0$)} if it contains a pair of the form\
$
\left(t,
\conf{\ldots \vdash \ldots ; E[\yultext{ASSERT}(0)]}_p\right)
$.\defEnd
\end{definition}

We next look at our game moves, categorised by the players involved in them: internal P-moves, moves between P and O, and visible moves between P and P.
Formally, a \textbf{\em move} is a transition in the game LTS we present below. A \textbf{\em move label} is a label that appears in a move, i.e.\ a transition label. Traces are sequences of move labels.
Here, and in the sequel, P and O is used as shorthand for Proponent and Opponent respectively. 
Finally, we revisit state maps $\mathcal A$ and addresses $\alpha$: internal addresses are owned by P, whereas external ones belong to O. We define domain sets
$\mathcal{A}\Pproj=\mathcal{A}^{-1}(\Accset)\ \text{ and }\
  \mathcal{A}\Oproj=\mathcal{A}^{-1}(\bullet)
$
to capture each of these address sets.

\begin{remark}[On secrets]\label{rem:secrets}
Given that EVM transactions are eventually recorded on the blockchain, and that hashing functions are public and deterministic, it is hard to keep secrets in Ethereum code. For example, we shall see in \cref{sec:PPmoves}, that addresses $\alpha$ can be dynamically created by Proponent in order to store newly created contracts in $\mathcal A$. Such addresses, even if not explicitly revealed to Opponent (e.g.\ by being passed as data in a call/return) can still be worked out by a resourceful Opponent (client+daemon) who studies the library contract code or simulates it on a test chain.
\\
On the other hand, it is reasonable to assume that Opponent cannot invert hashing functions, e.g.\ the contract below will be safe.
\begin{lstlisting}[language=Solidity]
 contract Foo {
   function bar(x) public {
     Yult.Assert(keccak256(x) != 0x084f026b9e198b90b1cd9eb3564f6e5733cc8d21cf3b70e71a5a891451e1b50b);
   }
 }
\end{lstlisting}
Our game model does not establish safety of the code above: Opponent can call \yultext{Foo.bar} with every possible argument, including the one leading to assertion failure. To capture such scenarios, a knowledge model would be needed, which is orthogonal to the treatment of open library code that our game semantics is targeted at. On the other hand, in our tool implementation, we indeed implement a heuristics-based knowledge model for Opponent. In general, note that even examples like the above are not easy to judge: the test value in the assertion is actually the hash of 42 so an Opponent could easily guess it by brute force attack.
\end{remark}

\begin{remark}[On client gas]\label{rem:opponent-gas}
  Computation on the EVM is based around the notion of \emph{gas}: almost every operation consumes gas. In contract function calls, the caller carves out an amount of gas from their current allowance and passes it on to the callee: this is the total gas that the callee would need to operate on for this call. In our semantics, as far as the Proponent is concerned (i.e.\ the library code), we have included precise gas calculations up to implementation of gas consuming operations. On the other hand, when control is passed to Opponent, the gas consumption is dependent on the computation steps occurring in client code. For that, we provide an upper bound: we assume that the client can operate with zero internal gas costs (i.e.\ internal client computations do not incur any such cost), but any calls that the client issues to library functions are costed as normal.
  This is a stringent bound, and leads to our semantics over-approximating actual library-client interactions,
  as typically clients would need to perform gas-costed internal operations before issuing certain library calls or returning certain data values. But, as there is no exact notion of what the least needed gas needed is in each such case, our assumption is necessary and, we argue, within reason.
\end{remark}

We discussed above how client gas consumption cannot be accurately assessed in an open semantics, and how our semantics over-approximates it. A consequence of the latter is that, in general, our semantics does not support control-flow branching that depends on gas.


We proceed to present game moves. In the rules below we use ellipsis ($\dots$) for configuration components that remain unchanged in a move (between source and target configuration).

\subsection{Internal Moves}
These are moves that happen within the Yul-EVM operational semantics and do not require the game apparatus. We cater for operations potentially changing the global state via custom $op$'s.

\[\begin{array}{@{}lll@{}}
\irule[Int][int]{
   \nbox{
    \conf{S \mid \conf{\mathcal{A};\ldots;g;\ldots;\memo};L ; \nameN} \to^* \conf{S' \mid \conf{\mathcal{A}';\ldots;g';\ldots;\memo'}; L' ; \nameN'}
    }
}{
  \nbox{
    \conf{ \mathcal{A}; \ldots;g;\ldots \vdash 
    \ldots ; \memo; S^L_\nameN }
    \trans{}
    \conf{ \mathcal{A}';\ldots;g';\ldots \vdash 
    \ldots ; \memo'; {S'}^{L'}_{\nameN'} }
  }
}
\end{array}\]
In the rule above, the target game configuration must be a stuck configuration.

\subsection{Proponent-Opponent Moves}
These are moves that pass control from Proponent to Opponent, and vice versa, by means of external contract calls and their corresponding returns. A special case arises when a library contract (i.e.\ P) returns when the stack is empty, that is, when there has been no precedent call by O. This occurs in a contract deployment phase, when executing the contract leads to its deployment in the state. From that point on, O can interrogate functions from the deployed contract (we refer to such calls as \emph{top level}), and each such call can initiate new calls from P to O, etc. 
\[\begin{array}{@{\!\!}c}
\irule[Deploy][p-deploy]{
   \mathcal{A}' = \mathcal{A}[\alpha_0.\mathsf{code} \mapsto bs]\\
    bs = \memo[v_1,v_1+v_2)
}{
  \nbox{
    \conf{\mathcal{A} ; \dots ; g; \varepsilon 
\vdash 
    \alpha_0 ; \memo ; E[\retop(v_1,v_2)]^L_\nameN}_p
    \trans{\deployl(\alpha_0,bs)}
    \conf{\mathcal{A}' ; \dots ; 0; \varepsilon 
   }_o\\
  }
}\\\\
    \irule[OCall][o-call]{
    m = \newCallMsg(g_c,\alpha_p,v, \alpha_o,g, bs)\\
    g = g_c\\
  m.\msgTarget=m.\msgCAddr=\alpha_p\in \mathcal A\Pproj\not\ni\alpha_o\\
  S = \mathcal{A}(\alpha_p).\accCode\\
    f \in \accabi(S)\\
    bs \in \{f.\abihash\}\times\bytesset\\
    \memo_{bs} = [\ioproj\mapsto bs]
}{
  \nbox{
    \conf{\mathcal{A};\ldots; g;K 
    }_o
    \trans{\ocall(m)}
    \conf{\mathcal{A}[ \alpha_p.\accBal \pluseq v][\alpha_o\mapsto\bullet];\ldots ;
    m.\msgGas; \CALL_o :: K 
    \vdash 
    \alpha_p;\memo_{bs};S^{\varnothing}_\varnothing}_p
  }
}\\\\
\irule[PORet][po-return]{
    bs = \memo[v_1,v_1+v_2)
}{
  \nbox{
    \conf{\ldots ; \CALL_o:: K 
    \vdash 
    \alpha ; \memo ; E[\retop(v_1,v_2)]^L_\nameN}_p
    \trans{\preturnl{o}(bs)}
    \conf{\ldots ; K 
    }_o
  }
}\\\\
\irule[POCall][po-call]{
   \vec{v} = g_c,\alpha_o,v,v_i,v_s,v_{o},v_z\\
   \memo',g' = \mathtt{extend}(\memo,\max(v_i+v_s,v_{o}+v_z))\\
   m = \newCallMsg(\vec v, \alpha, g, \memo')\\
    m.\msgTarget=m.\msgCAddr = \alpha_o \notin \mathcal{A}\Pproj\\
    \mathcal{A}' = \mathcal{A}[ \alpha.\accBal \minuseq v][\alpha_o\mapsto\bullet]\\
   g''\! = g-g'- m.\msgGas\geq0\\
    \memo'_\varepsilon=\memo'[\ioproj \mapsto \varepsilon] 
}{
  \nbox{
    \conf{\mathcal{A} ; T ; g ; K 
    \vdash 
    \alpha ; \memo ; E[\callop(\vec{v})]^L_\nameN}_p
    \trans{\pocalll(m)}
    \conf{\mathcal{A}' ; T ; m.\msgGas ; (\CALL^{v_o}_{v_z},  g'', \alpha , \memo'_\varepsilon, E^L_\nameN) :: K 
    }_o
  }
}\\\\
\irule[ORet][o-return]{
    {bs \in \bytesset}\\
    {\memo' = \memo\left[ [v_o,v_o+v_z) \mapsto bs,\ioproj\mapsto bs\right]}\\
    g''=g+g'
}{
  \nbox{
    \conf{\ldots; g; (\CALL^{v_o}_{v_z}, g'; \alpha , \memo , E^L_\nameN) :: K }_o
    \trans{\oreturnl(bs)}
    \conf{\ldots ; g''; K  \vdash 
    \alpha , \memo' , E^L_\nameN[1]}_p
  }
}\\[6mm]
\end{array}\]
We observe that \iref{p-deploy} and \iref{po-return} are almost identical, apart from the fact that \iref{p-deploy} updates the state by deploying new contract code and does not pop from the stack $K$ (actually, the rule is only applicable on an empty stack). Let us examine further the rules for calls and returns.
\begin{itemize}
  \item
    From an O-configuration, using \iref{o-call}, Opponent can call one of Proponent's functions (fetched from the ABI of one of P's contracts) and lead to a P-configuration that starts the evaluation of that function. Observe that function arguments are given as bytes and passed via the $\ioproj$ component of $\memo$.
    The move label constitutes a message $m$,
    which is built using the \texttt{callMsg} macro (itself left implicit here) out of: the call data $bs$, the current gas $g$ and the gas $g_c$ intended for the call, the addresses of the caller ($\alpha_o$) and callee ($\alpha_p$),\footnote{Technically, $m.\msgTarget$ is the message receiver while $m.\msgCAddr$ is the contract that is being called. These are the same for standard calls, but they differ if we include \texttt{\scriptsize delegatecall} in our model.}
    and the value $v$ to be transferred.
    Observe how we impose that $g=g_c$, i.e.\ Opponent must pass all available gas to Proponent. This is due to our assumption that Opponent does not incur internal gas costs (cf.\ \cref{rem:opponent-gas}), but also the fact that we aim for our semantics to explore all feasible paths (hence, Opponent should not hold back any gas from an O-call). The P-function is called with initial gas $m.\msgGas$, which is typically $g_c$ minus a gas cost for issuing the call.
When the called P-function completes its execution, it will return, using \iref{po-return}, to an O-configuration. In this case, the move label consists of the return data. 
  \item From a P-configuration, using \iref{po-call}, Proponent can also call an Opponent function. The function call information (call data, value, addresses, etc.) are embedded inside a newly created message (using again the \texttt{callMsg} macro) that becomes the move label. Observe the gas calculations ensuring that the correct amount of gas is assigned to the call and its continuation, and that there is enough gas overall. Note also that the memory is extended to accommodate the return data, but also the input data (in case either of the designated blocks is not fully allocated already).
A PO-call also stores the current continuation in $K$.
 Opponent can return a PO-call, via \iref{o-return}, leading execution to continue according to the stored continuation. The return data is stored in the previously designated output block (its interval being part of the continuation), and the I/O buffer. By convention, a default value 1 is inserted in the restored evaluation context $E$: we focus herein only on successful (non-reverting) executions leading to assertion violations.
\end{itemize}

\subsection{Proponent-Proponent Moves}\label{sec:PPmoves}
A stuck configuration may be also be calling a P-function, i.e.\ one stored in a contract belonging to Proponent. Such scenarios lead to another set of call and return rules, this time between P and P. These calls are `internal' to the contract, but we use visible moves for them since they are recorded on the blockchain; in contrast, the internal moves (without quotes) that we saw above remain externally invisible.
%
\[\begin{array}{@{\!\!\!\!\!\!\!}c}
\irule[PPCall][pp-call]{
   \vec{v} = g_c,\alpha_p,v,v_i,v_s,v_{o},v_z\quad
   \memo',g' = \mathtt{extend}(\memo,\max(v_i+v_s,v_{o}+v_z))\quad
   m = \newCallMsg(\vec v, \alpha, g, \memo')\\
   m.\msgTarget = m.\msgCAddr=\alpha_p \in {\mathcal A}\Pproj\quad
   \mathcal{A}' = \mathcal{A}[\alpha.\accBal \minuseq v][\alpha_p.\accBal \pluseq v]\quad
    g'' = g-g'- m.\msgGas\geq 0\\
    \memo'_\varepsilon = \memo'[\ioproj \mapsto \varepsilon]\\
    bs = \memo'[v_i,v_i+v_s)\\
    \memo_{bs}=[\ioproj\mapsto bs]\\
   S = \mathcal{A}(\alpha_p).\accCode
}{
  \nbox{
    \conf{\mathcal{A} ; T ; g ; K 
    \vdash 
    \alpha ; \memo ; E[\callop(\vec{v})]^L_\nameN}
    \trans{\ppcalll(m)}
    \conf{\mathcal{A}' ; T ; m.\msgGas ;  (\CALL^{v_o}_{v_z}, g'' , \alpha , \memo'_\varepsilon , E^L_\nameN)\!::\! K 
    \vdash 
    \alpha_p;\memo_{bs};S^{\varnothing}_\varnothing}
  }
}\\\\
\irule[Create][p-create]{
   \vec v = v,v_1,v_2\\
   \memo',g' = \mathtt{extend}(\memo,v_1+v_2)\\
   m = \newCreateMsg(\vec{v},\alpha,g,\memo',\alpha')\\
    \mathcal{A}' = \mathcal{A}[\alpha.\accNonce\pluseq1][ \alpha.\accBal \minuseq v] \uplus \left\{\alpha' \mapsto \conf{0 ,v ,\memo'[v_1,v_1+v_2) , \varepsilon}\right\}\\
   \alpha' = \mathtt{fresh}(\alpha,\mathcal A(a.\accNonce))\quad
   g'' = g-g'-m.\msgGas\geq0\quad
\memo'_\varepsilon=\memo'[\ioproj\mapsto\varepsilon]\quad
   S = \mathcal{A}'(\alpha').\accCode
}{
  \nbox{
    \conf{\mathcal{A} ; T ; g ; K 
    \vdash 
    \alpha ; \memo  ; E[\createop(\vec v)]^L_\nameN}
    \trans{\createl(m)}
    \conf{\mathcal{A}' ; T ; m.\msgGas; (\CREATE,  g'' , \alpha , \memo'_\varepsilon , E^L_\nameN)\! ::\! K 
    \vdash 
    \alpha';\bot;S^{\varnothing}_\varnothing}
  }
}\\\\
\irule[PPRet1][pp-return1]{
    bs = \memo[v_1,v_1+v_2)\\
    g''=g+g'\\
   \memo'' = \memo'\left[ [v_o,v_o+v_z) \mapsto bs,\ioproj\mapsto bs\right]\\
}{
  \nbox{
    \conf{\ldots ; g; (\CALL^{v_o}_{v_z}, g', \alpha' , \memo' , {E'}^{L'}_{\nameN'})\!::\! K 
    \vdash 
    \alpha ; \memo ; E[\retop(v_1,v_2)]^L_\nameN}
    \trans{\preturnl{p}}
    \conf{\ldots ; g'' ; K 
    \vdash 
    \alpha' , \memo'', {E'}^{L'}_{\nameN'}[1]}\\
  }
}\\\\
\irule[PPRet2][pp-return2]{
    g''=g+g'
}{
  \nbox{
    \conf{\ldots ; g ; (\CREATE, g', \alpha' , \memo' , {E'}^{L'}_{\nameN'})\!::\! K 
    \vdash 
    \alpha ; \memo ;E[\retop(v_1,v_2)]^L_\nameN}
    \trans{\preturnl{p}}
    \conf{\ldots ; g''; K 
\vdash 
    \alpha' , \memo' , {E'}^{L'}_{\nameN'}[\alpha]}\\
  }
}\\[6mm]
\end{array}\]
We notice that \textsc{Create} creates a new contract at address $\alpha'$, itself created by the deterministic function
\[
\mathtt{fresh}(\alpha,v)=\keccak(\rlp(\alpha, v))
\]
where $\rlp$ is Ethereum's serialization function.
Note that $\alpha'$ needs to be fresh, i.e.\ not in the domain of $\mathcal A$, for the rule to fire.
Otherwise, the rule is similar to \textsc{PPCall} and in particular it calls the code of the newly created contract. We briefly comment on these two rules and their corresponding returns.
\begin{itemize}
\item Each  \textsc{PPCall} and \textsc{Create} lead to calling P-code. The kind of rule used is registered in the stack frame that is pushed to accommodate for the different respective continuations. There are also differences in the arguments used in each case. Otherwise, the way that gas and state are updated is similar to the one for \textsc{POCall}.
\item Moves \textsc{PPRet1} and \textsc{PPRet2} match respectively \textsc{PPCall} and \textsc{Create}. We note that when returning from a PP-call the continuation value is 1, whereas a create returns with the new contract's address.
\end{itemize}

\subsection{Ending a Transaction and Waiting Moves}

EVM computation is transaction-based: contract calls need to be completed in a single transaction, and a block of completed transactions is then stored on the blockchain.
An Opponent configuration is \emph{top-level} when its stack $K$ is empty. Recall that e.g.\ \textsc{Deploy} can be followed by a top-level O-call, i.e.\ an O-call from a top-level configuration. A
Proponent return of a top-level call, which reaches a top-level configuration, may not necessarily complete a transaction. E.g.\ typically a client contract may call several library functions in a single transaction. Instead, it is Opponent who can end a transaction, using the following move.
%
\[\begin{array}{@{}lll@{}}
\irule[EndTx/Wait][wait]{
  i,g' \in \Uintset\\
i\geq 0
}{
  \nbox{
    \conf{\mathcal{A} ; T ; g ; \varepsilon }_o
    \trans{\waitl(i,g)}
    \conf{\mathcal{A} ; T+i ; g' ; \varepsilon}_o
  }
}
\end{array}\]
Observe that gas is updated to an arbitrary new value, which is to be used for a new transaction. Note also how time may increase arbitrarily, i.e.\ Opponent may wait before starting a new transaction. Waiting is an important feature in time-sensitive contracts, e.g.\ a certain time period is required between joining and voting in a DAO contract. Finally, note that \textsc{EndTx/Wait} is the only move that directly involves the daemon $\dD$ (cf.\ \Cref{fig:onchain-offchain-v3}): while it is the client code which ends the transaction, it is the daemon which determines the wait time.


\section{The \yult Tool}
\label{sec:tool}
\begin{figure}
    \centering
    \begin{tikzpicture}[
    node distance=0.6cm and 0.6cm,
    every node/.style={font=\scriptsize, align=center, inner sep=3.5pt, minimum height=0.95cm},
    solidbox/.style={draw=prborder, fill=process, rectangle, thick, minimum width=1.5cm, rounded corners=8pt},
    file/.style={draw, rectangle, thick, minimum width=1.5cm},
    dashedbox/.style={draw, dashed, rectangle, thick, minimum width=1.5cm},
    boldbox/.style={draw=prborder, fill=process, rectangle, line width=1.8pt, minimum width=1.1cm, rounded corners=8pt},
    arr/.style={-{Stealth[length=4pt]}, thick},
]

\definecolor{process} {RGB}{235,246,255}   
\definecolor{prborder}{RGB}{ 41,128,185}   

\node[file] (input)
{\textbf{INPUT}\\Solidity Sources};

\node[solidbox, right=of input] (compiler)
    {Solidity 0.8.x\\Compiler};

\node[solidbox, right=of compiler] (preproc)
    {Preprocessing};

\node[boldbox, right=of preproc, yshift=0cm] (tool)
    {\yult};

\node[right=of tool] (output)
{\textbf{OUTPUT}\\Witness Trace or\\Bound Exceeded};

\node[dashedbox, above=of input] (instr)
    {Instrumentation\\(optional)};

\node[solidbox, above=of compiler] (abiext)
    {ABI\\Extraction};

\node[dashedbox, above=of tool, rounded corners=8pt] (prune)
    {Manual ABI Pruning\\(optional)};

\node[dashedbox, above=of output] (cli)
    {CLI Parameters\\(optional)};

\draw[arr] (instr.south) --  (input.north);
\draw[arr] (input)  -- node[above, inner sep=0pt, yshift=-0.3cm] {.sol} (compiler);
\draw[arr] (abiext)  -- node[above, inner sep=0pt, yshift=-0.3cm] {.json} (prune);
\draw[arr] (compiler.north) -- node[right, inner sep=0pt, xshift=0.1cm] {.json} (abiext.south);
\draw[arr] (compiler) -- node[above, inner sep=0pt, yshift=-0.3cm] {.yul} (preproc);
\draw[arr] (prune.south) -- node[right, inner sep=0pt, xshift=0.1cm] {.json} (tool.north);
\draw[arr] (preproc.east) -- node[above, inner sep=0pt, yshift=-0.3cm] {.yul} (tool.west);
\draw[arr] (cli.south) -- ($(tool.north east)+(-0.09cm,-0.09cm)$);
\draw[arr] (tool) -- (output);

\end{tikzpicture}
    \caption{Overview of the \yult processing pipeline.}\label{graphic:pipeline}
\end{figure}

We have implemented the game semantics of the previous sections in a tool called \yult\footnote{YulTracer (v0.2.1). \url{https://doi.org/10.5281/zenodo.19868351}}, which sits in the heart of an analysis pipeline, summarised in \Cref{graphic:pipeline}.
The inputs are Solidity sources targeting a 0.8-series compiler.
Users may optionally instrument the sources with \emph{assertions}, facilitated by a |Yult| library.\footnote{See \texttt{GUIDE.md} in the \yult source code.}
calls to \yult hook functions expressing safety properties directly in Solidity,
and \emph{harnesses}, e.g.\ a deployer contract that constructs and initialises the system under test.

The instrumented sources are compiled with the Solidity compiler ($\geq$0.8.0), which produces two artefacts:
the Yul IR for each contract, and a JSON file containing the combined ABI of the project.
The Solidity compiler emits Yul IR with nested objects: the top-level contract
(e.g.\ the deployer) produces a single Yul object that contains all dependencies,
so \yult takes a single file as input for the entire project.
Before being passed to the core \yult engine, this file undergoes preprocessing,
including linking application libraries and a special |Yult| library (which provides an assertion function and other utilities).
For the ABI, the relevant entries are extracted from the JSON artefact.
Users may optionally \emph{prune} the ABI to retain only functions relevant to the analysis;
this reduces the branching factor of the game search and is a main lever for controlling analysis time on larger contracts.
The preprocessed Yul file and ABI are passed to the \yult engine together with \emph{command line interface} (CLI) parameters that set the exploration bounds. These are described in \Cref{tab:cli}.

\begin{table}[t]
    \centering
    \tiny
    \begin{tabularx}{0.95\textwidth}{@{} l l X | l l X @{}}
        \toprule
        \textbf{Parameter} & \textbf{Default} & \textbf{Description} &
        \textbf{Parameter} & \textbf{Default} & \textbf{Description} \\
        \midrule
        \multicolumn{3}{@{}l}{\itshape Opponent Call Settings} \\
        \midrule
        Call Bound             & 2           & Calls per proponent function.
        &
        Opponent Address Count & 1           & Opponent addresses.
        \\
        Call-Stack Bound       & 3           & Open O-to-P calls.
        &
        Opponent Balance       & 10 ETH      & Start balance per O-address.
        \\        
        Opponent Spending      & 1000 wei    & Value opponent may send per payable call (or zero).
        &        
        Opponent Uint256 Domain  & $\{0,1,1000\}$     & \multirow{2}{25mm}{\vrule~~\parbox{23.7mm}{Initial uint256 and  address values in O's knowledge; they grow in each trace.}}
        \\
        Opponent Return Values
                                 & false             & Whether opponent may choose return values.
        &
        \nbox{\text{Opponent Address Domain}\\\text{Observable Trace Bound}}  
        &
        \nbox{\text{empty}\\\text{20}}
        & 
        \nbox{\\\text{Observable trace moves.}}
        \\
        \addlinespace[1ex]
        \multicolumn{3}{@{}l}{\itshape Opponent Time Settings}
        &
        \multicolumn{3}{l}{\itshape Deployment Transaction Settings} 
        \\
        \midrule
        Wait Time              & 7 days               & Time added per “wait” move.
        &
        Deploy Gas             & 30000 ETH            & Gas limit for deployment.
        \\
        No Waiting             & false                & Disables waiting moves (equiv.\ to Wait Time = 0).
        &
        Deploy Address         & {0x0102…0A}   & Address where the top-level contract is deployed. 
        \\
        Wait First             & false                & Opponent explores waiting before other moves.
        &
        Deploy Value           & 123456789 ETH        & ETH sent with the deployment transaction.
        \\
        Max Wait               & 22 days              & Bound on total wait per trace.
        \\
        \bottomrule
    \end{tabularx}
    \caption{\yult Command Line Interface Parameters by category.}
    \label{tab:cli}
\end{table}

\yult performs a \emph{concrete}, not symbolic, exploration of game semantics traces, avoiding frequent expensive calls to SMT solvers.
During this exploration, the opponent calls all user-designated functions on any contract address that exposes them, including addresses deployed dynamically during the exploration itself.
The concrete values used in opponent calls are drawn from the |address| and |uint| domains specified via the CLI.
During each game trace, the |uint| domain grows with values observed from proponent calls,
and the address domain through calls to the library function |Yul.revealAddress|. 
This realises a pragmatic \emph{opponent knowledge model} (see motivating example in \cref{sec:example:secrets}).

\yult exhaustively explores all opponent moves in game traces within the above bounds.
It rejects a contract if it encounters a trace in which (i) a \texttt{Yult.Assert} present in the generated code is violated, (ii) the proponent attempts to send Ether from an address with insufficient balance, signalling the contract can be drained, or (iii) the proponent issues a \yulinline{delegatecall} to an opponent-controlled address,  signalling a fundamental access control error.
When a contract is rejected, \yult produces a witness trace demonstrating the violation; otherwise it terminates having exhausted the bounded search space.

\section{Evaluation}
\label{sec:evaluation}

We evaluated \yult on benchmarks and real-world case studies.
Although \yult is a general assertion reachability checker, here we focus on reentrancy, as it is a major class of real-world exploits~\cite{pcaversaccio-reentrancy-attacks}
and a challenging type of vulnerabilities to detect precisely.
In \cref{sec:beyond} we discuss the use of \yult to discover access control vulnerabilities.

The benchmarks are used to compare \yult against \textsc{Mythril}, \textsc{SMTChecker}, and \textsc{Slither}.
In the case studies we examine two historical reentrancy attacks
from a collection of known incidents~\cite{pcaversaccio-reentrancy-attacks}:
The DAO~\cite{dao-victim, dao-reentrancy-overview}
and Predy Finance~\cite{predy-victim, predyfinance-postmortem}.\footnote{%
We additionally examine dForce Lendf.Me~\cite{lendfme-victim, slowmist-lendfme-analysis} in \cref{sec:lendf} of the appendix.}
Unlike the benchmarks, these were large deployed systems that caused significant financial losses,
and they demonstrate how \yult can be applied in practice with domain-specific guidance.

The evaluation was conducted on a laptop with an Apple M1 CPU, 16\,GB RAM,
macOS 14.7.5, Solidity Compiler |0.8.34+commit.80d5c536.Darwin.appleclang|,
and OCaml 4.14.2.


\subsection{Gigahorse Benchmarks}
\label{sec:gigahorse}
We consider here the 23 Solidity source code benchmarks that involve reentrancy from the Gigahorse suite~\cite{grech-gigahorse-benchmarks},
a fork of the SmartBugs suite~\cite{smartbugs-framework}.
For each benchmark we
upgraded sources written in Solidity 0.4–0.5 to 0.8,\footnote{See~\cref{sec:upgrades} for the source code upgrades applied.}
compiled them with a recent version of the Solidity compiler ($\geq$0.8.0), and analysed them with the \yult toolchain.

\begin{table}
    \catcode`\_=12
    \footnotesize
    \centering
    \parbox{0.50\linewidth}{\begin{filecontents*}{experiments/gigahorse.csv}
Benchmark	Sol	Yul	ABI	Uint256 Dom.
0x01f8...d3f	44	1181	7	$10^{18}$
0x23a9...eb4	55	1113	5	$10^{18}{+}1$
0x4e73...106	68	577	8	$10^{18}$
0x561e...f31	70	577	8	1
0x627f...839	80	654	10	$10^{18}{+}1$
0x7541...615	63	1178	5	0, $2{\cdot}10^{18}{+}1$
0x7a87...782	65	1220	7	$10^{18}$
0x8c77...344	58	1112	5	$10^{18}{+}1$
0x941d...e9e	61	1123	5	$10^{18}{+}1$
0x96ed...78b	67	1195	6	1
0xaae1...0b8	70	577	8	1
0xb5e1...d12	58	1112	5	$10^{18}{+}1$
bonus	24	438	3	1
cross_function	28	452	3	1
dao	18	265	2	1
etherbank	16	299	3	$10^{18}{+}1$
etherstore	18	413	5	$10^{18}$
insecure	13	244	2	1
modifier_reentrancy	24	410	2	0x992\ldots{}282
reentrance	20	345	5	1
simple	15	296	3	1
simple_dao	14	316	3	1
spank_chain	662	7506	17	1
\end{filecontents*}
\csvautobooktabular[separator=tab]{experiments/gigahorse.csv}}\quad
    \parbox{0.30\linewidth}{\fbox{\parbox{\linewidth}{\footnotesize
          Sol and Yul LoC exclude comments and empty lines.
      ABI: number of functions.\\\\
          \textbf{Rest of CLI parameters:}\\
    Call Bound = 2;\\
    Stack Bound = 3;\\
    Trace Bound = 12;\\
    Deploy P-Balance = 0 ETH;\\
    O-Balance = 10 ETH;\\
    O-Calls value = $N$ wei;\\
    Wait time = 1 day;\\
    Max time delay = 1 day;\\
    where $N$ matches the non-zero value in the Uint256 domain.}}
    \\ \\ \\ \\ \\ \\ \\ \\ \\ \\ \\ \\
}
    \caption{Parameters of the vulnerable Solidity benchmarks involving reentrancy in the Gigahorse suite.}\label{tab:gigahorse}
\end{table}

\Cref{tab:gigahorse} summarises information about these benchmarks. The first three columns show measures of complexity:
22 benchmarks are relatively small programs (13--80 Solidity LoC) 
and one is a larger real-world case, SpankChain~\cite{spankchain-victim, spankchain-hack-explained} (662 LoC). 
Several benchmarks compile to over a thousand lines of Yul, the input language of \yult.
The number of ABI functions in the benchmarks is a measure of exploration complexity, determining the number of possible calls that the opponent can perform at each game move.
All but two examples had single-digit ABI functions.

The CLI parameters used for all benchmarks are shown in the box next to \cref{tab:gigahorse}.
These are bounds for opponent calls per function, call stack size, number of external moves in a trace, maximum time delays in a trace, as well as
initial Ether balance of the proponent and opponent contracts, and the amount of Ether that the opponent may use as transfer value in calls.
All benchmarks used the full ABI, as it was generally small.
Moreover, for each benchmark we briefly inspected the code to identify any specific hard-coded numbers needed to be provided in opponent calls,
including ETH amounts in wei ($1\text{ETH} = 10^{18}wei$). These are shown in the last column of \cref{tab:gigahorse}
and were provided to \yult by initialising the Uint256 domain via the CLI.

While most benchmarks were analysed directly after upgrading to Solidity 0.8, three benchmarks required minor manual modifications:
|reentrancy_insecure| and |reentrancy_cross_function| both implement a simple ETH bank with a vulnerable |withdrawBalance| method but no corresponding |deposit|, making the vulnerability unexploitable. We added a minimal \texttt{deposit} function to make the intended vulnerabilities exploitable.

Additionally, in \texttt{modifier\_reentrancy} a reentrancy vulnerability arises from an external call inside a |supportsToken| modifier:
\begin{lstlisting}[language=Solidity]
 contract ModifierEntrancy {
   ...
   // <yes> <report> REENTRANCY
   function airDrop() hasNoBalance supportsToken public {
     tokenBalance[msg.sender] += 20;
     Yult.Assert(tokenBalance[msg.sender] == 20);(*@\label{ln:benchME:3}@*)
   }
   modifier supportsToken() {
     Yult.printHex(keccak256(abi.encodePacked("Nu Token")));(*@\label{ln:benchME:2}@*)
     require(keccak256(abi.encodePacked("Nu Token")) == Bank(msg.sender).supportsToken()); _;(*@\label{ln:benchME:1}@*)
   }
   ...
 }
\end{lstlisting}
The |require| on line~\ref{ln:benchME:1} succeeds only if the opponent
(|msg.sender|) returns the Keccak hash of a specific string literal.
To satisfy this, we instrumented the code with the \yult print hook on line~\ref{ln:benchME:2},
revealed the expected value, and added it to the opponent's Uint256 domain via the CLI.
Moreover, the vulnerability lies purely in a balance invariant: an address should be recorded as holding either 0 or 20 tokens.
To capture this violation, we introduced the assertion on~\cref{ln:benchME:3}.

All other benchmarks were run through \yult unmodified. \yult detected all vulnerabilities either because of the assertions introduced in the benchmarks described above, or, in the remaining benchmarks, because the contracts attempted to transfer more Ether than they held due to their encoded vulnerabilities.
We examine \yult's performance in the following section, comparing it against that of other tools.


\subsection{Performance and Comparison with Other Tools}

To evaluate recall, precision, and runtime performance, we compare \yult against \textsc{Mythril}~\cite{mythril},
the Solidity compiler's built-in \textsc{SMTChecker}~\cite{smtchecker} in both CHC and BMC modes,
and \textsc{Slither}~\cite{slither}.
These tools represent complementary approaches to smart contract vulnerability analysis:
\textsc{Mythril} is a symbolic execution tool for EVM bytecode with embedded static analysis features;
\textsc{SMTChecker} in BMC and CHC modes represents bounded model checking and Horn-clause solving, respectively;
and \textsc{Slither} is a static analysis framework for Solidity source code.
All three tools are free and open-source, actively maintained by major organisations in the Ethereum ecosystem.
\textsc{Mythril} and \textsc{SMTChecker} additionally support \emph{assertion reachability checking}, which is
a main mode of analysis in \yult, while \textsc{Slither} relies solely on its built-in static detectors.

\subsubsection*{Methodology}
We modify the 23 Gigahorse reentrancy benchmarks in two variant sets:
One variant, used to evaluate recall across all tools, contains the \emph{vulnerable} contracts, as upgraded to Solidity~0.8 in \cref{sec:gigahorse}, annotated with matching Solidity and \yult assertions. These assertions allow assertion-checking tools such as \textsc{SMTChecker} to flag the vulnerability.
A second variant, used to evaluate precision, contains manually patched \emph{safe} versions of the programs in the first variant. In these programs the reentrancy
vulnerabilities were fixed by standard techniques---e.g.\ by applying the checks-effects-interactions
pattern or a reentrancy guard.

 \begin{table}[t]
    \catcode`\_=12
    \footnotesize
    \centering
    \resizebox{\linewidth}{!}{%
    \begin{tabular}{l ll ll ll ll ll}
      \toprule
      & \multicolumn{2}{c}{\textsc{Mythril}}
      & \multicolumn{2}{c}{\textsc{SMTChecker}-CHC}
      & \multicolumn{2}{c}{\textsc{SMTChecker}-BMC}
      & \multicolumn{2}{c}{\textsc{Slither}}
      & \multicolumn{2}{c}{\yult} \\
      \cmidrule(lr){2-3}\cmidrule(lr){4-5}\cmidrule(lr){6-7}%
      \cmidrule(lr){8-9}\cmidrule(lr){10-11}
      Benchmark
      & \multicolumn{1}{c}{vuln.} & \multicolumn{1}{c}{safe} 
      & \multicolumn{1}{c}{vuln.} & \multicolumn{1}{c}{safe} 
      & \multicolumn{1}{c}{vuln.} & \multicolumn{1}{c}{safe} 
      & \multicolumn{1}{c}{vuln.} & \multicolumn{1}{c}{safe} 
      & \multicolumn{1}{c}{vuln.} & \multicolumn{1}{c}{safe} \\
      \midrule
      \begin{filecontents*}{experiments/comparison.csv}
Benchmark,mythril (v),mythril (s),CHC (v),CHC (s),BMC (v),BMC (s),slither (v),slither (s),yultracer (v),yultracer (s)
0x01f...d3f,\rejected 16:36.1,\rejected 19:13.9,\rejected 00:01.5,\rejected 00:01.9,\rejected 00:00.2,\rejected 00:00.1,\rejected 00:00.9,\passed 00:00.5,\rejected 00:00.6,\passed 00:32.8
0x23a...eb4,\rejected 05:25.7,\rejected 05:44.6,\rejected 00:01.1,\rejected 00:01.4,\rejected 00:00.1,\rejected 00:00.1,\rejected 00:00.5,\passed 00:00.5,\rejected 00:00.1,\passed 00:02.3
0x4e7...106,\rejected 10:52.0,\rejected 10:25.5,\rejected 03:22.6,\passed 60:01.6,\rejected 00:00.1,\rejected 00:00.1,\rejected 00:00.5,\passed 00:00.5,\rejected 00:02.4,\passed 00:08.8
0x561...f31,\rejected 10:03.4,\rejected 11:07.4,\passed 60:01.3,\passed 60:01.2,\rejected 00:00.1,\rejected 00:00.1,\rejected 00:00.5,\passed 00:00.5,\rejected 00:02.4,\passed 00:08.8
0x627...839,\rejected 09:06.7,\rejected 08:51.3,\rejected 00:06.9,\passed 00:13.5,\rejected 00:00.1,\rejected 00:00.1,\rejected 00:00.5,\passed 00:00.5,\rejected 00:00.6,\passed 02:46.6
0x754...615,\passed 35:21.6,\passed 16:32.3,\rejected 00:01.0,\rejected 00:01.1,\rejected 00:00.1,\rejected 00:00.1,\rejected 00:00.5,\passed 00:00.5,\rejected 00:00.4,\passed 01:39.7
0x7a8...782,\rejected 06:28.7,\rejected 07:07.0,\rejected 00:01.2,\rejected 00:01.1,\rejected 00:00.1,\rejected 00:00.1,\rejected 00:00.5,\passed 00:00.5,\rejected 00:00.1,\passed 00:02.8
0x8c7...344,\rejected 08:42.1,\rejected 08:41.4,\rejected 00:01.0,\rejected 00:01.1,\rejected 00:00.1,\rejected 00:00.1,\rejected 00:00.5,\passed 00:00.5,\rejected 00:00.1,\passed 00:05.7
0x941...e9e,\rejected 09:15.9,\rejected 07:56.2,\rejected 00:01.0,\rejected 00:00.9,\rejected 00:00.1,\rejected 00:00.1,\rejected 00:00.5,\passed 00:00.5,\rejected 00:00.1,\passed 00:04.5
0x96e...78b,\rejected 23:47.6,\rejected 24:37.4,\rejected 00:01.1,\rejected 00:01.0,\rejected 00:00.1,\rejected 00:00.1,\rejected 00:00.5,\passed 00:00.5,\rejected 00:00.1,\passed 00:28.1
0xaae...0b8,\rejected 10:49.0,\rejected 10:20.0,\passed 60:00.7,\passed 60:02.4,\rejected 00:00.1,\rejected 00:00.1,\rejected 00:00.5,\passed 00:00.5,\rejected 00:02.4,\passed 00:08.8
0xb5e...d12,\rejected 09:01.4,\rejected 08:05.7,\rejected 00:01.1,\rejected 00:01.3,\rejected 00:00.1,\rejected 00:00.1,\rejected 00:00.5,\passed 00:00.5,\rejected 00:00.1,\passed 00:05.8
bonus,\rejected 01:25.1,\rejected 01:38.6,\passed 60:07.9,\passed 60:02.6,\rejected 00:00.1,\rejected 00:00.1,\rejected 00:00.5,\passed 00:00.5,\rejected 00:01.3,\passed 00:02.1
cross_function,\rejected 00:48.4,\rejected 00:35.2,\passed 00:00.9,\passed 00:01.1,\rejected 00:00.1,\rejected 00:00.1,\rejected 00:00.5,\passed 00:00.4,\rejected 00:02.7,\passed 00:07.7
dao,\rejected 00:17.1,\rejected 00:15.1,\passed 60:01.9,\passed 60:08.1,\rejected 00:00.1,\rejected 00:00.1,\rejected 00:00.5,\passed 00:00.4,\rejected 00:00.4,\passed 00:01.4
etherbank,\rejected 00:26.2,\rejected 00:20.8,\rejected 60:01.3,\passed 00:00.4,\rejected 00:00.1,\rejected 00:00.1,\rejected 00:00.5,\passed 00:00.4,\rejected 00:00.1,\passed 00:00.1
etherstore,\rejected 01:02.8,\rejected 00:45.7,\rejected 00:03.9,\passed 00:00.3,\rejected 00:00.1,\rejected 00:00.1,\rejected 00:00.5,\passed 00:00.4,\rejected 00:00.1,\passed 00:00.1
insecure,\rejected 00:15.6,\rejected 00:14.0,t/o,\passed 00:00.6,\rejected 00:00.1,\rejected 00:00.1,\rejected 00:00.5,\passed 00:00.4,\rejected 00:00.6,\passed 00:00.8
modifier_reentrancy,\rejected 00:09.3,\rejected 00:09.9,\rejected 60:00.5,\passed 00:00.5,\passed 00:00.1,\passed 00:00.1,\rejected 00:00.5,\rejected 00:00.4,\rejected 00:00.1,\passed 00:00.1
reentrance,\rejected 01:14.3,\rejected 01:17.4,\rejected 00:05.8,\passed 00:00.5,\rejected 00:00.1,\rejected 00:00.1,\rejected 00:00.4,\passed 00:00.6,\rejected 00:00.4,\passed 00:01.5
simple,\rejected 00:22.3,\rejected 00:19.3,\rejected 00:04.2,\passed 00:00.4,\rejected 00:00.1,\rejected 00:00.1,\rejected 00:00.5,\passed 00:00.4,\rejected 00:00.6,\passed 00:00.8
simple_dao,\rejected 00:58.3,\rejected 00:38.2,\rejected 00:03.1,\passed 00:00.3,\rejected 00:00.1,\rejected 00:00.1,\rejected 00:00.5,\passed 00:00.4,\rejected 00:00.1,\passed 00:00.3
spank_chain,t/o,t/o,t/o,t/o,\rejected 00:02.4,\rejected 00:02.4,\rejected 00:01.7,\rejected 00:01.7,\rejected 00:00.4,\passed 00:02.2
Rejected,91.3\%,91.3\%,69.6\%,34.8\%,95.7\%,95.7\%,100.0\%,8.7\%,100.0\%,0.0\%
Passed,4.3\%,4.3\%,21.7\%,60.9\%,4.3\%,4.3\%,0.0\%,91.3\%,0.0\%,100.0\%
Timeout,4.3\%,4.3\%,8.7\%,4.3\%,0.0\%,0.0\%,0.0\%,0.0\%,0.0\%,0.0\%
Average time,07:23.2,06:35.3,17:20.5,13:40.1,00:00.2,00:00.2,00:00.5,00:00.5,00:00.7,00:17.0
Total time,162:30.7,144:58.0,364:10.8,300:44.3,00:05.9,00:05.8,00:13.4,00:13.0,00:17.2,06:32.8
\end{filecontents*}
\csvreader[
        separator=comma,
        late after line=\\,
        late after last line=\\\midrule,
        filter test={\ifnumless{\thecsvinputline}{25}}
      ]{experiments/comparison.csv}%
      {1=\benchmark, 2=\mythrilv,   3=\mythrilsafe,
       4=\smtchcv,   5=\smtchcsafe,
       6=\smtbmcv,   7=\smtbmcsafe,
       8=\slitherv,  9=\slithersafe,
       10=\yultv,   11=\yultsafe}%
      {\benchmark & \mythrilv & \mythrilsafe
                  & \smtchcv  & \smtchcsafe
                  & \smtbmcv  & \smtbmcsafe
                  & \slitherv & \slithersafe
                  & \yultv    & \yultsafe}
      \csvreader[
        separator=comma,
        late after line=\\,
        late after last line=\\\bottomrule,
        filter test={\ifnumgreater{\thecsvinputline}{24}}
      ]{experiments/comparison.csv}%
      {1=\benchmark, 2=\mythrilv,   3=\mythrilsafe,
       4=\smtchcv,   5=\smtchcsafe,
       6=\smtbmcv,   7=\smtbmcsafe,
       8=\slitherv,  9=\slithersafe,
       10=\yultv,   11=\yultsafe}%
      {\benchmark & \mythrilv & \mythrilsafe
                  & \smtchcv  & \smtchcsafe
                  & \smtbmcv  & \smtbmcsafe
                  & \slitherv & \slithersafe
                  & \yultv    & \yultsafe}
    \end{tabular}}%
    \caption{Tool comparison on the 23 Gigahorse reentrancy benchmarks.
             \rejected = vulnerability reported; \passed = no issue; runtimes in min:sec;
             t/o = 60\,min timeout.
             Safe variants test for false positives.}\label{tab:comparison}
  \end{table}

We ran each tool on the vulnerable and safe variants and recorded their runtimes and output.
The tool versions used were: \textsc{Mythril}~v0.24.8, \textsc{Slither}~v0.11.5, \textsc{SMTChecker} as shipped with \texttt{solc}~v0.8.34, and Z3~v4.15.2 as the underlying SMT solver.
\yult was run with the same parameters described in \cref{sec:gigahorse}.
To speed up execution, \textsc{Mythril} was run only with the relevant detection modules for these benchmarks:
\texttt{UnexpectedEther}, \texttt{UserAssertions}, \texttt{Exceptions}, \texttt{StateChangeAfterCall}.
Similarly, \textsc{Slither} was run with only relevant detectors: \texttt{reentrancy-eth}, \texttt{reentrancy-no-eth}, \texttt{reentrancy-balance}.
\textsc{SMTChecker} was run twice on each benchmark: once with the BMC and once with the CHC engine enabled.
A 60-minute timeout was enforced for each run.

We used the output messages of each run to determine whether the benchmark \emph{passed} the checks or was \emph{rejected} by the tool due to the vulnerability.
Specifically for \textsc{Mythril}, we consider a run to be rejected when the output contains the message \texttt{State access after external call}, disregarding benign reentrancy messages.
For \textsc{SMTChecker}, we consider a run rejected when the output contains the message \texttt{Assertion violation happens here}.
For \textsc{Slither}, a run is rejected when the output contains the markers of the used modules: \texttt{reentrancy-eth}, \texttt{reentrancy-no-eth}, \texttt{reentrancy-balance}.
\yult rejects a run with a counterexample trace; we consider a run passed when it terminates without producing such a trace.

\subsubsection*{Results}
 \cref{tab:comparison} contains evaluation results.
\yult and \textsc{Slither} both flagged all 23 benchmarks as vulnerable (100\,\%), with no timeouts.
\textsc{SMTChecker}-BMC achieved 95.7\,\% recall (22/23), missing only \texttt{modifier\_reentrancy}.
\textsc{Mythril} achieved 91.3\,\% (21/23), timing out on \texttt{spank\_chain} and failing to detect \texttt{0x754...615}.
\textsc{SMTChecker}-CHC had the lowest recall at 69.6\,\% (16/23), passing 5 benchmarks and timing out on 2.

\yult produced zero false positives on all 23 safe benchmarks (0\,\% false-positive rate).
\textsc{Slither} was the next most precise, falsely flagging 2 safe contracts (8.7\,\%). 
\textsc{SMTChecker}-CHC flagged 8 out of 23 safe contracts as vulnerable (34.8\,\%), with 14 true negatives and 1 timeout.
\textsc{Mythril} flagged 21 out of 23 safe contracts as vulnerable (91.3\,\%), correctly clearing only one benchmark, 
and a timeout on \texttt{spank\_chain}.
\textsc{SMTChecker}-BMC performed worst on precision, flagging 22 out of 23 safe contracts as vulnerable (95.7\,\%), correctly clearing only \texttt{modifier\_reentrancy}.

\textsc{SMTChecker}-BMC was the fastest tool, averaging 0.2\,s per benchmark (approx 6\,s total per variant set).
\textsc{Slither} completed every analysis in 0.5\,s on average (approx.\ 13\,s total per variant set).
\yult completed every vulnerable-variant analysis in 0.7\,s, on average (approx.\ $17\,s$ in total),  with every benchmark finishing in under 3~seconds.
On safe variants, \yult required 17\,s on average (approx.\ 6\,m 33\,s in total), reflecting the higher cost of exhaustively exploring the bounded statespace compared to finding a single witness.
\textsc{Mythril} and \textsc{SMTChecker}-CHC were the slowest tools.
The former tool averaged approx.\ 7\,m 23\,s on vulnerable and 6\,m 35\,s on safe benchmarks (approx.\ 163 and 145 minutes total, respectively, excluding timed-out benchmarks), for a combined runtime of over 5~hours.
The latter tool averaged approx.\ 17\,m 20\,s on vulnerable and 13\,m 40\,s on safe benchmarks (approx.\ 364 and 301 minutes total, respectively, excluding timed-out benchmarks) for a total runtime of over 11~hours.

The results show that \yult uniquely combines high recall with no false positives, with runtimes over these benchmarks close to those of the fastest tools.
Absence of false positives is \yult's unique benefit --  a direct consequence of the theoretical foundations presented in the previous sections --
and a practically valuable property: contrary to other tools, it requires no manual triage of every alarm.
\yult shares 100\,\% recall with \textsc{Slither}, but unlike \textsc{Slither} it accompanies each finding with a concrete exploit trace.
This precision comes without significant runtime cost on these benchmarks: on the vulnerable benchmarks, \yult is faster than all tools except \textsc{Slither} and \textsc{SMTChecker}-BMC, neither of which achieves zero false positives.
However, \yult performs a combinatorial exploration constrained by its bounding parameters. Choosing these parameters carefully is
important for scaling to the real-world case studies in the following sections.

\subsection{The DAO}
\label{sec:dao}

We now return to our first real-world case study, the DAO project~\cite{dao-victim}, discussed in \cref{sec:dao:explain}.
We analysed three versions.
In the original on-chain version upgraded to Solidity 0.8 code\footnote{See \cref{sec:upgrades} for the source code upgrades applied.} (\textbf{splitDAO}),
\yult
found a reentrancy trace that exploited |splitDAO|, as explained in \cref{sec:dao:explain}.
In a second version (\textbf{transfer}), we added a |nonReentrant| modifier to |splitDAO|.
Here, \yult discovered a different attack trace,
involving |splitDAO| followed by a callback into DAO's \texttt{transfer}.
In the final version (\textbf{safe}), we additionally
moved the external call in |splitDAO| after all state updates, following the 
checks-effects-interactions pattern~\cite{code_patterns}.
With this fix, \yult found no exploit within the analysis bounds,
indicating that this small change would have prevented the DAO attack.
Parameters for all three versions are given in \cref{tab:dao}.
\begin{table}
	\catcode`\_=12
	\scriptsize
	\centering
	\parbox{0.78\linewidth}{\begin{filecontents*}{experiments/dao.csv}
DAO Version	Sol	Yul	ABI	Uint256	O-Addr	Trace	Time (m:s)
splitDAO	1354	12384	4 / 22 (+27)	0, 604800	1	10 / 45	00:00.6
transfer	1361	12504	5 / 22 (+27)	0, 604800	2	15 / 58	01:57.4
safe	1358	12484	5 / 22 (+27)	0, 604800	2	--	t/o
\end{filecontents*}
\csvautobooktabular[separator=tab]{experiments/dao.csv}
          \smallskip
          
	{\scriptsize Sol and Yul LoC exclude comments and empty lines.
	ABI: X / Y (+Z) = X out of Y non-view (+Z view) functions. 
Runtime is unchanged by adding view functions.
Uint256: integer domain.
O-Addr : number of opponent addresses.
Trace: number of opponent (o-call/o-ret/wait) / total moves. t/o = 60\,min timeout.}}\quad
      \fbox{\parbox{0.18\linewidth}{\tiny	\textbf{Rest of CLI Parameters:}\\
  Call Bound = 2;\\
  Stack Bound = 3;\\
  Trace Bound = 30;\\
  Deploy P-Balance = 0;\\
  O-Balance = 10 ETH;\\
  O-Calls value = \num{1000} wei;\\
Wait Time = 7 days;\\
Max time delay = 22 days;\\
Wait First = true.}}

	\caption{DAO Experiments}\label{tab:dao}
\end{table}


\subsubsection*{The Deployer contract.}
To analyse the DAO, \yult must deploy it, but its constructor requires several arguments.
To supply these, we created a small \emph{Deployer} contract that \yult executes at the start of the analysis.
\begin{lstlisting}[language=Solidity]
 contract Deployer {
   constructor() payable {
     DAO_Creator daoCreator = new DAO_Creator();                  // 1. Deploy DAO_Creator
     address curator = address(0x908070605040302010);             // 2. Hard-coded parameters
     uint256 proposalDeposit = 7;
     uint256 minTokensToCreate = 10;
     uint256 closingTime = block.timestamp + 15 days;
     address privateCreation = address(0);
     new DAO(curator, daoCreator, proposalDeposit,                // 3. Deploy DAO contract
             minTokensToCreate, closingTime, privateCreation);
 } }
\end{lstlisting}
This contract performs the necessary steps to initialise and deploy a valid DAO.
It provides the DAO constructor with:
the address of a |DAO_Creator| contract,
a curator address (we provided a dummy one),
protocol parameters such as the proposal deposit cost (in wei) and the minimum number of DAO tokens to mint during the funding period,
the closing time of that period, and a flag (|address(0)| if the DAO is not private).
Every Proponent address created during analysis is callable by the Opponent if the contract is listed in the exploration ABI.
Thus, in the above code, the new DAO address need not be stored; the Opponent automatically discovers it from the exploration ABI.

\subsubsection*{Results.}
As shown in \cref{tab:dao}, \yult detected vulnerabilities in both \textbf{splitDAO} and \textbf{transfer} versions, and found none (within a 60-min bound) in \textbf{safe}.
Analysis time scaled with the size of the exploration ABI and the number of Opponent addresses: runtime grew from about 0.6s for \textbf{splitDAO} to almost 2 minutes for \textbf{transfer}.
These results demonstrate that \yult can reliably expose vulnerabilities while keeping analysis practical, given modest user guidance via CLI parameters to control state-space growth.
They also confirm its precision: \yult produced no false positives on \textbf{safe}.
Below we show the key parts of the trace produced for the \textbf{splitDAO} version.

\begin{lstlisting}[language=EVMTrace]
 [new opponent address: <0x..30>]
 create(object:<DAO_3087> , address:<0x..aa>) -> ... ->
 deploy(object:<Deployer_62_deployed>) ->
 o-wait -> o-wait ->(*@\label{ln:dao:trace:1}@*)
 o-call(caller:<0x..30>, target:<DAO_3117_deployed>, sig:<fallback()>, value:<1000>) -> ... -> po-ret([1]) ->(*@\label{ln:dao:trace:2}@*)
 o-wait ->(*@\label{ln:dao:trace:3}@*)
 o-call(caller:<0x..30>, target:<DAO_3117_deployed>, sig:<newProposal(..)>, args:<[..]>) -> po-ret(1) ->(*@\label{ln:dao:trace:4}@*)
 o-call(caller:<0x..30>, target:<DAO_3117_deployed>, sig:<vote(uint256,bool)>, args:<[..1..]>) -> po-ret() ->(*@\label{ln:dao:trace:5}@*)
 o-wait ->(*@\label{ln:dao:trace:6}@*)
 o-call(caller:<0x..30>, target:<DAO_3117_deployed>, sig:<splitDAO(uint256,address)>, args:<[..1..]>) -> ... ->(*@\label{ln:dao:trace:7}@*)
 po-call(caller:<0x..aa>, target:<0x..30>) ->(*@\label{ln:dao:trace:8}@*)
 o-call(caller:<0x..30>, target:<DAO_3117_deployed>, sig:<splitDAO(uint256,address)>, args:<[..1..]>)(*@\label{ln:dao:trace:9}@*)
 ERROR! sender 0x..aa has insufficient balance (0) to transfer 666
\end{lstlisting}
In this trace, the Opponent:
        waits two weeks (ln.~\ref{ln:dao:trace:1});
        joins the DAO by calling the fallback with 1000 wei (ln.~\ref{ln:dao:trace:2});
        waits another week (ln.~\ref{ln:dao:trace:3});
        creates a new proposal that gets assigned ID 1 (ln.~\ref{ln:dao:trace:4});
        votes for the proposal with ID 1 (ln.~\ref{ln:dao:trace:5});
        waits a week for the debating period (ln.~\ref{ln:dao:trace:6});
        calls split (ln.~\ref{ln:dao:trace:7});
        and, after the DAO calls back to transfer rewards (ln.~\ref{ln:dao:trace:8}), reenters split (ln.~\ref{ln:dao:trace:9}), triggering the violation.

\begin{remark}
The trace above illustrates why the DAO exploit resists precise automated detection.
Reaching the reentrant call requires navigating a specific multi-step protocol: funding the DAO, creating and voting on a proposal, and waiting for the debating period to close.
Tools that over-approximate interactions cannot account for this sequencing:
\textsc{Slither} reports almost identical reentrancy warnings on both the vulnerable \textbf{splitDAO} and the patched \textbf{safe} versions, unable to distinguish the two;
\textsc{SMTChecker} in BMC mode reports that the assertion can be violated in both versions, producing a false positive on the safe variant;
and \textsc{Mythril} and \textsc{SMTChecker} in CHC mode time out at 60\,min on both versions.
\end{remark}



\subsection{PredyPool}
\label{sec:predy}

To analyse the PredyPool case study from \cref{sec:predy:explain}, we first deploy and initialise the contract.
\begin{lstlisting}
contract Deployer {
  constructor() payable {
      WETH9  weth   = new WETH9();(*@\label{ln:predy:deployer:1}@*)
      AERC20 tokenA = AERC20(new TokenA());
      AERC20 tokenB = AERC20(new TokenB());
      __Yult__Toolbox toolbox = new __Yult__Toolbox(weth, tokenA, tokenB);(*@\label{ln:predy:deployer:2}@*)
      // Participant configuration parameters
      uint moveAmount = 10;
      uint24 fee1 = 3000;
      uint24 fee2 = 10000;
      // Create Participant Contracts
      PredyParticipant participant1 = new PredyParticipant(tokenA, tokenB, fee1, toolbox);(*@\label{ln:predy:deployer:3}@*)
      PredyParticipant participant2 = new PredyParticipant(tokenA, tokenB, fee2, toolbox);(*@\label{ln:predy:deployer:4}@*)
      // Give tokens to participants
      tokenA.transfer(address(participant1), 2 * moveAmount * (10 ** tokenA.decimals()));(*@\label{ln:predy:deployer:5}@*)
      tokenB.transfer(address(participant1), 2 * moveAmount * (10 ** tokenB.decimals()));
      tokenA.transfer(address(participant2), 2 * moveAmount * (10 ** tokenA.decimals()));
      tokenB.transfer(address(participant2), 2 * moveAmount * (10 ** tokenB.decimals()));(*@\label{ln:predy:deployer:6}@*)
      // Force a failure if Predy tries to transfer tokenA to participant1 without enough balance
      tokenA.setCriticalTransferAddress(address(participant1), true);(*@\label{ln:predy:deployer:7}@*)
} }
\end{lstlisting}
This involves creating ERC20 tokens, as Predy operates on ERC20-compliant assets:
Wrapped Ether (WETH) required for Uniswap operations, and two tokens to be used in trading pairs within PredyPool.
The constructor of | __Yult__Toolbox| (not shown here) deploys boilerplate Uniswap v3 contracts that PredyPool depends on
and the PredyPool itself.
%

Since interacting with the PredyPool API is complex and intractable at the low level, we avoid requiring the Opponent to discover valid API call sequences. Instead, we instantiate two mock participant addresses using the |PredyParticipant| contract (ll.\ref{ln:predy:deployer:3}–\ref{ln:predy:deployer:4} above) and fund them via the |Deployer| (ll.\ref{ln:predy:deployer:5}–\ref{ln:predy:deployer:6}). The values for |moveAmount|, |fee1| and |fee2| are for participant configuration.
The Opponent is given the simpler |PredyParticipant| ABI to perform valid high-level operations, such as creating and registering a Uniswap pool, supplying or withdrawing liquidity, and trading.
This reduces the number of traces explored by \yult, compared to interacting directly with PredyPool, while still reflecting typical test setups.
The key difference with testing is that \yult explores \emph{all} possible traces over these high-level actions.

On line~\ref{ln:predy:deployer:7}, the |Deployer| sets up a critical \yult assertion in |TokenA| to detect an imbalance in PredyPool’s internal accounting.
If at any point PredyPool attempts to refund TokenA to |participant1| without holding a sufficient TokenA balance, the \yult exploration fails and the trace is reported. This indicates that PredyPool’s internal accounting overestimates its TokenA holdings\,---\,a clear sign that the contract has been drained of TokenA's. Due to symmetry in tokens and participants, this single assertion is sufficient to detect exploit traces.

Below are the main parts of the |PredyParticipant| contract, which exposes functions the Opponent can call to trigger high-level participant actions.
The key one is |trade| (ll.~\ref{ln:predy:participant:4}--\ref{ln:predy:participant:5}) which records the Opponent's address (ln.~\ref{ln:predy:participant:1}). When PredyPool calls the participant's post-trade callback (ln.~\ref{ln:predy:participant:2}), the contract forwards the call to the stored address (ln.~\ref{ln:predy:participant:3}), allowing the Opponent to enumerate all possible moves from that point.
\begin{lstlisting}
contract PredyParticipant is IHooks {
  address controller = address(0);
  ...
  function createPredyPool() public { ... }
  function supplyLiquidity(bool toQuoteAsset) external { ... }
  function withdrawLiquidity(bool toQuoteAsset) external { ... }
  function take(bool toQuoteAsset) external { ... }
  function trade() external {(*@\label{ln:predy:participant:4}@*)
    require(predyPoolId != 0);
    controller = msg.sender; // Opponent(*@\label{ln:predy:participant:1}@*)
    toolbox.trade(predyPoolId);
  }(*@\label{ln:predy:participant:5}@*)
  function predyTradeAfterCallback(IPredyPool.TradeParams memory tradeParams,(*@\label{ln:predy:participant:2}@*)
                                   IPredyPool.TradeResult memory tradeResult) external {
    IHooks(controller).predyTradeAfterCallback(tradeParams, tradeResult); // Relay callback to Opponent(*@\label{ln:predy:participant:3}@*)
} }
\end{lstlisting}


\begin{table}
	\catcode`\_=12
	\scriptsize
	\centering
	\parbox{0.8\linewidth}{\begin{filecontents*}{experiments/predy.csv}
Code	Sol	Yul	ABI	Uint256	O-Addr	Trace	Time (m:s)
PredyPool	4589 / 27079	87832	5 ; 22 (+8)	1	1	9 / 375	00:06.1
\end{filecontents*}
\csvautobooktabular[separator=tab]{experiments/predy.csv}\smallskip

	{\scriptsize Sol and Yul LoC exclude comments and empty lines;
Sol: $a$ / $b$ = $a$ LoC in project files over $b$ LoC in all files including external libraries.
ABI: 5 functions in \lstinline{PredyParticipant} explored by \yult; \lstinline{PredyPool} itself has 22 non-view and 8 view functions. 
Uint256: integer domain.
O-Addr : number of opponent addresses.
Trace: number of opponent (o-call/o-ret/wait) / total moves.
 t/o = 60\,min timeout.}}\quad
      \parbox{0.15\linewidth}{\fbox{\parbox{\linewidth}{\tiny\textbf{CLI Parameters:}\\
Call Bound = 2;\\
Stack Bound = 6;\\
Trace Bound = 20;\\
Deploy P-Balance = 0;\\
O-Balance = 10 ETH;\\
O-Calls value = \num{1} wei;\\
No Waiting.}}\\\\
}
	\caption{PredyPool Experiment}\label{tab:predy}
\end{table}

The results of this case study appear in \cref{tab:predy}.
With the described instrumentation, \yult finds an exploit trace involving 9 high-level participant moves (375 total game moves) in about 6 seconds (we omit the trace due to lack of space).
PredyPool alone exposes 22 non-view functions (90 when including all library contracts).
\yult explored the 5 functions of our |PredyParticipant| wrapper, which internally interacts with 6 PredyPool functions (|supply|, |withdraw|, |take|, |trade|, |registerPair|, |updateWhitelistAddress|) and 3 Uniswap functions (|createPool|, |initialize|, |mint|).
Without this domain-specific instrumentation, exploring the full 90-function ABI via low-level calls would be intractable.
As noted in \cref{sec:predy:explain}, adding the |nonReentrant| modifier to |trade| causes \yult to exhaustively explore the bounded statespace within the 60\,min bound without finding an assertion violation, indicating that this prevents the attack.

\begin{remark}[Instrumentation]
  In all examples, instrumentation amounted to tuning CLI parameters (cf.~\cref{tab:gigahorse,tab:dao,tab:predy}).
  In the Gigahorse benchmarks, we instrumented |ModifierEntrancy| to reveal to the Opponent the hashed value of a string literal.
  In DAO and PredyPool we additionally included a deployer contract for initial setup, reflecting typical test setups.
  In PredyPool, we provided the Opponent with the simpler |PredyParticipant| ABI to perform valid high-level operations.
  This is an instance of test harnessing (cf.~\cref{sec:tool}), as used in e.g.~\textsc{Echidna}~\cite{echidna}, and \textsc{Foundry}~\cite{foundry-forge}.
  In PredyPool and one Gigahorse benchmark we added assertions to identify exploits relating to ERC-20 token balances.
  Adding such assertions is standard practice in specification-based tools, including \textsc{SMTChecker}~\cite{solc-smtchecker}, \textsc{Certora}~\cite{certora}, and \textsc{Halmos}~\cite{halmos}.
\end{remark}


\section{Beyond Reentrancy}
\label{sec:beyond}
The evaluation in \cref{sec:evaluation} focused on reentrancy, but \yult is not inherently limited to this class of vulnerabilities.
Since the game semantics models all interactions between a contract and its environment (\cref{sec:games}), any \emph{safety property}~\cite{lamport} --- one whose violation is witnessed by a finite execution trace --- can be checked by expressing it as a Solidity assertion (\cref{sec:tool}).

Not all vulnerability classes fall within this scope.
\emph{Frontrunning}, for example, depends on transaction ordering in the mempool, determined by miners and validators, which lies outside the execution model.
Moreover, \emph{denial-of-service} vulnerabilities often violate \emph{liveness} properties (e.g., that honest users can always withdraw their funds); liveness violations have no finite witnessing trace and are therefore also out of scope.

Many important vulnerability classes, however, do reduce naturally to assertion reachability.
\emph{Access control} vulnerabilities~\cite{atzeiSurveyAttacksEthereum2017a,kushwahaSystematicReviewSecurity2022} provide a clear illustration.
Consider the following annotated \texttt{multiowned\_vulnerable} benchmark in the Gigahorse suite~\cite{grech-gigahorse-benchmarks}:
\begin{lstlisting}[language=Solidity]
contract MultiOwnable {
  address public root;
  mapping (address => address) public owners; // owner => parent of owner
  ...
  modifier onlyOwner() {
    require(owners[msg.sender] != address(0));
    _;
  }
  ...
  function newOwner(address _owner) external returns (bool) {
    require(_owner != address(0));
    owners[_owner] = msg.sender;
    return true;
  }
  ...
}
contract TestContract is MultiOwnable {
  function withdrawAll() external onlyOwner {
    // yultracer: opponent should not withdraw the balance
    Yult.Assert(msg.sender != address(0x4f505f414444524553535f30));
    payable(msg.sender).transfer(address(this).balance);
  }
  ...
}
\end{lstlisting}
The |newOwner| function is missing the |onlyOwner| modifier, allowing any caller to register as an owner.
The assertion in |withdrawAll| checks that \yult's Opponent --- address \texttt{0x4f505f...30}, a deterministic constant --- never satisfies the |onlyOwner| guard.
\yult finds a trace in which the Opponent calls |newOwner| to register itself, then calls |withdrawAll|, triggering the assertion.
We ran \yult on all 16 access control benchmarks from the Gigahorse suite and detected violations in all 16.
A systematic evaluation of further vulnerability classes is left to future work.


\section{Limitations}
\label{sec:limitations}

We now discuss the theoretical and practical limitations of the approach.
\paragraph{Gas.}
 The game semantics is sound up to gas-dependent control flow (cf.~\cref{rem:opponent-gas}): if $\lL$ branches on the remaining gas, 
the games may not find the failing trace.
Client gas usage is idealised, so the games may produce traces realisable only by an unrealistically gas-efficient $\cC$; in particular, traces requiring substantial gas on the $\cC$ side
(e.g. multiple internal calls or memory-expansion attacks) may not be realisable.
However, since the gas given to Opponent is an upper bound, distinctions such as |transfer| versus |call{value: ...}("")| are preserved: with a stipend of 2300, $\cC$ cannot reenter $\lL$.
At the implementation level, the tool is additionally imprecise in some of $\lL$'s gas calculations, since Yul abstracts away the low-level stack operations and their gas costs, which may lead to potential gas over-approximations.

\paragraph{Omniscience.}
To prove completeness, Opponent is omniscient (cf.~\cref{rem:secrets}), meaning it can in principle guess any value including $\keccak$ preimages.
This is not a practical limitation: the implementation uses knowledge sets containing only values the contract has explicitly produced or revealed, and is therefore unable to guess secrets
in contracts that rely on this (cf.~\cref{sec:example:secrets}).
As a minor consequence of working at the Yul level rather than bytecode, $\keccak$ hashes of Yul code computed by the interpreter may differ from their EVM values. A contract that depends on hard-coded $\keccak$ hashes may need to update those to match the values the tool reports

\paragraph{Exceptional control flow.}
When $\lL$ makes an external call that fails, \yult does not explore the failure-handling branch: execution along the revert path is pruned.
As a consequence, behaviours that depend on $\lL$ recovering from a failed call are not captured.

\section{Conclusions}
\label{sec:conclusions}
We have introduced the first game semantics for open-world inter-contract reachability in the Ethereum Virtual Machine (\cref{sec:games}), targeting the Yul intermediate language as the vehicle of study (\cref{sec:lang}). We present a notion of soundness and completeness, up to realisability of the gas model, for our game model (\cref{thm:sound:complete}) and provide a proof of this statement (\cref{sec:sound}) that relies on the compositionality (\cref{lem:compositionality}) and definability (\cref{lem:definability}) of our games; proving the latter additionally provides us with a method to realise every trace found as a concrete Yul object. This theory has yielded \yult (\cref{sec:tool}), the first bounded game-semantics safety checker for Ethereum smart contracts. Being based on a sound and complete open-world theory, \yult offers soundness and bounded-completeness guarantees that, up to the limitations discussed in \cref{sec:limitations}, have allowed us to examine smart contracts with 100\% recall and zero false positives; a unique combination among similar tools, while also comparing favourably in performance on the challenging task of detecting real reentrancy errors (\cref{sec:gigahorse}). In particular, besides providing such guarantees, we show that \yult is applicable to non-trivial real-world contracts by examining the original DAO (\cref{sec:dao}) and Predy Finance (\cref{sec:predy}); the latter being a particularly large use case consisting of 27,111 lines of Solidity code (87,913 lines of Yul). Lastly, we leverage \yult's capability as a safety (reachability) checker to detect all access-control violations in the Gigahorse suite (\cref{sec:beyond}), showcasing the application of reachability beyond reentrancy problems.


\ifblind \else
\fi
\begin{acks}
	This work was funded in part by:
	\grantsponsor{RI}{Research Ireland}{https://research.ie/} under Grants~\grantnum{IRC}{GOIPD/2024/809} and ~\grantnum{SFI}{13/RC/2094\_2};
	the Cisco University Research Program Fund, a corporate advised fund of the \grantsponsor{SVCF}{Silicon Valley Community Foundation}{https://www.svcf.org/};
	and the \grantsponsor{Ethereum}{Ethereum Foundation}{https://ethereum.foundation/} under Grant~\grantnum{Ethereum}{FY23-1127}.
We thank Ellen Clarke for her study of versions of the access control benchmarks used here with YulTracer, carried out as part of her MCS dissertation at Trinity College Dublin in 2025–26.
We are also grateful to the anonymous reviewers for their thoughtful feedback, which substantially improved this paper.
\end{acks}

\ifblind
%
%
%

\fi

\bibliographystyle{ACM-Reference-Format}
\bibliography{references}

\ifblind
\newpage
\appendix
\noindent

\section{Soundness and Completeness}
\label{sec:sound}
\newcommand{\msgDataOffset}{\ma{\mathsf{offset_{data}}}}
\newcommand{\msgDataSize}{\ma{\mathsf{size_{data}}}}
\newcommand{\msgMemo}{\ma{\mathsf{memory}}}
\newcommand{\msgTerm}{\ma{\mathsf{term}}}
\newcommand{\innerl}{\ma{\mathsf{inner}}}

\begin{figure}
\[\begin{array}{@{\!\!\!\!\!}c}
\irule[C-Int][C-int]{
   \nbox{
    \conf{S \mid \conf{\mathcal{A};\ldots;g;\ldots;\memo};L ; \nameN} \to^* \conf{S' \mid \conf{\mathcal{A}';\ldots;g';\ldots;\memo'}; L' ; \nameN'}
    }
}{
  \nbox{
    \conf{ \mathcal{A}; \ldots;g;\ldots \vdash 
    \ldots ; \memo; S^L_\nameN }
    \trans{}
    \conf{ \mathcal{A}';\ldots;g;\ldots \vdash 
    \ldots ; \memo'; {S'}^{L'}_{\nameN'} }
  }
}\\\\
    \irule[C-OCall][C-o-call]{
K\neq\varepsilon\\
    m = \newCallMsg(g_c,\alpha_p,v, \alpha_o,g', bs)\\
    g'\leq g\\
  m.\msgTarget=m.\msgCAddr=\alpha_p\in \mathcal A\Pproj\not\ni\alpha_o\\
  S = \mathcal{A}(\alpha_p).\accCode\\
    f \in \accabi(S)\\
    bs \in \{f.\abihash\}\times\bytesset\\
    \memo_{bs} = [\ioproj\mapsto bs]
}{
  \nbox{
    \conf{\mathcal{A};\ldots; g;K 
    }_o
    \trans{\ocall(m)}
    \conf{\mathcal{A}[ \alpha_p.\accBal \pluseq v][\alpha_o\mapsto\bullet];\ldots ;
    m.\msgGas; \CALL_o :: K 
    \vdash 
    \alpha_p;\memo_{bs};S^{\varnothing}_\varnothing}_p
  }
}\\\\
\irule[C-POCall][C-po-call]{
   \vec{v} = g_c,\alpha_o,v,v_i,v_s,v_{o},v_z\\
   \memo',\_ = \mathtt{extend}(\memo,\max(v_i+v_s,v_{o}+v_z))\\
   m = \newCallMsg(\vec v, \alpha, g, \memo')\\
    m.\msgTarget=m.\msgCAddr = \alpha_o \notin \mathcal{A}\Pproj\\
    \mathcal{A}' = \mathcal{A}[ \alpha.\accBal \minuseq v][\alpha_o\mapsto\bullet]\\
   g''\! = g- m.\msgGas\geq0\\
    \memo'_\varepsilon=\memo'[\ioproj \mapsto \varepsilon] 
}{
  \nbox{
    \conf{\mathcal{A} ; T ; g ; K 
    \vdash 
    \alpha ; \memo ; E[\callop(\vec{v})]^L_\nameN}_p
    \trans{\pocalll(m)}
    \conf{\mathcal{A}' ; T ; m.\msgGas ; (\CALL^{v_o}_{v_z},  g'', \alpha , \memo'_\varepsilon, E^L_\nameN) :: K 
    }_o
  }
}\\\\
\irule[C-ORet][C-o-return]{
    {bs \in \bytesset}\\
    {\memo' = \memo\left[ [v_o,v_o+v_z) \mapsto bs,\ioproj\mapsto bs\right]}\\
    g''\leq g+g'
}{
  \nbox{
    \conf{\ldots; g; (\CALL^{v_o}_{v_z}, g'; \alpha , \memo , E^L_\nameN) :: K }_o
    \trans{\oreturnl(bs)}
    \conf{\ldots ; g''; K  \vdash 
    \alpha , \memo' , E^L_\nameN[1]}_p
  }
}\\\\
\irule[C-PPCall][C-pp-call]{
   \vec{v} = g_c,\alpha_p,v,v_i,v_s,v_{o},v_z\quad
   \memo',\_ = \mathtt{extend}(\memo,\max(v_i+v_s,v_{o}+v_z))\quad
   m = \newCallMsg(\vec v, \alpha, g, \memo')\\
   m.\msgTarget = m.\msgCAddr=\alpha_p \in {\mathcal A}\Pproj\quad
   \mathcal{A}' = \mathcal{A}[\alpha.\accBal \minuseq v][\alpha_p.\accBal \pluseq v]\quad
    g'' = g- m.\msgGas\geq 0\\
    \memo'_\varepsilon = \memo'[\ioproj \mapsto \varepsilon]\\
    bs = \memo'[v_i,v_i+v_s)\\
    \memo_{bs}=[\ioproj\mapsto bs]\\
   S = \mathcal{A}(\alpha_p).\accCode
}{
  \nbox{
    \conf{\mathcal{A} ; T ; g ; K 
    \vdash 
    \alpha ; \memo ; E[\callop(\vec{v})]^L_\nameN}
    \trans{\ppcalll(m)}
    \conf{\mathcal{A}' ; T ; m.\msgGas ;  (\CALL^{v_o}_{v_z}, g'' , \alpha , \memo'_\varepsilon , E^L_\nameN)\!::\! K 
    \vdash 
    \alpha_p;\memo_{bs};S^{\varnothing}_\varnothing}
  }
}\\\\
\irule[C-Create][C-p-create]{
   \vec v = v,v_1,v_2\\
   \memo',\_ = \mathtt{extend}(\memo,v_1+v_2)\\
   m = \newCreateMsg(\vec{v},\alpha,g,\memo',\alpha')\\
    \mathcal{A}' = \mathcal{A}[\alpha.\accNonce\pluseq1][ \alpha.\accBal \minuseq v] \uplus \left\{\alpha' \mapsto \conf{0 ,v ,\memo'[v_1,v_1+v_2) , \varepsilon}\right\}\\
   \alpha' = \mathtt{fresh}(\alpha,\mathcal A(a.\accNonce))\quad
   g'' = g-m.\msgGas\geq0\quad
\memo'_\varepsilon=\memo'[\ioproj\mapsto\varepsilon]\quad
   S = \mathcal{A}'(\alpha').\accCode
}{
  \nbox{
    \conf{\mathcal{A} ; T ; g ; K 
    \vdash 
    \alpha ; \memo  ; E[\createop(\vec v)]^L_\nameN}
    \trans{\createl(m)}
    \conf{\mathcal{A}' ; T ; m.\msgGas; (\CREATE,  g'' , \alpha , \memo'_\varepsilon , E^L_\nameN)\! ::\! K 
    \vdash 
    \alpha';\bot;S^{\varnothing}_\varnothing}
  }
}\\\\
\irule[EOCall][eoa-call]{
    S = \mathcal{A}(\amain).\accCode\\
}{
  \nbox{
    \conf{\mathcal{A};\ldots; \varepsilon}_o
    \trans{\eocall()}
    \conf{\mathcal{A};\ldots ; \CALL_o :: \varepsilon \vdash 
    \amain;\bot;S^{\varnothing}_\varnothing}_p
  }
}
  \end{array}\]
\caption{Client game semantics. Missing rules are identical to the ones for the standard (library) game semantics in \Cref{sec:games}.}\label{fig:client-semantics}
\end{figure}
\subsection{Client Semantics}
Following the library-client-daemon setting of \Cref{fig:onchain-offchain-v3},
let $\lL$ denote a \emph{library}, namely a set of smart contract accounts with known code, and $\cC$ a \emph{client}, namely a set of smart contract accounts that may call contracts in $\lL$, along with a \emph{daemon} $\dD$.
More specifically, we formalise the Opponent as consisting of two components:
\begin{enumerate}
  \item an \emph{on-chain component} (the client), represented by a closed Yul object $\cC$ whose evaluation deploys a set of client contracts, including a main contract $\cmain$; and
  \item an \emph{off-chain component} (the daemon), represented by an \textbf{externally owned account} $\dD$ that initiates non-programmatic calls from outside the blockchain by calling $\cmain$ after, optionally, waiting any amount of seconds.
\end{enumerate}
We stipulate that the client $\cC$ has a distinguished contract $\cmain$ with address $\amain$ which functions as an entry point to $\cC$ in its interaction with the daemon $\dD$ (in this library-client-daemon setting, computation is triggered by the daemon).
This setup is perhaps better conveyed through a Yul-level general example.
\begin{example}
Suppose
\yultext{"C"} denotes a Yul object representing the client $\cC$. Its sub-objects $C_1,\dots,C_n$ correspond to the contract accounts owned by $\cC$ that may interact with $\lL$, and their deployed addresses \yultext{c\_addr\_1},\dots,\yultext{c\_addr\_n} are intended to match opponent addresses $\alpha_1,\dots,\alpha_n$ in the LTS produced by $\gamesem{\lL}$. We additionally distinguish a sub-object \yultext{"Main"}, representing the entry-point contract $\cmain$ deployed with an initial balance of $v_\cC$, and write $\amain$ for the address \yultext{c\_addr\_main}. A daemon $\dD$ may interact with $\cC$ by calling $\cmain$ (at address $\amain$); this in turn dispatches to a suitable contract $C_i$ (at $\alpha_i$), which then interacts with $\lL$.
\begin{lstlisting}[language=Yul]
object "C" {
    code {
        mstore(64, memoryguard(128))
        let _1 := allocate_unbounded()
        codecopy(_1, dataoffset("C_1"), datasize("C_1"))
        let c_addr_1 := create(0, _1, datasize("C_1"))
        ...
        codecopy(_1, dataoffset("C_n"), datasize("C_n"))
        let c_addr_n := create(0, _1, datasize("C_n"))
        codecopy(_1, dataoffset("Main"), datasize("Main"))
        let c_addr_main := create(v_C, _1, datasize("Main"))
        return(0, 0)
    }
    object "C_1" { 
        // contents of C_1
    }
    ...
    object "C_n" { 
        // contents of C_n
    }
    object "Main" { 
        // contents of Main
    }
}
\end{lstlisting}
\end{example}

The game semantics we presented in \cref{sec:games} is for the library $\lL$. This means that the Proponent in the rules consists of contracts included in $\lL$, or dynamically created within it, whereas the Opponent represents the client $\cC$ along with the daemon $\dD$. In this section, we extend the game model to the client $\cC$. We call this the \emph{client semantics}. The game moves for the client are presented in \Cref{fig:client-semantics}. They are identical to those in \cref{sec:games}, apart from the aspects we enumerate below.
\begin{enumerate}
  \item In \iref{C-o-call} we stipulate that $K\neq\varepsilon$. This is because the top-level interactions for $\cC$ are exclusively with $\dD$, while the interactions between $\cC$ and $\lL$ are nested within. This dynamics is better depicted in \Cref{fig:LCD}.
\item We also add an additional move \iref{eoa-call} to cater for $\dD$ calling the main contract of $\cC$;
  this is an Opponent call from an empty stack and allows $\dD$ to call the main object of $\cC$.
\item We have tampered with the gas calculations in the listed moves (apart from \iref{eoa-call}). This is to impose our assumption that the client consumes no gas in internal steps/calls.
  Moreover, in \iref{C-o-call} we allow for Opponent (i.e.\ the library) to use gas before the call (so $g'\leq g$) and pick a cost $g_c\neq g'$. Similarly, Opponent can use gas before returning with \iref{C-o-return}.
\end{enumerate}
Note that in client semantics the Proponent is the client $\cC$, while the Opponent is $\lL$ in all moves apart from \textsc{EndTx/Wait}, where the Opponent is the daemon $\dD$.

\begin{figure}[t]
\centering
\scalebox{0.7}{\begin{tikzpicture}[
  >=Stealth,
  arrstyle/.style={
    line width=1.6pt,
    draw={rgb,255:red,27;green,38;blue,49},
    shorten >=4pt, shorten <=4pt
  }
]

\definecolor{liborg}  {RGB}{243,156, 18}   
\definecolor{liborgdk}{RGB}{176, 98,  0}   
\definecolor{cligrn}  {RGB}{ 88,177, 88}   
\definecolor{cligrndk}{RGB}{ 28,115, 28}   
\definecolor{daered}  {RGB}{213, 78, 71}   
\definecolor{daereddk}{RGB}{148, 25, 18}   
\definecolor{onbg}    {RGB}{235,246,255}   
\definecolor{onborder}{RGB}{ 41,128,185}   
\definecolor{navycol} {RGB}{ 27, 38, 49}   



\draw[gray!55, line width=1pt, dash pattern=on 6pt off 4pt]
    (7.5, 4.8) -- (7.5,-4.4);

\node[
  rectangle, rounded corners=8pt,
  minimum width=2.2cm, minimum height=8cm, inner sep=4pt,
  top color=liborg, bottom color=liborgdk,
  text=white, font=\bfseries, align=center,
  drop shadow={shadow xshift=2.5pt, shadow yshift=-2.5pt,
               opacity=0.45}
] (L) at (0,0) {\Huge $\mathcal L$\\[4pt]\small Library};

\node[
  rectangle, rounded corners=8pt,
  minimum width=2.2cm, minimum height=8cm, inner sep=4pt,
  top color=cligrn, bottom color=cligrndk,
  text=white, font=\bfseries, align=center,
  drop shadow={shadow xshift=2.5pt, shadow yshift=-2.5pt,
               opacity=0.45}
] (C) at (5.5,0) {\Huge $\mathcal C$\\[4pt]\small Client};

\node[
  rectangle, rounded corners=8pt,
  minimum width=2.2cm, minimum height=8cm, inner sep=4pt,
  top color=daered, bottom color=daereddk,
  text=white, font=\bfseries, align=center,
  drop shadow={shadow xshift=2.5pt, shadow yshift=-2.5pt,
               opacity=0.45}
] (D) at (11,0) {\Huge $\mathcal D$\\[4pt]\small Daemon};

\draw[arrstyle, ->] ([yshift=32mm]L.east) -- node[above] {$\deployl()$} ([yshift=32mm]C.west);
\draw[arrstyle, <-] ([yshift=30mm]D.west) -- node[above] {$\deployl()$} ([yshift=30mm]C.east);
\node at (8.25, 26mm) {$\langle\mathsf{wait}\rangle$};
\draw[arrstyle, ->] ([yshift=22mm]D.west) -- node[below] {$\mathsf{call}(\alpha_{\rm main})$} ([yshift=22mm]C.east);
\draw[arrstyle, <-] ([yshift=20mm]L.east) -- node[above] {$\mathsf{call}()$} ([yshift=20mm]C.west);
\node at (2.75, 18.5mm) {\bf $\vdots$};
\draw[arrstyle, ->] ([yshift=10mm]L.east) -- node[above] {$\mathsf{ret}(\dots)$} ([yshift=10mm]C.west);
\node at (2.75, 6.5mm) {\bf $\vdots$};
\draw[arrstyle, <-] ([yshift=-4mm]L.east) -- node[above] {$\mathsf{call}()$} ([yshift=-4mm]C.west);
\node at (2.75, -5.5mm) {\bf $\vdots$};
\draw[arrstyle, ->] ([yshift=-14mm]L.east) -- node[above] {$\mathsf{ret}()$} ([yshift=-14mm]C.west);
\draw[arrstyle, <-] ([yshift=-16mm]D.west) -- node[above] {$\mathsf{ret}()$} ([yshift=-16mm]C.east);
\node at (8.25, -20mm) {$\langle\mathsf{wait}\rangle$};
\draw[arrstyle, ->] ([yshift=-24mm]D.west) -- node[below] {$\mathsf{call}(\alpha_{\rm main})$} ([yshift=-24mm]C.east);
\draw[arrstyle, <-] ([yshift=-26mm]L.east) -- node[above] {$\mathsf{call}()$} ([yshift=-26mm]C.west);
\node at (2.75, -30mm) {\bf $\vdots$};
\node at (8.25, -32mm) {\bf $\vdots$};

\node[font=\bfseries\large, text=darkgray, anchor=south west]
    at (1.75, 4.4) {On-Chain};

\node[font=\bfseries\large, text=darkgray, anchor=south west]
    at (10, 4.4) {Off-Chain};



\end{tikzpicture}}
\caption{Game semantics setup between Library $\mathcal L$, Client $\cC$ and Daemon $\mathcal D$. The game starts with $\mathcal L$ and $\cC$ deploying their root contracts. Thereon, $\mathcal D$ leads the game by triggering the main contract of $\cC$ and applying a delay between each new such trigger.}
\label{fig:LCD}
\end{figure}



It is useful to restrict our attention to clients (and libraries)
of a specific kind, e.g.\ to avoid trivial assertion violations caused in client code. We will also place some restrictions on allowable moves before a contract is deployed.
We first
partition game move labels into \emph{initial}, \emph{inner} and \emph{outer}:
\begin{itemize}
  \item initial labels are produced by deploy moves;
\item inner labels are produced by create and PP-moves (\createl, \ppcalll, \preturnl{p}, and their client-semantics counterparts);
\item outer labels are produced by all other moves (excluding \iref{int} and \iref{C-int}). 
\end{itemize}
We may write $\innerl$ to represent inner move labels.

\begin{definition}[Deployment and Good Clients]\label{def:good:clients}
  For a Yul object $O$, we say the \emph{deployment} of $O$ is the production of an initial configuration $\rho_0 \in \gamesem{O}$, defined by the evaluation of $O$ until a top-level opponent configuration is reached via the \iref{p-deploy} rule. More formally:
\[\conf{[\alpha_0\mapsto\conf{0;v_0;\varepsilon;\varepsilon}]; T_0 ; g_0 ; \varepsilon \vdash 
\alpha_0 ; \bot ; O^\varnothing_\varnothing}_p 
\to^*.\trans{\deployl(\alpha_0,l_0)} \rho_0\]
We say a client $\cC$ is \emph{valid} if:
\begin{enumerate}
\item it contains no assertions (i.e.\ the opcode \yultext{ASSERT} does not appear in $\cC$); and
\item during deployment it creates a designated main contract $\amain$.
\end{enumerate}
A valid client $\cC$ is called \emph{good} if, in addition, in all applications of \iref{C-po-call} and \iref{C-pp-call} it allocates all available gas to the call (i.e.\ $g_c=g$).
\defEnd
\end{definition}
We write $\rho_0$ and variants in either $\gamesem{\lL}$ or $\gamesem{\cC}$ to define the initial configuration obtained by deploying $\lL$ or $\cC$ respectively.

\begin{definition}[Good Traces]\label{def:good:traces}
  We say a trace $(t,\rho) \in \gamesem{O}$ for some object $O$ is \emph{good} if it consists only of inner move labels or it can be split into traces $t_1$ and $t_2$ such that:
  \begin{itemize}
  \item $t_2$ starts with an initial label (i.e.\ \deployl) and $t_1$ contains only inner labels,
  \item there are no consecutive wait labels in $t_2$. 
    \defEnd
  \end{itemize}
\end{definition}


Given \cref{def:good:clients,def:good:traces}, we shall consider in this section only good clients that produce good traces, and similarly only libraries that produce good traces.
Note that the requirement for no consecutive wait labels is a technical one as consecutive waits can be merged into one by summing up their waiting times and using just the last gas value.

\subsection{Soundness and Completeness}
We prove in this section that our trace semantics for EVM inter-contract interactions is sound and complete. This is dependent on infrastructure proven in later sections. Intuitively, completeness states that every assertion violation reachable by concretely interacting with a given library $\lL$ will be found by our games, whereas soundness means that every violation reached by our games is \emph{real} (i.e.\ concretely realisable by a client+daemon).


\begin{definition}[Syntactic Composition]\label{def:syn:comp}
Consider a library $\lL$ and a client $\cC$, each defined by a Yul object. Let the \emph{syntactic composition} of the two Yul objects $\lL$ and $\cC$ to be a single object $\syncomp{\lL}{\cC}$ that deploys $\lL$ first and then $\cC$.
We define $\syncomp{\lL}{\cC}$ by the object \yultext{"L\_union\_C"} only for $\lL$ and $\cC$ such that the addresses obtained by deploying $\cC$ after $\lL$ do not collide.
\begin{lstlisting}[language=Yul]
object "L_union_C" {
    code {
        mstore(64, memoryguard(128))
        let _1 := allocate_unbounded()
        codecopy(_1, dataoffset("L"), datasize("L"))
        let l_deployer_addr := create(v_L, _1, datasize("L"))
        codecopy(_1, dataoffset("C"), datasize("C"))
        let c_deployer_addr := create(v_C, _1, datasize("C"))
        return(0, 0)
    }
    object "L" {
        // contents of L
    }
    object "C" { 
        // contents of C 
    }
}
\end{lstlisting}
Note that $\lL$ and $\cC$ are deployed with some initial balances $v_L$ and $v_C$ respectively, which $\syncomp{\lL}{\cC}$ is implicitly parameterised by.
The game semantics $\gamesem{\lL\cup\cC}$ is that of a library (cf.\ \Cref{sec:games}, using moves \iref{p-deploy}, \iref{p-create}, \iref{pp-call}, \iref{pp-return1}, \iref{pp-return2} and \iref{po-return}) with the addition of the \iref{eoa-call} move. 

\defEnd
\end{definition}

We next formalise the notion of a library and a client producing compatible traces, that is, traces obtained by playing each against the other. 

\begin{definition}[Trace Compatibility]\label{def:trace:compat}
  Given a trace $t$:
  \begin{itemize}
  \item Let $t^\circ$ be the \emph{observable projection} of $t$, obtained by removing all initial and inner move labels, as well as all EO-calls (for client semantics).
  \item Given a subsequence $t'$ of $t$, let $\overline{t'}$ denote the \emph{polarity reversal} of $t'$, obtained by exchanging $\mathsf{p}$ and $\mathsf{o}$ in every move label:
 each $\pocalll(m)$ becomes $\ocall(m)$ (and viceversa), each $\preturnl{}(bs)$ becomes $\oreturnl{(bs)}$ (and viceversa), while wait labels remain unchanged.
  \end{itemize}
For any library $\lL$ and good client $\cC$, $(t_1,\rho_1) \in \gamesem{\lL}$ and $(t_2,\rho_2) \in \gamesem{\cC}$, we say traces $t_1$ and $t_2$ are \emph{compatible}, written $t_1 \compatible t_2$, when $t_1^\circ = \overline{t_2^\circ}$. We say that $\cC$ is \emph{compatible with $\lL$ on $t_1$} if $t_1\compatible t_2$ for some $(t_1,\rho_1) \in \gamesem{\lL}$ and $(t_2,\rho_2) \in \gamesem{\cC}$.
\defEnd
\end{definition}

The main results are the following. The proofs of \cref{lem:compositionality,lem:definability} occupy the remainder of \cref{sec:sound}. Below, we say that $\gamesem{\lL\cup\cC}$ \emph{weakly fails} if it fails in a trace produced by applying moves also from the client semantics (i.e.\ ignoring internal gas costs).

\begin{lemma}[L--C Compositionality]\label{lem:compositionality}
  For any library $\lL$ and good client $\cC$, $(1)\Rightarrow(2)\Rightarrow(3)$:
  \begin{enumerate}
  \item $\gamesem{\syncomp{\lL}{\cC}}$ fails;
    \item there exist $(t_1,\rho_1)\in\gamesem{\lL}$ and $(t_2,\rho_2)\in\gamesem{\cC}$ such that $t_1\compatible t_2$ and $\rho_1$ failed;
  \item $\gamesem{\syncomp{\lL}{\cC}}$ weakly fails.
  \end{enumerate}
\end{lemma}

\begin{lemma}[Definability]\label{lem:definability}
Let $(t,\rho)\in\gamesem{\lL}$ for any library $\lL$. There exists a good client $\cC$ that is compatible with $\lL$ on trace $t$.
\end{lemma}

\begin{theorem}[Soundness and Completeness]\label{thm:sound:complete}
For any library $\lL$, we have $(1)\Rightarrow(2)\Rightarrow(3)$:
\begin{enumerate}
\item there exists a valid compatible client $\cC$ such that $\gamesem{\syncomp{\lL}{\cC}}$ fails;
\item $\gamesem{\lL}$ fails (reaches an assertion violation);
\item there exists a valid compatible client $\cC$ such that $\gamesem{\syncomp{\lL}{\cC}}$ weakly fails.
\end{enumerate}
\begin{proof}
(2 to 3) Let $(t,\rho) \in \gamesem{\lL}$ for some trace $t$ and failed $\rho$. By \cref{lem:definability} we have a good client $\cC$ that realises $t$. By \cref{lem:compositionality}, we have that $\gamesem{\syncomp{\lL}{\cC}}$ weakly fails. 

(1 to 2) Let $\gamesem{\syncomp{\lL}{\cC}}$ fail for some valid client $\cC$.
Without loss of generality, since good clients enable more interactions, we may assume that $\cC$ is good.
By \cref{lem:compositionality} there exist $(t_1,\rho_1) \in \gamesem{\lL}$ and $(t_2,\rho_2) \in \gamesem{\cC}$ such that $t_1\compatible t_2$ and $\rho_1$ failed.
\end{proof}
\end{theorem}

In the next 4 sections we embark on studying composition of a library with a client. We define two kinds of composition: \emph{external}, where we compose their semantics; and \emph{internal}, where we merge their configurations and produce a combined new semantics.
The main result the is showing that the two kinds of composition give bisimilar LTS's. From the latter, we obtain L-C compositionality (\cref{lem:compositionality}) by relating internal composition of $\lL$ and $\cC$ with the (game-free) Yul-EVM semantics of $\lL\cup\cC$.
Finally, definability (\cref{lem:definability}) is proven in \cref{sec:definability} by defining, for each given trace $t$, a client $\cC$ that simulates precisely the moves in $t$. 


\subsection{External Composition}
We introduce here the notion of \emph{external composition}, which to pair configurations semantically.
First, let us consider a slight modification to the annotations of proponent stack frames $F_p$.
\[
F_p ::= ({\CALL}^{v_o}_{v_z},g,\alpha,\memo,E^L_\nameN)_{po} \mor
 ({\CALL}^{v_o}_{v_z},g,\alpha,\memo,E^L_\nameN)_{pp} \mor
 (\CREATE,g,\alpha,\memo,E^L_\nameN)_{pp}
\]
Proponent frames now have either $pp$ or $po$ annotations depending on whether they were created by $pp$ moves or $po$ moves. More specifically, \iref{po-call} now produces $F_{po}$ frames, whereas \iref{pp-call} and \iref{p-create} produce $F_{pp}$ frames. With these annotations, we can now describe what it means for two configurations to be compatible\,---\,i.e.\ whether they can be composed\,---\,starting with their stacks.
\begin{definition}[Stack Compatibility]\label{def:stack:compat}
Stacks $K_1$ and $K_2$ are \emph{compatible}, written $K_1 \compatible K_2$, when:
\begin{itemize}
\item $K_1 = K_2 = \varepsilon$;
\item $K_1 = \CALL_o :: \varepsilon$ and $K_2 = F_{po} :: F_{pp} :: \CALL_o :: \varepsilon$;
\item $K_1 = F_{pp}^* :: \CALL_o :: K_1'$ and $K_2 = F_{po} :: K_2'$,
and $K_1' \compatible K_2'$;
\item $K_1 = F_{po} :: K_1'$ and $K_2 = F_{pp}^* :: \CALL_o :: K_2'$,
and $K_1' \compatible K_2'$.
\end{itemize}
where $F_{pp}^*$ refers to zero or more instances of any frame $F$ produced by PP moves.
\defEnd
\end{definition}

Next, let us say that a configuration $\rho$ is \emph{post-deployment} if it is not a top-level P-configuration.

\begin{definition}[Compatible Game Configurations]
We say configurations $\rho_1 \in \gamesem{\lL}$ and $\rho_2 \in \gamesem{\cC}$ are \emph{compatible}, written $\rho_1 \compatible \rho_2$, if they are both post-deployment and:
\begin{itemize}
\item they are of opposite polarity (p-o and o-p) with compatible stacks $\rho_1.K\compatible\rho_2.K$, or
\item they are both top-level O-configurations (i.e. with an empty stack $\rho_1.K=\rho_2.K=\varepsilon$);
\end{itemize}
and their environments are compatible:
\begin{itemize}
\item their closed address space is disjoint, i.e., $\rho_1.\mathcal{A}\Pproj \cap \rho_2.\mathcal{A}\Pproj = \varnothing$; 
\item they complement their open addresses, i.e., $\rho_1.\mathcal{A}\Oproj \subseteq \rho_2.\mathcal{A}\Pproj$ and $\rho_2.\mathcal{A}\Oproj \subseteq \rho_1.\mathcal{A}\Pproj$; and
\item they agree on the time, i.e.\ $\rho_1.T = \rho_2.T$.
  \defEnd
\end{itemize}
\end{definition}
With this notion of compatibility, we define how configurations can be paired up. Intuitively, compatible configurations may be run in parallel. We consider the paring of every configuration in $\gamesem{\lL}$ with every compatible configuration in $\gamesem{\cC}$ to be their external composition.

Let $\rho_1,\rho_1'\in\gamesem{\lL}$ and $\rho_2,\rho_2'\in\gamesem{\cC}$ be game configurations. The following rules define the external composition of two compatible configurations $\rho_1\compatible\rho_2$.
\begin{align*}
\begin{array}{@{}lll@{}}
\irule[Int1][int1]{
  \nbox{
    \rho_1 \trans{} \rho_1'
  }\quad
  \nbox{
    \rho_2 = \rho_2'
  }
}{
  \nbox{
    \rho_1 \semcomp \rho_2
    \trans{}'
    \rho_1' \semcomp \rho_2'
  }
}
\end{array}
\quad
&\begin{array}{@{}lll@{}}
\irule[Int2][int2]{
  \nbox{
    \rho_2 \trans{} \rho_2'
  }
    \quad
  \nbox{
    \rho_1 = \rho_1'
  }
}{
  \nbox{
    \rho_1 \semcomp \rho_2
    \trans{}'
    \rho_1' \semcomp \rho_2'
  }
}
\end{array}
\\  
\begin{array}{@{}lll@{}}
\irule[Ret1][ret1]{
  \nbox{
    \rho_1 \trans{\preturnl{o}(\evmOut)} \rho_1'
  }
    \quad
  \nbox{
    \rho_2 \trans{\oreturnl(\evmOut)} \rho_2'
  }
}{
  \nbox{
    \rho_1 \semcomp \rho_2
    \trans{}'
    \rho_1' \semcomp \rho_2'
  }
}
\end{array}
\quad
&\begin{array}{@{}lll@{}}
\irule[Ret2][ret2]{
  \nbox{
    \rho_1 \trans{\oreturnl(\evmOut)} \rho_1'
  }
    \quad
  \nbox{
    \rho_2 \trans{\preturnl{o}(\evmOut)} \rho_2'
  }
}{
  \nbox{
    \rho_1 \semcomp \rho_2
    \trans{}'
    \rho_1' \semcomp \rho_2'
  }
}
\end{array}
\\
\begin{array}{@{}lll@{}}
\irule[Call1][call1]{
  \nbox{
    \rho_1 \trans{\pocalll(l)} \rho_1'
  }
    \quad
  \nbox{
    \rho_2 \trans{\ocall(l)} \rho_2'
  }
}{
  \nbox{
    \rho_1 \semcomp \rho_2
    \trans{}'
    \rho_1' \semcomp \rho_2'
  }
}
\end{array}
\quad
&\begin{array}{@{}lll@{}}
\irule[Call2][call2]{
  \nbox{
    \rho_1 \trans{\ocall(l)} \rho_1'
  }
    \quad
  \nbox{
    \rho_2 \trans{\pocalll(l)} \rho_2'
  }
}{
  \nbox{
    \rho_1 \semcomp \rho_2
    \trans{}'
    \rho_1' \semcomp \rho_2'
  }
}
\end{array}
\\
  \begin{array}{@{}lll@{}}
\irule[Inn1][inner1]{
  \nbox{
    \rho_1 \trans{\innerl} \rho_1'
  }
    \quad
  \nbox{
    \rho_2 = \rho_2'
  }
}{
  \nbox{
    \rho_1 \semcomp \rho_2
    \trans{}'
    \rho_1' \semcomp \rho_2'
  }
}
\end{array}
\quad
&\begin{array}{@{}lll@{}}
\irule[Inn2][inner2]{
  \nbox{
    \rho_2 \trans{\innerl} \rho_2'
  }
    \quad
  \nbox{
    \rho_1 = \rho_1'
  }
}{
  \nbox{
    \rho_1 \semcomp \rho_2
    \trans{}'
    \rho_1' \semcomp \rho_2'
  }
}
\end{array}
  \\
\begin{array}{@{}lll@{}}
\irule[Wait][wait]{
  \nbox{
    \rho_1 \trans{\waitl(i,g)} \rho_1'
  }
    \quad
  \nbox{
    \rho_2 \trans{\waitl(i,g)} \rho_2'
  }
}{
  \nbox{
    \rho_1 \semcomp \rho_2
    \trans{}'
    \rho_1' \semcomp \rho_2'
  }
}
\end{array}
\quad
&\begin{array}{@{}lll@{}}
\irule[EOCall2][eoa]{
  \nbox{
    \rho_1 = \rho_1'
  }
    \quad
  \nbox{
    \rho_2 \trans{\eocall} \rho_2'
  }
}{
  \nbox{
    \rho_1 \semcomp \rho_2
    \trans{}'
    \rho_1' \semcomp \rho_2'
  }
}
\end{array}
\end{align*}

\begin{definition}[External Composition]\label{sem:comp:lib}
For a library $\lL$ and compatible client $\cC$, their \emph{external composite semantics} is:
\[\gamesem{L}\semcomp\gamesem{C} = \{\rho \mid \rho_0 \semcomp \rho_0' \to'^* \rho\}\]
where $\to'^*$ is the reflexive transitive closure of $\to'$, and $\rho_0 \in \gamesem{L}$ and $\rho_0' \in \gamesem{C}$ are the respective initial configurations obtained by deploying $\lL$ and $\cC$.
  \defEnd
\end{definition}

\subsection{Internal Composition}
We now define the \emph{internal composition} of two game configurations $\rho_1 \in \gamesem{\lL}$ and $\rho_2 \in \gamesem{\cC}$, written $\rho_1 \intcomp \rho_2$. Intuitively, $\rho_1 \intcomp \rho_2$ is obtained by merging $\rho_1$ and $\rho_2$ into a single configuration of a composite semantics.
Since compatible configurations close each other, the composite semantics shall be based on the closed fragment of the client semantics, while still allowing external moves. In particular, it excludes the P-O moves \iref{po-call}, \iref{po-return}, \iref{o-call}, and \iref{o-return}, but retains \iref{eoa-call}. We write $\to_{1,2}$ for the transition relation of this composite semantics. 

Let configurations for our composite semantics be of the form
\begin{align*}
  \conf{{\mathcal{A}} ; T ; \vec g ; K
  \vdash \alpha ; \memo ; S^L_\nameN}_{p}
\qquad
  \conf{{\mathcal{A}} ; T ; \vec g ; K
  }_{eo}
\end{align*}
where $\vec g = (g_1,g_2)$ and components labelled $1$ and $2$ according to ownership by the library and the client, respectively. Similarly, all addresses $\alpha \in \mathcal{A}$ shall be annotated $\alpha^{\conf{i}}$ for $i = 1,2$; for economy, we shall omit these annotations where they are not used or can be inferred. Since the semantics is closed except for the EOA, state $\mathcal{A}$ includes only proponent addresses, i.e., $\mathcal{A}\Oproj = \varnothing$, and opponent configurations occur only when the EOA has control, i.e., when $K = \varepsilon$.
Next, we define the transition relation $\to_{1,2}$.

\[\begin{array}{@{}c}
\irule[Int][composite-int]{
    \conf{S \mid \conf{\mathcal{A};T;g_i;\memo};L ; \nameN} \to^* \conf{S' \mid \conf{\mathcal{A}';T;g_i';\memo'}; L' ; \nameN'}\\
    (i,g_i'') \in\{(1,g_1'),(2,g_2)\}
}{
  \nbox{
    \conf{{\mathcal{A}}; T ; \vec{g} ; K \vdash 
    \alpha^{\conf{i}} ; \memo; S^L_\nameN }_p
    \trans{}_{1,2}
    \conf{ {\mathcal{A}}'; T ; \vec{g}[i\mapsto g''_i] ; K \vdash 
    \alpha^{\conf{i}} ; \memo'; {S'}^{L'}_{\nameN'} }_p
  }
}\\\\
\end{array}\]
\[\begin{array}{@{}lll@{}}
\irule[Call][composite-call]{
   \vec{v} = g_c,\alpha_p,v,v_i,v_s,v_{o},v_z\\
   \memo',g_i' = \mathtt{extend}(\memo,\max(v_i+v_s,v_{o}+v_z))\\
   m = \newCallMsg(\vec v, \alpha, g_i, \memo')\\
   \alpha_p = m.\msgTarget = m.\msgCAddr \in {\mathcal A_j}\\
   {\mathcal{A}}' = {\mathcal{A}}[\alpha.\accBal \minuseq v][\alpha_p.\accBal \pluseq v]\\
   (i,g_i'') \in \{(1,g_i'), (2,0)\}\\
    g'' = g_i-g_i''- m.\msgGas\geq 0\\
    \memo'_\varepsilon = \memo'[\ioproj \mapsto \varepsilon]\\
    bs = \memo'[v_i,v_i+v_s)\\
    \memo_{bs}=[\ioproj\mapsto bs]\\
   F = (\CALL^{v_o}_{v_z}, g'' , \alpha , \memo'_\varepsilon , E^L_\nameN)\\
    S = \mathcal{A}(m.\msgCAddr).\accCode\\
    (i,g'_i)\in\{(1,g'),(2,0)\}
}{
  \nbox{
    \conf{{\mathcal{A}} ; T ; \vec g ; K 
    \vdash 
    \alpha^{\conf{i}} ; \memo ; E[\callop(\vec{v})]^L_\nameN}_p
    \trans{}_{1,2} 
    \conf{{\mathcal{A}}' ; T ; \vec g[i,j \mapsto m.\msgGas] ;  F\!::\! K 
    \vdash 
    \alpha_p;\memo_{bs};S^{\varnothing}_\varnothing}_p
  }
}\\\\
\end{array}\]
\[\begin{array}{@{}lll@{}}
\irule[Create][composite-create]{
   \vec v = v,v_1,v_2\\
   \memo',g_i' = \mathtt{extend}(\memo,v_1+v_2)\\
   m = \newCreateMsg(\vec{v},\alpha,g,\memo',\alpha')\\
    \mathcal{A}' = \mathcal{A}[\alpha.\accNonce\pluseq1][ \alpha.\accBal \minuseq v] \uplus \left\{\alpha' \mapsto \conf{0 ,v ,\memo'[v_1,v_1+v_2) , \varepsilon}\right\}\\
   \alpha' = \mathtt{fresh}(\alpha,\mathcal A(a.\accNonce))\\
   (i,g_i'') \in \{(1,g_i'), (2,0)\}\\
   g'' = g_i-g_i''- m.\msgGas\geq 0\\
\memo'_\varepsilon=\memo'[\ioproj\mapsto\varepsilon]\\
   F = (\CREATE,  g'' , \alpha , \memo'_\varepsilon , E^L_\nameN)\\
    S = \mathcal{A}'(\alpha').\accCode\\
    (i,g_i')\in\{(1,g'),(2,0)\}
}{
  \nbox{
    \conf{\mathcal{A} ; T ; \vec g ; K 
    \vdash 
    \alpha^{\conf{i}} ; \memo  ; E[\createop(\vec v)]^L_\nameN}_p
    \trans{}_{1,2} 
    \conf{\mathcal{A}' ; T ; \vec g[i \mapsto m.\msgGas]; F\! ::\! K 
    \vdash 
    \alpha'^{\conf{i}};\bot;S^{\varnothing}_\varnothing}_p
  }
}\\\\
\end{array}\]
\[\begin{array}{@{}lll@{}}
\irule[Ret1][composite-return1]{
    F=(\CALL^{v_o}_{v_z}, g', \alpha'^{\conf{j}} , \memo' , {E'}^{L'}_{\nameN'})\\
    bs = \memo[v_1,v_1+v_2)\\
    g''=g_i+g'\\
   \memo'' = \memo'\left[ [v_o,v_o+v_z) \mapsto bs,\ioproj\mapsto bs\right]\\
}{
  \nbox{
    \conf{\ldots ; \vec  g;F\!::\! K 
    \vdash 
    \alpha^{\conf{i}} ; \memo ; E[\retop(v_1,v_2)]^L_\nameN}_p
    \trans{}_{1,2} 
    \conf{\ldots ; \vec  g[i,j\mapsto g''] ; K 
    \vdash 
    \alpha' , \memo'', {E'}^{L'}_{\nameN'}[1]}_p
  }
}\\\\
\end{array}\]
\[\begin{array}{@{}lll@{}}
\irule[Ret2][composite-return2]{
    F = (\CREATE, g', \alpha' , \memo' , {E'}^{L'}_{\nameN'})\\
    g''=g_i+g'
}{
  \nbox{
    \conf{\ldots ; \vec g ; F\!::\! K 
    \vdash 
    \alpha^{\conf{i}} ; \memo ;E[\retop(v_1,v_2)]^L_\nameN}_p
    \trans{}_{1,2} 
    \conf{\ldots ; \vec g[i\mapsto g'']; K 
\vdash 
    \alpha'^{\conf{i}} , \memo' , {E'}^{L'}_{\nameN'}[\alpha]}_p
  }
}\\\\
\end{array}\]
\[\begin{array}{@{}lll@{}}
\irule[EOCall][composite-eoa-call]{
    S = \mathcal{A}(\amain).\accCode\\
}{
  \nbox{
    \conf{\mathcal{A};\ldots; \varepsilon}_{eo}
    \trans{}_{1,2} 
    \conf{\mathcal{A};\ldots ; \CALL_o :: \varepsilon \vdash 
    \amain^{\conf{2}};\bot;S^{\varnothing}_\varnothing}_p
  }
}\\\\
\end{array}\]
\[\begin{array}{@{}lll@{}}
\irule[EndTx][composite-wait]{
  T' = T + i\\
  i \geq 0\\
  g' \in \Uintset
}{
  \nbox{
    \conf{\mathcal{A} ; T ; \vec g ; \varepsilon }_{eo}
    \trans{}_{1,2} 
    \conf{\mathcal{A} ; T' ; (g',g') ; \varepsilon}_{eo}
  }
}\\\\
\end{array}\]

\begin{definition}[Internal Composition]\label{def:ext:comp}
For a library $\lL$ and compatible client $\cC$, their \emph{internal composite semantics} is:
\[\gamesem{L}\intcomp\gamesem{C} = \{\rho \mid \rho_0 \intcomp \rho_0' \to^*_{1,2} \rho\}\]
where $\to_{1,2}^*$ is the reflexive transitive closure of $\to_{1,2}$, and $\rho_0 \in \gamesem{L}$ and $\rho_0' \in \gamesem{C}$ are the respective initial configurations obtained by deploying $\lL$ and $\cC$.
\defEnd
\end{definition}

The merging of configurations $\rho \intcomp \rho'$ used above is defined as follows. 

\begin{definition}\label{def:int:comp}
Given compatible game configurations $\rho_1\compatible\rho_2$, let their {internal composition} be defined as follows:

\textbf{PO Configurations:}
\begin{flalign*}
\rho_1 &= 
\conf{{\mathcal{A}_1} ; T ; g_1 ; K_1
  \vdash \alpha ; \memo ; S^L_\nameN}_{p}&\\
\rho_2 &= 
\conf{{\mathcal{A}_2} ; T ; g_2 ; K_2}_{o}\\
\qquad\rho_1 \intcomp \rho_2 &= 
\conf{\mathcal{A}^{\conf{1}}_1 \uplus \mathcal{A}^{\conf{2}}_2 ; T ; (g_1,g_2) ; K_1 \intcomp K_2
  \vdash \alpha ; \memo ; S^L_\nameN}_{p}
\end{flalign*}

\textbf{OP Configurations:}
\begin{flalign*}
\rho_1 &= 
\conf{{\mathcal{A}_1} ; T ; g_1 ; K_1}_{o}\\
\rho_2 &= 
\conf{{\mathcal{A}_2} ; T ; g_2 ; K_2
  \vdash \alpha ; \memo ; S^L_\nameN}_{p}&\\
\qquad\rho_1 \intcomp \rho_2 &= 
\conf{\mathcal{A}^{\conf{1}}_1 \uplus \mathcal{A}^{\conf{2}}_2 ; T ; (g_1,g_2) ; K_1 \intcomp K_2
  \vdash \alpha ; \memo ; S^L_\nameN}_{p}
\end{flalign*}

\textbf{EOA Configurations:}
\begin{flalign*}
\rho_1 &= 
\conf{{\mathcal{A}_1} ; T ; g_1 ; K_1}_{o}&\\
\rho_2 &= 
\conf{{\mathcal{A}_2} ; T ; g_2 ; K_2}_{o}\\
\qquad\rho_1 \intcomp \rho_2 &= 
\conf{\mathcal{A}^{\conf{1}}_1 \uplus \mathcal{A}^{\conf{2}}_2 ; T ; (g_1,g_2) ; K_1 \intcomp K_2}_{eoa}
\end{flalign*}
where $\mathcal{A}^{\conf{i}}_i$ means every $\alpha\in\mathcal{A}_i$ has been annotated to be $\alpha^{\conf{i}}$ for $i=1,2$, $\mathcal{A}_1 \uplus \mathcal{A}_2$ is the union ${\mathcal{A}_1}\Pproj \uplus {\mathcal{A}_2}\Pproj$, and $K_1 \intcomp K_2$ is a single stack resulting from the merging of compatible stacks $K_1 \compatible K_2$, defined as follows:
\begin{align*}
\varepsilon &\intcomp \varepsilon &&= \varepsilon&&&&
\\
(\CALL_o :: \varepsilon) &\intcomp (F_{po} :: F_{pp} :: \CALL_o :: \varepsilon) &&= F_{po} :: F_{pp} :: \CALL_o :: \varepsilon
\\
(F_{pp}^* :: \CALL_o :: K_1') &\intcomp (F_{po} :: K_2') &&= F_{pp}^* :: F_{po} :: (K_1' \intcomp K_2')
\\
(F_{po} :: K_1') &\intcomp (F_{pp}^* :: \CALL_o :: K_2') &&= F_{pp}^* :: F_{po} :: (K_1' \intcomp K_2')
\end{align*}
where we write $F_{pp}^*$ for a block of consecutive PP frames.
\defEnd
\end{definition}


\subsection{Bisimilarity of External and Internal Composition}

First, we define bisimilarity between the external and internal composition.

\begin{definition}[Bisimulation]
Let $\rR$ be a relation with elements of the form $(\rho_1,\rho_2)$, where $\rho_1$ and $\rho_2$ are configurations arising from external and internal composition respectively. We say \rR is a \emph{bisimulation} if for all $(\rho_1,\rho_2) \in \rR$:
\begin{itemize}
\item if $\rho_1\to\rho_1'$ then $\rho_2\to^*_{1,2}\rho_2'$ and $(\rho_1',\rho_2')\in\rR$;
\item if $\rho_2\to^*_{1,2}\rho_2'$ then $\rho_1\to\rho_1'$ and $(\rho_1',\rho_2')\in\rR$.
\end{itemize}
%
We write $\rho \rR \rho'$ if $(\rho,\rho')\in\rR$. We say two game configurations $\rho$ and $\rho'$ are \emph{bisimilar}, written $\rho\sim\rho'$, if there exists a bisimulation \rR such that $\rho\rR\rho'$.
\defEnd
\end{definition}

\begin{lemma}[External--Internal Bisimilarity]\label{lem:int:ext:bisim}
For game configurations $\rho\compatible\rho'$, it is the case that $(\rho\semcomp\rho')\sim(\rho\intcomp\rho')$.
\begin{proof}
The bisimilarity between internal and external composition is largely by design. We illustrate this correspondence on a representative case (\iref{po-call}); the remaining cases follow uniformly from the definitions and are omitted.

Let $\rho_1\compatible\rho_2$ be game configurations. Consider the case where $\rho_1$ makes a \pocalll. We want to show:
\begin{enumerate}
\item if $\rho_1\trans{\pocalll(m)}\rho_1'$ and $\rho_2\trans{\ocall(m)}\rho_2'$ via \iref{call1}, then $(\rho_1\intcomp\rho_2)\to_{1,2}(\rho_1'\intcomp\rho_2')$; and
\item if $(\rho_1\intcomp\rho_2)\trans{\calll(m)}_{1,2}(\rho_1'\intcomp\rho_2')$ via \iref{composite-call}, then $\rho_1\trans{\pocalll(m)}\rho_1'$ and $\rho_2\trans{\ocall(m)}\rho_2'$.
\end{enumerate} 
Now we prove this case. Firstly, we have:
\begin{flalign*}
\qquad
\rho_1 &= 
    \conf{\mathcal{A}_1 ; T ; g_1 ; K_1
    \vdash 
    \alpha ; \memo ; E[\callop(\vec{v})]^L_\nameN}_p
&\\
\rho_2 &= 
    \conf{\mathcal{A}_2 ; T ; g_2 ; K_1}_o
&\\
\rho_1 \intcomp \rho_2 &=
    \conf{\mathcal{A}_{1,2} ; T ; \vec g ; K_{1,2}
    \vdash 
    \alpha^{\conf{1}} ; \memo ; E[\callop(\vec{v})]^L_\nameN}_p
\end{flalign*}
such that $\mathcal{A}_{1,2} = \mathcal{A}^{\conf{1}}_1\uplus\mathcal{A}^{\conf{2}}_2$, $\vec g = g_1, g_2$ and $K_{1,2}=K_1 \intcomp K_2$.

For (1), we know from \iref{call1} in the external composition such that $(\rho_1\semcomp\rho_2)\to'(\rho_1'\semcomp\rho_2')$:
\begin{flalign*}
\qquad
\rho_1' &= 
    \conf{\mathcal{A}_1[ \alpha.\accBal \minuseq v][\alpha'\mapsto\bullet] ; T ; m.\msgGas ; (\CALL^{v_o}_{v_z},  g'', \alpha , \memo'_\varepsilon, E^L_\nameN) :: K_1}_o
&\\
\rho_2' &= 
    \conf{\mathcal{A}_2[\alpha'.\accBal \pluseq v][\alpha\mapsto\bullet];T;g_2; \CALL_o :: K_2
    \vdash 
    \alpha';\memo_{bs};S^{\varnothing}_\varnothing}_p
\shortintertext{such that the following composite configuration can be constructed:}
\rho_1' \intcomp \rho_2' &=
    \conf{\mathcal{A}_{1,2}[ \alpha^{\conf{1}}.\accBal \minuseq v][\alpha'^{\conf{2}}.\accBal \pluseq v];T;\vec g'; (\CALL^{v_o}_{v_z},  g'', \alpha , \memo'_\varepsilon, E^L_\nameN) :: K_{1,2}
    \vdash 
    \alpha';\memo_{bs};S^{\varnothing}_\varnothing}_p
\end{flalign*}
where $\vec g' = (m.\msgGas,g_2) = \vec g[1 \mapsto m.\msgGas]$. We can immediately see from \iref{composite-call} that: \[(\rho_1\intcomp\rho_2)\trans{\calll(m)}_{1,2}(\rho_1'\intcomp\rho_2')\]
Thus, case (1) holds.
For (2), we know from \iref{composite-call} such that $(\rho_1\intcomp\rho_2)\trans{\calll(m)}_{1,2} \rho'_{1,2}$:
\begin{flalign*}
\qquad
\rho_{1,2}' &= 
    \conf{{\mathcal{A}_{1,2}}[\alpha^{\conf{1}}.\accBal \minuseq v][\alpha'^{\conf{2}}.\accBal \pluseq v] ; T ; \vec g' ;  (\CALL^{v_o}_{v_z}, g'' , \alpha , \memo'_\varepsilon , E^L_\nameN)\!::\! K_{1,2}
    \vdash 
    \alpha';\memo_{bs};S^{\varnothing}_\varnothing}_p
\shortintertext{where $\vec g' [1 \mapsto m.\msgGas] = (m.\msgGas,g_2)$. We can split it into the following configurations:}
\rho_1' &= 
    \conf{\mathcal{A}_1[ \alpha.\accBal \minuseq v][\alpha'\mapsto\bullet] ; T ; m.\msgGas ; (\CALL^{v_o}_{v_z},  g'', \alpha , \memo'_\varepsilon, E^L_\nameN) :: K_1}_o
&\\
\rho_2' &= 
    \conf{\mathcal{A}_2[\alpha'.\accBal \pluseq v][\alpha\mapsto\bullet];T;g_2; \CALL_o :: K_2
    \vdash 
    \alpha';\memo_{bs};S^{\varnothing}_\varnothing}_p
\end{flalign*}
by definition of $\mathcal{A}_{1,2}$, $\vec g'$, and $K_{1,2}$. We can similarly see from \iref{call1} that:
\[(\rho_1\intcomp\rho_2)\trans{\calll(m)}_{1,2}(\rho_1'\intcomp\rho_2')\]
Thus, case (2) holds. Additionally, case where $\rho_2$ makes the \pocalll holds by symmetry on the current case using the client semantics. Other rules follow similarly.
\end{proof}
\end{lemma}


\subsection{Library-Client Compositionality}
\label{sec:compositionality}

We can now address \cref{lem:compositionality} and prove library-client compositionality for our game semantics. We show that syntactic composition (\cref{def:syn:comp}) can be obtained from an external (semantic) composition (\cref{def:ext:comp}), and vice versa. Since we have shown that internal and external composition are bisimilar (\cref{lem:int:ext:bisim}), and there is an equivalence between internal composition (\cref{def:int:comp}) and syntactic composition, we are able to obtain compositionality. 
\begin{proof}[Proof (\cref{lem:compositionality})]
We need to show that for any library $\lL$ and compatible good client $\cC$, $(1)\Rightarrow(2)\Rightarrow(3)$:
\begin{enumerate}
\item $\gamesem{\syncomp{\lL}{\cC}}$ fails;
\item there exists a $(t_1,\rho_1) \in \gamesem{\lL}$ and $(t_2,\rho_2) \in \gamesem{\cC}$ such that $t_1 \compatible t_2$ and $\rho_1$ is a failed configuration;
\item $\gamesem{\syncomp{\lL}{\cC}}$ weakly fails. 
\end{enumerate} 
We prove each direction:
\paragraph{$(1 \implies 2)$:}
\begin{enumerate}[label=\arabic*.]
\item Let $\gamesem{\syncomp{\lL}{\cC}}$ fail with some good client $\cC$, for some initial deployment balances $v_L$ and $v_C$ respectively for $\lL$ and $\cC$.
\item By our composite semantics, and since $\cC$ is good, we know $\gamesem{\lL}\intcomp\gamesem{\cC}$ also fails.
\item By external-internal bisimilarity (\cref{lem:int:ext:bisim}), we know $(\gamesem{\lL}\intcomp\gamesem{\cC}) \sim (\gamesem{\lL}\semcomp\gamesem{\cC})$, so $\gamesem{\lL}\semcomp\gamesem{\cC}$ also fails.
\item By external composition (\cref{def:ext:comp}), we know $(t_1,\rho_1)\in\gamesem{\lL}$ and $(t_2,\rho_2)\in\gamesem{\cC}$ such that $t_1 \compatible t_2$ and $\rho_1$ failed.
\end{enumerate}
\paragraph{$(2 \implies 3)$:}
\begin{enumerate}[label=\arabic*.]
\item Let $(t_1,\rho_1)\in\gamesem{\lL}$ and $(t_2,\rho_2)\in\gamesem{\cC}$ such that $t_1 \compatible t_2$ and $\rho_1$ failed.
\item By external composition (\cref{def:ext:comp}), we know that $\gamesem{\lL}\semcomp\gamesem{\cC}$ fails.
\item By external-internal bisimilarity (\cref{lem:int:ext:bisim}), $\gamesem{\lL}\intcomp\gamesem{\cC}$ also fails.

\item By our composite semantics, $\gamesem{\syncomp{\lL}{\cC}}$ weakly fails, for the given initial balances $v_L$ and $v_C$ of $\lL$ and $\cC$ respectively. Note that weak failure is needed in order to reconcile for client-semantics rules in $\gamesem{\cC}$ (that consume no gas in internal computations) becoming library-semantics ones in $\gamesem{\syncomp{\lL}{\cC}}$. \qedhere
\end{enumerate}
\end{proof}


\subsection{Definability}\label{sec:definability}
In this section we show that given a library $\lL$ every trace $t$ in $\gamesem{\lL}$ has a corresponding good client $\cC$, {enabled by a daemon $\dD$}, that realises the same trace in its semantics. For convenience, we shall assume in this section that there is a Yul representation for Solidity (typically obtained via compilation), and write all code fragments in Solidity.

Let $(t,\rho) \in \gamesem{\lL}$. Given $t$, a realised opponent comprises the following components:
\begin{itemize}
\item proxy objects $C_i$ for each opponent address $\alpha_i \in \mathcal{A}\Oproj$ occurring in $t$;
\item a \cmain object at a reserved address $\amain \notin \dom{\mathcal A}$ that acts as the central coordinator among the relevant addresses $\alpha_i$ so as to realise the programmatic actions in the trace; 
\item an Externally Owned Account (EOA) $\dD$ that is able to perform the top-level non-programmatic opponent moves in the trace, which include calling $\cmain$ or waiting.
\end{itemize}
{Note that the EOA $\dD$ can be any off-chain entity or process that is able to call client contracts. It is not a Yul or Solidity program, and as such it is not part of our definability arguments, but we include it here for expository reasons.}

\begin{figure}[t]
\centering
\begin{lstlisting}[language=Solidity]
contract Proxy {
	Main main;
	
	function setMain(Main m) external {
		main = m;
	}
	
	function call(address a, bytes memory d) public payable returns (bytes memory) {
		(bool ok, bytes memory r) = a.call{value: msg.value}(d);
		require(ok);
		return r;
	}
	
	fallback(bytes calldata) external payable returns (bytes memory) {
		return main.receiveCall{value: msg.value}(msg.sender, msg.data);
	}
}

\end{lstlisting}
\caption{\solinline{Proxy} contract, deployed at each $\alpha_i\in\mathcal{A}\Oproj$, that forwards calls and returns between \cmain and $\lL$.}
\label{fig:Proxy-contract}
\end{figure}

\begin{figure}[t]
\centering
\begin{lstlisting}[language=Solidity]
contract Main {
	Step[] t; uint k; // trace and step counter
	struct Step {bool isOp ; bool isRetOrWait; address caller; address target; uint256 value; bytes data;}
	
	// EOA D uses this to set the trace
	function setTrace(Step[] calldata trace) external {
		for (uint i = 0; i < trace.length; i++) {
			t.push(trace[i]); // copy to non-volatile storage
		}
		k = 0;
	}
	
	// performs next move in the trace
	function run() public payable returns (bytes memory) {
		Step storage s = t[k++];
		require(s.isOp);
		if (s.isRetOrWait) return s.data;
		bytes memory r = Proxy(payable(s.caller)).call{value: s.value}(s.target, s.data);
		Step storage s_ret = t[k++];
		require(!s_ret.isOp);
		require(s_ret.isRetOrWait);
		require(keccak256(s_ret.data) == keccak256(r));
		return run();
	}
	
	// receives calls from the library (forwarded by a Proxy)
	function receiveCall(address caller, bytes memory d) public payable returns (bytes memory) {
		Step storage s = t[k++];
		require(!s.isOp);
		require(!s.isRetOrWait);
		require(s.caller == caller);
		require(s.target == address(this));
		require(s.value == msg.value);
		require(keccak256(s.data) == keccak256(d));
		return run();
	}
}
\end{lstlisting}
\caption{$\cmain$ contract, deployed at $\amain\notin\dom{\mathcal{A}}$, that stores the trace $t$ and replays it step by step.}
\label{fig:main-contract}
\end{figure}

\begin{figure}[t]
\centering
\begin{lstlisting}[language=Python]
# We consider an object C where:
# - t is a trace of wait, o-call, o-ret, po-call, po-ret, pp-call, pp-ret, create.
# - t_main is t restricted to:
#      o-call(m), o-ret(bs), po-call(m), po-ret(bs) and wait(i).
# - t_eoa is t restricted to wait(i) and o-call(m) whose call depth is 0, 
# 	where depth is computed by well-bracketing.
# - (T,g,A) is an EVM environment where C is deployed:
# 	- C.main is deployed as an instance of the Main contract; and
# 	- C.proxies are deployed as instances of the Proxy contract
# 	- C.main is funded with ||t||.
# - g is gas starting at the initial gas of t

for c in C.proxies:
    c.setMain(C.main.address)

C.main.setTrace(t_main)

for m in t_eoa:
    match m:
        case call(_):
            (T,g,A) = begin_tx{T,g,A}(C.main.run()) # start new transaction main.run() with env (T,g,A)
        case wait(i,g_new):
            (T,g,A) = (T+i,g_new,A)                 # increment time T and update gas
\end{lstlisting}
\caption{Pseudocode depicting $\dD$ that initialises the deployed contracts and replays the top-level (external) moves.}
\label{fig:demon:pseudocode}
\end{figure}

We now give a sketch, in Solidity, of the contracts implementing the $\cmain$ and $C_i$ components of $\cC$; see \Cref{fig:Proxy-contract,fig:main-contract}, together with the EOA pseudocode in \Cref{fig:demon:pseudocode}.
The overall idea is that $\cC$ interacts with $\lL$ by having $\cmain$ replay the programmatic parts of the trace through a family of proxy contracts $C_i$; this is a deterministic interaction that follows the given trace, or fails otherwise, which we consider to have diverged. For every address $\alpha_i \in \mathcal A \setminus O_{\mathit{proj}}$ occurring in $t$, a \solinline{Proxy} contract is deployed at $\alpha_i$. Each such \solinline{Proxy} acts only as a bridge between \solinline{Main} and the addresses owned by $\lL$: calls entering a \solinline{Proxy} are redirected to \solinline{Main}, while calls issued by \solinline{Main} to a \solinline{Proxy} are forwarded onward to the corresponding $\lL$-owned target, with returns propagated back along the same path. For this, a \solinline{Main} contract is deployed at an address $\amain \notin \dom{\rho.\mathcal A}$ that does not appear in $t$. This contract coordinates the replay of the trace by storing a pruned trace $t_{main} = t^\circ$, consisting only of moves observable to Opponent, and by advancing through it one step at a time: each call to \solinline{run()} either causes \solinline{Main} to invoke the appropriate \solinline{Proxy}, which then calls into $\lL$, or returns the next recorded return value to its own caller, which may be either a \solinline{Proxy} or the EOA. Finally, the non-programmatic EOA is realised as a script external to the EVM, written here in Python-style pseudocode. It is given the full trace $t$, derives from $t$ both $t_{main}$ and the top-level trace $t_{eoa}$, initialises the \solinline{Main} and \solinline{Proxy} contracts appropriately, and then replays $t_{eoa}$ by either calling \solinline{Main} or waiting a required number of seconds. Note that \solinline{Main} is deployed with an initial balance $v_\cC$


Let us write $\|t\|$ for the total value of the trace $t$, i.e.\ the sum of $m.\msgVal$'s of all messages (appearing in call labels) in $t$, and ${\mid}t{\mid}$ for the length of $t$.

\begin{definition}[Trace-Realising Clients]\label{def:c:t}
Let $(t,\rho) \in \lL$. Let $\cC_t$ be the client constructed from $t$. It defines a Yul object whose deployment includes the contract \cmain (\Cref{fig:main-contract}) at address $\amain$ (with initial balance $\|t\|$) and, for every $\alpha_o \in \rho.\mathcal A\Oproj$ occurring in $t$, an instance of the \solinline{Proxy} contract (\Cref{fig:Proxy-contract}) at address $\alpha_o$ (initial balance 0).
\defEnd
\end{definition}
\begin{remark}
We assume that for every opponent address appearing in the fixed trace, there exists a salt that causes \cmain to deploy a proxy at that address. Since the trace is finite, the opponent can choose such salts for all required proxies. This is only required for the proof of completeness and is not a statement of feasibility. A more practical client would be limited to deploying fresh addresses such that it can only replay traces under permutation.
\end{remark}
%

Now we define the object code for the proxy and the main contracts. Let $S_P$ be the object code for the \yultext{Proxy} contract. Consider the function implementing \solinline{call} in \solinline{Proxy}:
\begin{lstlisting}[language=solidity]
function call(address a, bytes memory d) public payable returns (address, bytes memory) {
	(bool ok, bytes memory r) = a.call{value: msg.value}(d);
	require(ok);
	return r;
}
\end{lstlisting}
We can consider the continuation after running \yultext{call} to be:
\begin{lstlisting}[language=solidity]
(bool ok, bytes memory r) = [.];
require(ok);
return r;
\end{lstlisting}
In Yul, this corresponds to two nested evaluation contexts: (1)~an out dispatcher and return context ${E_P}_1$, (2)~and inner context ${E_P}_2$ that evaluates the actual external call.
For~(1):
\begin{lstlisting}[language=yul]
let ret_0 := [.]
let memPos := allocate_unbounded()
let memEnd := abi_encode_tuple_t_bytes_memory_ptr__to_t_bytes_memory_ptr__fromStack(memPos , ret_0)
return(memPos, sub(memEnd, memPos))
\end{lstlisting}
which gets a \yultext{ret\_0} memory pointer from a function call that becomes the evaluation context~(2):
\begin{lstlisting}[language=yul]
let expr_34_component_1 := [.]
let expr_34_component_2_mpos := extract_returndata()
let var_ok_25 := expr_34_component_1
let var_r_27_mpos := expr_34_component_2_mpos
let _6 := var_ok_25
let expr_37 := _6
require_helper(expr_37)
let _7_mpos := var_r_27_mpos
let expr_40_mpos := _7_mpos
var__22_mpos := expr_40_mpos
leave
\end{lstlisting}
With these, we shall write ${E_P} = {E_P}_1[{E_P}_2]$ for the full continuation for \yultext{call()}. Since we are considering only the Solidity representation, we shall omit the $L$ and $\nameN$ annotations henceforth.

\yultext{Proxy} also defines a fallback function:
\begin{lstlisting}[language=solidity]
fallback(bytes calldata) external payable returns (bytes memory) {
	return main.receiveCall{value: msg.value}(msg.sender, msg.data);
}
\end{lstlisting}
Note that the field \yultext{main} resides in the account storage (i.e. $\alpha_P.\stateSto$).
The continuation (in Solidity) after running the fallback function would be:
\begin{lstlisting}[language=solidity]
return [.];
\end{lstlisting}
In Yul, this corresponds to two contexts: (1)~an outer context ${E_{fall}}_{1}$ for dispatching and returning, and (2)~and inner context ${E_{fall}}_{2}$ for evaluating \yultext{fallback} and processing its return data. For~(1):
\begin{lstlisting}[language=Yul]
let retval := [.]
return(add(retval, 0x20), mload(retval))
\end{lstlisting}
where \yultext{retval} points to a block of bytes in memory at position \yultext{add(retval, 0x20)} of length \yultext{mload(retval)}. For~(2):
\begin{lstlisting}[language=Yul]
let _13 := [.]
if iszero(_13) { revert_forward_1() }
let expr_59_mpos
if _13 {
	let _14 := returndatasize()
	returndatacopy(_11, 0, _14)
	finalize_allocation(_11, _14)
	expr_59_mpos := abi_decode_tuple_t_bytes_memory_ptr_fromMemory(_11, add(_11, _14))
}
var__48_mpos := expr_59_mpos
leave
\end{lstlisting}
We thus write $E_{fall} = {E_{fall}}_{1}[{E_{fall}}_{2}]$ for the full continuation for \yultext{fallback()}.

On the other hand, $S_{main}$ corresponds to the Yul object code for the \cmain contract with trace $t$ at some step $k$. Recall the \solinline{run()} function:
\begin{lstlisting}[language=solidity]
function run() public payable returns (bytes memory) {
	Step storage s = t[k++];
	require(s.isOp);
	if (s.isRetOrWait) return s.data;
	bytes memory r = Proxy(payable(s.caller)).call{value: s.value}(s.target, s.data);
	Step storage s_ret = t[k++];
	require(!s_ret.isOp);
	require(s_ret.isRetOrWait);
	require(keccak256(s_ret.data) == keccak256(r));
	return run();
}
\end{lstlisting}
we can similarly consider a continuation for \yultext{run}:
\begin{lstlisting}[language=solidity]
bytes memory r = [.];
Step storage s_ret = t[k++];
require(!s_ret.isOp);
require(s_ret.isRetOrWait);
require(keccak256(s_ret.data) == keccak256(r));
return run();
\end{lstlisting}
with corresponding Yul representation ${E_{main}}$, which we shall leave out for brevity.
Notice that \yultext{t} and \yultext{k} are stored in the account storage (i.e. $\amain.\stateSto$), since their values must persist across transactions.
  
Given a trace $t\in\gamesem{\lL}$ we 
next define intermediate configurations that produce suffixes of $t^\circ$ using the client semantics. These will be stepping stones in our definability argument later. 
Recall  that $t^\circ$ comprises all the observable moves in $t$.
We shall split it as $t_1t_2$ where $t_1$ is the portion of $t$ that has been played so far, and $t_2$ the one that remains to be played. In the following we shall assume that all P-moves in $t$ are annotated with the gas of their source configuration: e.g.\ we write $\preturnl{o}^{\conf{g}}$.\footnote{This does not require us to change the game rules; given $\lL$ and the trace $t$, we can run the game semantics of ${\lL}$ and obtain the sequence of configurations producing $t$.}
We also write $t^{\lnot wait}$ for the subsequence of $t$ obtained by removing all wait labels from $t$.


\begin{definition}[Client O-Configurations]\label{def:client:o:confs}
  Let $t\in\gamesem{\lL}$ and suppose  $t^\circ=t_1t_2$ where ${\mid}t_1^{\lnot wait}{\mid}$ is odd.
We define the set of opponent configurations $\mathbb{C}^{t_1}_{t_2}$ that play (in the client game semantics) the remainder trace $t_2$ of $t^\circ$ to contain all tuples of the form
\[
\conf{\mathcal{A}_{t_1} ; T_{t_1} ; g_{t_1} ; K_{t_1}}_o
\]
where:
\begin{itemize}
\item $\mathcal{A}_{t_1}$ is built by adding $[\amain\mapsto \conf{0 ,\|t\|,S_{main}, [\yultext{t} \mapsto t^\circ, \yultext{k} \mapsto {\mid}t_1{\mid}-1]}]$ and, in addition:
  \begin{itemize}
    \item adding/subtracting $m.\msgVal$ from $\amain.\accBal$ for every $\pocalll(m)$/$\ocall(m)$ in $t_1$, 
    \item
      adding $[m.\msgCaller\mapsto\bullet]$ for every $\pocalll(m)$ and 
      $[m.\msgTarget\mapsto\bullet]$ for every $\ocall(m)$ in $t_1$,
      \item adding $[m.\msgTarget\mapsto \conf{0 ,0,S_P, [\yultext{main}\mapsto \amain]}]$ for every $\pocalll(m)$ and $[m.\msgCaller\mapsto \conf{0 ,0,S_P, [\yultext{main}\mapsto \amain]}]$ for every $\ocall(m)$ in $t^\circ$;
      \end{itemize}
\item $T_{t_1}$ is the result of adding the time of every \waitl move in $t_1$ to its initial time;
\item $g_{t_1}$ is either:
    \begin{itemize}
    \item $g_{t_1} = m.\msgGas$ if the last move in $t_1$ is $\ocall(m)$, or
    \item $g_{t_1} = g$ if the last move in $t_1$ is $\waitl(i,g)$, or
    \item $g_{t_1} = g$ otherwise, where the last P-move in $t_1$ has gas annotation $g$;
    \end{itemize}
\item $K_{t_1}$ is the stack produced by starting from $\CALL_o::\varepsilon$ and evaluating every move $t_1$ by pushing:
    \begin{itemize}
    \item $(\CALL^{0}_{0}, 0 , m.\msgTarget , \_\, , E_{fall})::\CALL_o$ for every $\pocalll(m)$, and
    \item $(\CALL^{0}_{0}, 0 , m.\msgCaller , \_\, , {E_P})::(\CALL^{0}_{0}, 0 , \amain , \_\, , {E_{main}})$ for every $\ocall(m)$,
    \end{itemize}
    and by popping for every $\oreturnl{}$ and $\preturnl{o}$ accordingly (the memory component in the stack frames is left unspecified). \defEnd
  \end{itemize}
\end{definition}


Recall below that the dual trace $\overline{t}$ of $t$ is obtained by swapping p and o polarities in moves (see \cref{def:trace:compat}).

\begin{lemma}\label{lem:replayability}
Let $t$ be a good $\lL$ trace such that $t^\circ$ is of the form $t_1t_2$ where ${\mid}t_1^{\lnot wait}{\mid}$ is odd.
For any configuration $\rho\in \mathbb{C}^{t_1}_{t_2}$, $\rho$ produces (using the client semantics) some $\hat t_2$ such that $\hat t_2^\circ=\overline{t_2}$.
\end{lemma}
\begin{proof}
  We do induction on ${\mid}t_2{\mid}$. The base case for ${\mid}t_2{\mid}=0$ is trivial.
  For the inductive step, suppose that $\eta$ is the first move in $t_2$.
  This corresponds to:
  \begin{itemize}
    \item[-]
      either a library move (i.e.\ a P-move in $t_2$, and an O-move in $\overline{t_2}$), and therefore $\overline{\eta}$ can be played by $\rho$, as opponent can always play a given move so long as the stack discipline is respected;
    \item[-] or a wait move (i.e.\ an O-move in both $t_2$ and $\overline{t_2}$), and therefore $\overline{\eta}=\eta$ can be played by $\rho$ as its stack must be empty by definition and the fact that $t$ is a trace for $\lL$. 
    \end{itemize}
    If $\eta$ is the last move in $t_2$ then we are done. Otherwise, suppose that we start from configuration $\rho=\conf{\mathcal{A}_{t_1} ; T_{t_1} ; g_{t_1} ; K_{t_1}}_o$.
    If $\eta=\waitl(i,g)$ then, 
    by performing $\eta$, $\rho$ leads to $\conf{\mathcal{A}_{t_1} ; T_{t_1}+i ; g ; K_{t_1}}_o=\conf{\mathcal{A}_{t_1\eta} ; T_{t_1\eta} ; g_{t_1\eta} ; K_{t_1\eta}}_o$ and we use the inductive hypothesis to conclude. 
Next, let us suppose that $\eta$ is not a wait label and $t_2=\eta\,\zeta\,t_2'$, 
and do a case analysis on ${\eta}$. 
\begin{asparaitem}
\item $\eta=\pocalll(m)$. Performing $\overline{\eta}=\ocall(m)$ from $\rho$ (using \iref{C-o-call}), we move to
  \[
    \rho'=\conf{\mathcal{A}_{t_1}[\alpha_p.\accBal\pluseq v][\alpha_o\mapsto\bullet] ; T_{t_1} ; m.\msgGas ; \CALL_o::K_{t_1}\vdash\alpha_p;\memo_{bs};S_\varnothing^\varnothing}_p\]
  where $\alpha_p=m.\msgTarget$, $\alpha_o=m.\msgCaller$, $S=\mathcal{A}_{t_1}(\alpha_p).\accCode=S_P$ and $bs=m.\msgData$. Next, the proxy contract invokes its fallback function, which results into a PP-call to $\amain$ (via \iref{C-pp-call}), leading to configuration
  \[
    \rho''=\conf{\mathcal{A}_{t_1}[\amain.\accBal\pluseq v][\alpha_o\mapsto\bullet] ; T_{t_1} ; m.\msgGas ; (\CALL^0_0, 0 , \alpha_p , \_\, , E_{fall})\,{::}\,\CALL_o\,{::}\,K_{t_1}\vdash\amain;\memo_{\hat{bs}};S_{main\,\varnothing}^\varnothing}_p\]
where $\hat{bs} = ((\text{\solinline{Main.receiveCall.selector}}),(m.\msgCaller,bs))$, at which point \yultext{Main.receiveCall} is invoked.
After checking that $\eta$ is indeed the next label in the trace, and increasing the counter $\yultext{k}$, the function calls \yultext{Main.run}. The latter
increases $\yultext{k}$ and
fetches the next move $\zeta$ from storage and if it is a return label then it returns, otherwise it issues a call to the corresponding proxy contract. We
set $\mathcal{A}'=\mathcal{A}_{t_1}[\amain.\accBal\pluseq v][\alpha_o\mapsto\bullet][\alpha_{main}.\stateSto(\yultext{k})\pluseq2]$ and
  examine each case separately:
  \begin{itemize}
  \item If $\zeta=\oreturnl(bs')$ then $\amain$ returns (using \iref{pp-return1}) and we reach (for some $\memo'$):
  \[
    \rho'''=\conf{\mathcal{A}' ; T_{t_1} ; m.\msgGas ; \CALL_o::K_{t_1}\vdash\alpha_p;\memo'_{bs'};E_{fall}[1]}_p\]
    This, in turn, returns (using \iref{po-return}) and we reach:
  \[
    \hat\rho=\conf{\mathcal{A}' ; T_{t_1} ; m.\msgGas ; K_{t_1}}_o\]
  \item If $\zeta=\ocall(m')$ then note that \iref{o-call} (for $t$) imposes that $m'.\msgGas=m.\msgGas$. Now, $\amain$ issues a call to the proxy contract with address $\alpha_p'=m'.\msgCaller$ and, using $v'=m'.\msgVal$, $bs'=m'.\msgData$ and $K'=(\CALL^0_0, 0 , \alpha_p , \_\, , E_{fall})::\CALL_o::K_{t_1}$, we reach (via \iref{C-pp-call}):
    \[
      \rho'''=\conf{\mathcal{A}'[\amain.\accBal\minuseq v'][\alpha'_p.\accBal\pluseq v'] ; T_{t_1} ; m'.\msgGas ; (\CALL^0_0, 0 , \alpha_{main} , \_\, , E_{main})::K'\vdash\alpha_p';\memo_{bs'};S_{P\,\varnothing}^\varnothing}_p
    \]
    which, in turn, issues the call to $\alpha_o'=m'.\msgTarget$ (via \iref{C-po-call}) and we obtain:
    \[
      \hat\rho=\conf{\mathcal{A}'[\amain.\accBal\minuseq v'] ; T_{t_1} ; m'.\msgGas ; (\CALL^0_0, 0 , \alpha_{p}' , \_\, , E_P)::(\CALL^0_0, 0 , \amain , \_\, , E_{main})::K'}_o
    \]
  \end{itemize}
  In both cases, we can verify that $\hat\rho=\conf{\mathcal{A}_{t_1'};T_{t_1'};g_{t_1'};K_{t_1'}}$, where $t_1'=t_1\,\eta\,\zeta$.
\item $\eta=\preturnl{o}(bs)^{\conf{g}}$.
  Since $\rho=\conf{\mathcal{A}_{t_1} ; T_{t_1} ; g_{t_1} ; K_{t_1}}_o$, by definition we have that $K_{t_1}=(\CALL^{0}_{0}, 0 , \alpha_p , \_\, , {E_P})::(\CALL^{0}_{0}, 0 , \amain , \_\, , {E_{main}})::K'$, where $\ocall(\hat m)$ is the matching previous call move in $t_1$ and $\alpha_p=\hat m.\msgCaller£$.
  Performing $\overline{\eta}=\oreturnl(bs)$ (using \iref{C-o-return}), we move (for some $\memo'$) to
  \[
    \rho'=\conf{\mathcal{A}_{t_1}; T_{t_1} ; g ; (\CALL^{0}_{0}, 0 , \amain , \_\, , {E_{main}})::K'\vdash\alpha_p;\memo'_{bs};E_P[1]}_p\]
and in the continuation $E_P$ of the proxy contract. The latter returns to $\alpha_{main}$ (via \iref{pp-return1}), leading (for some $\memo''$) to configuration
  \[
    \rho''=\conf{\mathcal{A}_{t_1}; T_{t_1} ; g ; K''\vdash\amain;\memo''_{{bs}};E_{main}[1]}_p\]
At this point, we are back to
\yultext{Main.run},
at the continuation after the call (in line~5),
and check that $\eta$ is indeed the next label in the trace, increase the counter $\mathtt{k}$, and recursively call \yultext{Main.run}.
The latter
increases \yultext{k} and 
 fetches the next move $\zeta$ from storage and if it is a return label then it returns, otherwise it issues a call to the corresponding proxy contract. We
set $\mathcal{A}'=\mathcal{A}_{t_1}[\alpha_{main}.\stateSto(\yultext{k})\pluseq2]$ and continue as in the previous case.
\end{asparaitem}
In both cases above, using the inductive hypothesis we obtain that $\rho$ produces some $\hat t_2$ which, by projection and complement, matches $t_2$. 
\end{proof}
With the infrastructure above and the replayability of traces (\Cref{lem:replayability}), we can move on to proving definability. Recall \Cref{lem:definability}, which states that, given any library $\lL$, for any $(t,\rho)\in\gamesem{\lL}$ there exists a good client that is compatible with $\lL$ on trace $t$.
\begin{proof}[Proof (\Cref{lem:definability})]
  Let $(t,\rho)\in\gamesem{\lL}$ and consider the good client $\cC_t$ of \cref{def:c:t}. Let $\rho_0=\conf{\mathcal{A}_0;T_0; g_0; \varepsilon}_o$ be the configuration reached by $\gamesem{\cC}$ (using the client semantics) right after deployment.
Here, $T_0,g_0$ are taken from the first move in $t$.
  We also assume that the initial balance for $\cC$ is $\|t\|$ and is passed on to $\alpha_{main}$ (i.e.\ $\mathcal{A}_0(\alpha_{main}).\accBal=\|t\|$).
  If the first move of $t^\circ$ is some $\waitl(i,g)$ then $\cC_t$ can make the move:
\[\rho_0
    \trans{\waitl(i,g)}
    \conf{\mathcal{A}_0;T_0+i ; g;\varepsilon}_o
  \]
So, assume WLOG that $t^\circ = t_1 t_2$ with $t_1=\ocall(m)$. Then, $\cC_t$ can make the move:
\[
  \rho_0
    \trans{\eocall()}
    \rho_0'=\conf{\mathcal{A}_0;T_0 ; g_0;\CALL_o :: \varepsilon \vdash 
    \amain;\bot;S^{\varnothing}_{main\,\varnothing}}_p
\]
From $\rho_0'$, the main contract will invoke \yultext{run}, which in turn will set \yultext{k} to 1 and check the first move in $t^\circ$, i.e.\ $\ocall(m)$. Reasoning as in the proof of \cref{lem:replayability}, we will issue moves $\ppcalll(m)\,\pocalll(m)$ and reach a configuration:
    \[
      \rho_1=\conf{\mathcal{A}_1 ; T_0 ; m.\msgGas ; (\CALL^0_0, 0 , \alpha_{p} , \_\, , E_P)::(\CALL^0_0, 0 , \amain , \_\, , E_{main})::\CALL_o}_o
    \]
    where
$\mathcal{A}_1=\mathcal{A}_0[\amain.\accBal\minuseq v][\alpha_{main}.\stateSto(\yultext{k})=1]$,
    $v=m.\msgVal$ and $\alpha_p=m.\msgTarget$. We now note that, by definition, $\rho_1=\conf{\mathcal{A}_{t_1},T_{t_1},g_{t_1},K_{t_1}}$. Using \cref{lem:replayability}, we have that $\rho_1$ can produce a trace $\hat t_2$ such that $\hat t_2^\circ=\overline{t_2}$, which concludes the proof.
\end{proof}


\section{Motivating Example for an Opponent Knowledge Model}
\label{sec:example:secrets}
\begin{lstlisting}[float,
                   caption={Blind Auction (Solidity 0.8+).},
                   label={lst:simple-blind-auction}]
contract SimpleBlindAuction {
    address payable public immutable beneficiary; // who receives the money
    uint public immutable biddingEnd; // time after which no more bid() calls are accepted
    uint public immutable revealEnd; // time after which no more reveal() calls can be made

    // auction accounting
    struct Info { bytes32 b; uint d; } // blinded bid + total deposit
    mapping(address => Info) private info; // last blinded bid + total deposit per bidder
    mapping(address => uint) private refund; // how much each address can withdraw

    // auction results
    address public highestBidder; // winning account so far
    uint public highestBid; // highest bid revealed so far
    bool public ended; // true if auctionEnd() has already been successfully called
    
    constructor(uint bidTime, uint revTime, address payable bene) {
        beneficiary = bene;
        biddingEnd = block.timestamp + bidTime;
        revealEnd = biddingEnd + revTime;
    }

    // place a bid as keccak(value, secret)
    function bid(bytes32 blindedBid) external payable {
        require(block.timestamp < biddingEnd);
        info[msg.sender].b = blindedBid;
        info[msg.sender].d += msg.value;
    }

    // reveals and sets refund; may need revealing again if outbid
    function reveal(uint v, bytes32 s) external {
        require(block.timestamp >= biddingEnd && block.timestamp < revealEnd);
        Info storage x = info[msg.sender];
        if (x.b != keccak256(abi.encodePacked(v, s))) return;
        if (x.d >= v && v > highestBid) {
            if (highestBidder != address(0)) refund[highestBidder] += highestBid;
            highestBid = v;
            highestBidder = msg.sender;
            refund[msg.sender] = x.d - v;
        } else refund[msg.sender] = x.d;
        x.b = bytes32(0); x.d = 0;
    }

    // refund whatever is set to refund; needs calling reveal() beforehand
    function withdraw() external {
        if (refund[msg.sender] == 0) return;
        refund[msg.sender] = 0;
        (bool ok,) = payable(msg.sender).call{value:refund[msg.sender]}("");
        require(ok);
    }

    // sets auction to end and sends highest bid value to beneficiary
    function auctionEnd() external {
        require(block.timestamp >= revealEnd && !ended);
        ended = true;
        (bool ok,) = beneficiary.call{value:highestBid}("");
        require(ok);
    }
}
\end{lstlisting}

We present a motivating example for the need of an opponent knowledge model in a realistic contract setting. We consider the Solidity contract in \cref{lst:simple-blind-auction}, a simplified blind auction that uses the \emph{commit-reveal} pattern.

The contract has two phases. During the bidding phase, which lasts until |biddingEnd|, a bidder calls |bid(blindedBid)| and submits a commitment of the form |keccak256(abi.encodePacked(value, secret))| together with a deposit. For each bidder, the contract stores only the latest commitment and the total deposited amount. During the reveal phase, which lasts from |biddingEnd| until |revealEnd|, a bidder calls |reveal(v, s)| to open the commitment by supplying a bid value |v| and a secret |s|. If the opening matches the stored commitment and the deposit is sufficient, the bid is accepted and may become the current highest bid. After the reveal phase, |auctionEnd()| finalises the auction and transfers the winning amount to the beneficiary, while bidders recover funds through |withdraw()|.

Assuming $\keccak$ is a cryptographically secure hash function, the contract's behaviour depends on values not available to Opponent. During the bidding phase, the daemon $\dD$ can observe the commitment |blindedBid| and the deposited amount by inspecting publicly available blockchain data (transaction messages, account states, execution traces, etc.), but cannot determine the underlying pair |(value, secret)| that produced them: Opponent can instruct $\cC$ to replay any hash observed on-chain, but cannot produce a value that, when hashed, matches it.

The example also shows that the opponent model must be multi-transactional. Opponent cannot be limited to a client $\cC$ operating within a single transaction; it must also include a daemon $\dD$ acting outside the EVM's programmatic scope. The daemon can inspect public blockchain data, observe the outcome of earlier transactions, wait for time to pass and the contract state to evolve, and then submit new transactions in response. In the blind auction, for example, $\dD$ may observe commitments during the bidding phase, wait until the reveal phase, and only then decide whether and how to interact. Our game semantics captures precisely this scenario: Opponent combines an on-chain client $\cC$ with an off-chain daemon $\dD$ that orchestrates interactions across multiple transactions.

\section{Modelling EVM-Yul in \yult}
\label{sec:yultracer:evm}

We describe the EVM dialect of Yul used by \yult, which builds on the operational semantics and interpreter of Yul presented in~\cite{yul:semantics}.

\subsection{Supporting Yul Objects}
A Yul program comprises \emph{objects} which contain
a code block of Yul statements and potentially nested member objects,
addressed by dot-separated paths from the current object to the target.
A representative object produced by the Solidity compiler is shown next.

\begin{lstlisting}[language=Yul]
/// @use-src 0:"src/DAO.sol"
object "DAO_3087" {
    code {
        ... // memory initialisation and checks
        let _1, _2, _3, _4, _5, _6 := copy_arguments_for_constructor_1235_object_DAO_3087()
        constructor_DAO_3087(_1, _2, _3, _4, _5, _6)
        let _7 := allocate_unbounded()
        codecopy(_7, dataoffset("DAO_3087_deployed"), datasize("DAO_3087_deployed"))(*@\label{ln:dataoffset}@*)
        return(_7, datasize("DAO_3087_deployed"))
        // ... auxiliary functions ...
    }
    /// @use-src 0:"src/DAO.sol"
    object "DAO_3087_deployed" {
        code {
          ...(*@
          %   /// @src 0:29429:47003  "contract DAO is DAOInterface, TokenCreation ..."
          %   mstore(64, memoryguard(128))
          %   if iszero(lt(calldatasize(), 4)) {
          %       let selector := shift_right_224_unsigned(calldataload(0))
          %       switch selector
          %       // ... other cases ...
          %       case 0x82661dc4 {
          %           // splitDAO(uint256,address)
          %           external_fun_splitDAO_2352()
          %       }
          %       // ... other cases ...
          %       default { /* ... payable fallback ... */ }
          %   }
          %   // ... auxiliary functions ...
          @*)
        }
        data ".metadata" hex"a264697..."
    }
}
\end{lstlisting}
In Yul we do not have the bytecode representation for each object, so \yult identifies each object by a unique 256-bit value (\yulinline{uint256}) that it can copy into memory.
In the above example, the constructor code copies arguments from the message that called it to memory and calls an auxiliary function \yulinline{constructor_DAO_n}. Line~\ref{ln:dataoffset} uses object-specific instructions:
\yulinline{dataoffset("DAO_3087_deployed")} yields the object’s ID in \yult, \yulinline{datasize(...)} returns \num{32} (the ID size), and \yulinline{codecopy(_7,...)} copies the 32-byte ID into memory at \yulinline{_7}. The subsequent \yulinline{return} passes this ID to our runtime environment, which looks up the object by ID and deploys it at the current address.

\subsection{Our EVM Dialect and Instructions Supported}
\label{sec:evm}

Our EVM dialect comprises a substantially complete custom implementation of the EVM execution layer -- originally based on the Ethereum Shanghai hard fork upgrade as accessed in October 2023 (corresponding to Solidity v0.8.22\cite{solidity-0.8.22-release}), but updated to reflect minor changes in the specifications -- and a Game Semantics for EVM-flavoured Yul. We have not yet implemented the Transient Storage\cite{solidity-transient-storage} (EIP-1153\cite{eip1153}) of the Cancun upgrade\cite{solidity-0.8.24-release}. Since Yul is relatively stable across versions and only differs in the opcodes available to it through its EVM dialect, we can in principle cover Solidity beyond v0.8.30 so long as transient storage is not used.

\begin{remark}
  Transient storage is a volatile variant of regular storage that does not persist across transactions. It is cleared at the end of each transaction, meaning it persists through nested calls within the same transaction. From the point of view of \yult, this is effectively another storage field that is cleared whenever the opponent performs a wait move (which ends transactions and lets time pass). While this is a simple change, it is not in scope for this paper as none of the benchmarks or real-world examples tested use transient storage.
\end{remark}

Our implementation ports the Python client execution specification~\cite{ethereum-execution-specs} to OCaml, preserving line-by-line fidelity where possible.
\yult supports the full range of opcode categories defined in the specifications~\cite{ethereum-execution-specs}: Arithmetic, Bitwise, Block, Comparison, Control-flow, Message Environment, Keccak, Memory, Stack, Storage and System.
It also supports Yul-specific \emph{data} opcodes for objects (see previous section), \emph{immutable values} and \emph{linking}.
Opcodes irrelevant to Yul (e.g.~certain stack operations) or with no semantic effect are implemented as no-ops. Next, we highlight some details.

\emph{Gas} costs (the EVM execution cost model) are computed concretely in \yult using \yulinline{uint256} arithmetic for supported opcodes.
However, because Yul replaces explicit stack and jump instructions with structured control flow, gas computations are under-approximations of actual cost.

\emph{Memory} is a symbolic-capable byte-indexable total map from concrete |uint256| indexes to potentially symbolic 8-bit |byte| values. Thus, memory acts like a large byte array that is implicitly zero-initialised where accessing higher addresses incurs substantially higher gas costs. Loading from and storing to memory requires combining and splitting |bytes| into |uint256| words and vice-versa, respectively. Doing so accumulates constraints in the symbolic environment.

\emph{Storage} is a symbolic-capable word-indexable map from |uint256| words to potentially symbolic 256-bit EVM words. Because it is word-indexable, words can be loaded and stored whole, so accumulating symbolic constraints is not needed.

\emph{The linker} 
allows users to provide placeholders for libraries to be replaced with addresses during compilation. In Yul, they are another feature officially specified using a non-standard instruction -- the compiler produces Yul objects containing a Yul-specific \yulinline{linkersymbol("library_id")} opcode that takes a string containing an ID for the library to be linked (replaced with an address). Because we start from an empty initial configuration, we model the linker by first extending the EVM's environment component with a |linker| component. We then add another custom opcode \yulinline{SETLINKER("name",address)} that records into the |linker| component that \yulinline{"name"} maps to |address|. As part of the pipeline, we run linking scripts that preprocess all libraries to be linked by deploying them at the top of the top-level object with blocks of the following form (e.g. for a library "Lib5").
 \begin{lstlisting}[language=Yul]
 let lib_5 := mload(64)
 codecopy(lib_5, dataoffset("Lib5_22223"), datasize("Lib5_22223"))
 let lib_5_address := create(0, lib_5, datasize("Lib5_22223"))
 SETLINKER("Lib5_22223", Lib5_22223)
 \end{lstlisting}
 With these set, the getter opcode, e.g. \yulinline{linkersymbol("Lib5_22223")}, simply looks up the value of \yulinline{"Lib5_22223"} -- again converted into |uint256| -- using the |linker| map.

\emph{Reverts} prune the current path -- we consider all \yulinline{revert} instructions to be aborting the current transaction and do not explore behaviours after the \yulinline{revert}. This means we partially handle errors in the EVM as we do not handle safely recovering from exceptions.

We omit the implementation of opcodes that are not listed in the official documentation for Yul\cite{yul} (e.g. stack operations handled by Yul's higher-level syntax); the Cancun instructions \yulinline{tload}, \yulinline{tstore}, \yulinline{blobbasefee}, and \yulinline{blobhash}; and all the opcodes in the Log category (e.g. compiled from |emit| statements), as these do not affect the behaviour of contracts.

\begin{remark}
\yulinline{memoryguard} is a Yul-specific instruction available in the EVM dialect with objects, typically found as the first instruction in Yul objects. In the Solidity compiler, it is used to reserve memory for the optimiser if necessary and initialise the free pointer. That is, a statement |let ptr := memoryguard(size)| tells us that the optimiser will only use memory in the range [|size|,|ptr|), where the scratch-pad portion of memory is everything under |size|. Since the optimiser is only used within the Solidity compiler's pipeline, once the compiler lowers Yul into bytecode, the reserved memory vanishes and the instruction effectively evaluates to |ptr|. As such, the reserved optimiser memory does not exist at runtime, so we model \yulinline{memoryguard} in \yult as an identity function.
\end{remark}

\paragraph{Custom opcodes} 
\yult also defines a set of custom opcodes (written in uppercase) used exclusively for analysis,
inserted post-compilation in the place of hooks written by the user in the program. The main custom opcodes include:
\yulinline{IMPERSONATECALL}, a variant of the \yulinline{call} opcode with an additional address argument specifying the \yulinline{msg.sender} of the call message;
\yulinline{ASSERT}, which halts the analysis and reports the Game Semantics trace;
\yulinline{WAIT}, which advances time by the predefined interval;
\yulinline{REVEAL_UINT} and \yulinline{REVEAL_ADDR}, which add values to the Opponent's known-value sets;
\yulinline{EXT_FUND}, directly increasing the balance of an address;
\yulinline{PRINT}, \yulinline{PRINT_signed}, \yulinline{PRINT_hex}, etc.\ which output values to the terminal.
(\yulinline{MK_SYMBOL}) -- creates a symbolic value, currently unused in our experiments; 
\yulinline{START_ANALYSIS} -- starts the game by passing control to the opponent;

 \begin{remark}
 \yult was originally designed to be a symbolic execution engine. For this reason, the models in  it are compatible with symbolic reasoning using symbolic bit-vectors. Types in the EVM include only |uint256| for words, and |bytes| otherwise (e.g. handled via memory), which can both be modelled by splitting and combining bit-vectors. However, in our experiments, symbolic execution was intractable so we do not present the usage of symbolic values. Instead, we perform an exhaustive games exploration with a bounded domain of values.
 \end{remark}

 \paragraph{Data instructions} 
 \label{sec:yul:data}
 are Yul-specific opcodes that enable programmatic access of objects nested within the current running object. Using our model for bytecode, the model for data instructions becomes very simple:
 \begin{itemize}
 	\item \yulinline{datasize}: always returns 32 (i.e. the byte length for |uint256|);
 	\item \yulinline{dataoffset}: the identity function -- at the level of Yul semantics, before being passed to the EVM, the input string is replaced with the object's UID, which is then returned as its offset;
 	\item \yulinline{datacopy}: identical to \yulinline{codecopy} -- if the offset is negative, this is the identity function; otherwise, it copies a portion from the EVM code component (e.g. arguments).
 \end{itemize}

 \paragraph{Immutable variables} 
 \label{sec:yul:immutable}
 are officially specified in Yul by two opcodes \yulinline{setimmutable} and \yulinline{loadimmutable}. These take string arguments and thus do not map cleanly into Yul (strings are not a word type in the EVM dialect). We therefore model immutables by first converting these string arguments into |uint256| words (e.g. by their ASCII representation or by hashing). We then extend the account triple with an |immutables| component -- a map from concrete |uint256| numbers to EVM words. Finally, we implement the immutable instructions:
 \begin{itemize}
 	\item \yulinline{setimmutable(_,"name",value)}: updates |account.immutables|[\yulinline{"name"}$\mapsto$|value|].
 	\item \yulinline{loadimmutable("name"}: looks up |account.immutables|[\yulinline{"name"}].
 \end{itemize}

 \begin{remark}
 The Solidity immutable feature allows declaring state variables whose values are set once within the constructor and cannot be changed thereafter. Immutable variables are typically stored in the code component of accounts, appended after object code -- i.e. they compile into \yulinline{codecopy} and \yulinline{mload} instructions and are managed programmatically in bytecode. This differs from Yul, which defines two non-standard opcodes and exposes the feature to Solidity's inline assembly.
 \end{remark}


\section{Solidity Upgrades}
\label{sec:upgrades}
\lstset{
  basicstyle=\ttfamily\small,
  breaklines=true,
  keepspaces=true,
}

\newcommand{\toarrow}{\ensuremath{\;\to\;}}

\noindent
Here we describe the main source-level changes we applied to upgrade older Solidity source code to 0.8.x.
For each kind of change we provide an example and a file name from our Gigahorse and DAO experiments where this change occurred.

\paragraph{Pragma Version}

The version pragma was updated to target Solidity~0.8.x.

\begin{center}
  \lstinline!^0.4.2! \toarrow{} \lstinline!^0.8.0! \quad
  (from \texttt{SimpleDAO.sol})
\end{center}

\paragraph{Integer Type Disambiguation}
Many
upgrades make the bit-width explicit for clarity and for stricter ABI-encoding
compatibility.
This was by far the most frequent change (113~occurrences across the Gigahorse
benchmarks alone).

\begin{center}
  \lstinline!uint! \toarrow{} \lstinline!uint256! \quad
  (from \texttt{SimpleDAO.sol})
\end{center}

\paragraph{Deprecated Global Variable \texttt{now}}

The global alias \lstinline!now! for the current block timestamp was deprecated in
favour of the explicit \lstinline!block.timestamp!.

\begin{center}
  \lstinline!now! \toarrow{} \lstinline!block.timestamp! \quad
  (from \texttt{EtherStore.sol})
\end{center}

\paragraph{Low-Level Call Syntax}

The old chained-call notation \lstinline!addr.call.value(x)()! was replaced by the
\emph{call-options} syntax.
The return value (a boolean success flag and optional return data) must now be
explicitly captured in a tuple.

\begin{center}
  \lstinline!msg.sender.call.value(x)()! \toarrow{}
  \lstinline!(bool ok,) = msg.sender.call{value: x}("")! \quad
  (from \texttt{EtherStore.sol})
\end{center}

\paragraph{Fallback and Receive Functions}

In Solidity~0.8.x 
the old unnamed \lstinline!function()! syntax is 
replaced by
two special functions: \lstinline!receive()! and
\lstinline!fallback()!.

\begin{center}
  \lstinline!function() public! \toarrow{} \lstinline!receive() external! \quad
  (from \texttt{Reentrance.sol}) \\[4pt]
\end{center}

\paragraph{Constructor Syntax}

Named constructor functions (which shared the contract name) were replaced by the
\lstinline!constructor! keyword.

\begin{center}
  \lstinline!function PrivateDeposit(...) public! \toarrow{} \lstinline!constructor(...)! \quad
  (from \texttt{PrivateDeposit.sol})
\end{center}

\paragraph{\texttt{address()} Wrapping}

Several implicit conversions to \lstinline!address! that were permitted in~0.4.x
now require an explicit cast.

\begin{center}
  \lstinline!this! \toarrow{} \lstinline!address(this)! \quad
  (from \texttt{LedgerChannel.sol}) \\[4pt]
  \lstinline!0x0! \toarrow{} \lstinline!address(0)! \quad
  (from \texttt{LedgerChannel.sol}) \\[4pt]
  \lstinline!extraBalance.balance! \toarrow{}
  \lstinline!address(extraBalance).balance! \quad
  (from \texttt{TokenCreation.sol})
\end{center}

\paragraph{\texttt{payable()} Casting}

Addresses that receive Ether via \lstinline!.transfer()! or \lstinline!.send()!
must be explicitly cast to \lstinline!payable!.

\begin{center}
  \lstinline!msg.sender! \toarrow{} \lstinline!payable(msg.sender)! \quad
  (from \texttt{TokenCreation.sol}) \\[4pt]
  \lstinline!addr.transfer(x)! \toarrow{} \lstinline!payable(addr).transfer(x)! \quad
  (from \texttt{LedgerChannel.sol})
\end{center}

\paragraph{Function Visibility and Mutability Modifiers}

The \lstinline!constant! modifier was split into \lstinline!view! (does not modify
state) and \lstinline!pure! (does not read state).
Functions with no visibility modifiers which were implicitly public, must have an explicit modifier in 0.8.x.
Additionally, overridable functions in inheritance hierarchies require explicit
\lstinline!virtual! and \lstinline!override! annotations.

\begin{center}
  \lstinline!function balanceOf(address _owner) constant returns (uint256 balance);!\\
  \toarrow{}\\
  \lstinline!function balanceOf(address _owner) public view virtual returns (uint256 balance);!
  \\ \quad
  (from \texttt{DAO.sol})
\end{center}

\paragraph{Error Handling: \texttt{throw} to \texttt{require}/\texttt{revert}}

The \lstinline!throw! statement was removed.
The idiomatic replacement is \lstinline!require()! for input validation (returns
remaining gas) or \lstinline!revert()! for unconditional failure.

\begin{center}
  \lstinline!if (intitalized) throw! \toarrow{}
  \lstinline|require(!intitalized)| \quad
  (from \texttt{PENNY\_BY\_PENNY.sol})
\end{center}

\paragraph{\texttt{sha3} to \texttt{keccak256}}

The \lstinline!sha3()! global was renamed to \lstinline!keccak256()!.
Furthermore, 0.8.x requires arguments to be explicitly ABI-packed via
\lstinline!abi.encodePacked()! rather than passed directly.

\begin{center}
  \lstinline!sha3(a, b)! \toarrow{}
  \lstinline!keccak256(abi.encodePacked(a, b))! \quad
  (from \texttt{DAO.sol})
\end{center}

\paragraph{\texttt{memory} Keyword for Reference Types}

Reference types (\lstinline!bytes!, \lstinline!string!, arrays, structs) used as
function parameters or local variables must now carry an explicit data-location
keyword: \lstinline!memory!, \lstinline!storage!, or \lstinline!calldata!.

\begin{center}
  \lstinline!string! \toarrow{} \lstinline!string memory! \quad
  (from \texttt{LedgerChannel.sol})
\end{center}

\paragraph{Dynamic Array Push}

The old idiom of extending a dynamic storage array via \lstinline!arr.length++!
is no longer supported; \lstinline!arr.push()! must be used instead.

\begin{center}
  \lstinline!proposals.length++! \toarrow{} \lstinline!proposals.push()! \quad
  (from \texttt{DAO.sol})
\end{center}

\paragraph{Byte Literal Casting}

Integer literals used as byte values must be explicitly converted through
\lstinline!bytes1(uint8(...))!.
The old \lstinline!byte! type alias was also renamed to \lstinline!bytes1!.

\begin{center}
  \lstinline!48! \toarrow{} \lstinline!bytes1(uint8(48))! \quad
  (from \texttt{LedgerChannel.sol})
\end{center}

\paragraph{ABI Encoding for Function Selectors}

Hard-coded function selectors built with \lstinline!bytes4(sha3(...))! must be
replaced with \lstinline!abi.encodeWithSignature(...)!, which computes the selector
from the human-readable signature string.

\begin{center}
  \lstinline!bytes4(sha3("f(address,uint256)"))! \toarrow{}
  \lstinline!abi.encodeWithSignature("f(address,uint256)")! 
  \\\qquad
  (from \texttt{reentrancy\_0x627f\ldots.sol})
\end{center}

\paragraph{\texttt{var} Keyword Removal}

The \lstinline!var! keyword for local type inference was removed.
All variables must be explicitly typed; storage pointers additionally require the
\lstinline!storage! keyword.

\begin{center}
  \lstinline!var acc = Acc[msg.sender]! \toarrow{}
  \lstinline!Holder storage acc = Acc[msg.sender]! 
  \\\quad
  (from \texttt{PENNY\_BY\_PENNY.sol})
\end{center}

\section{Lendf.Me}
\label{sec:lendf}

\begin{table}
	\catcode`\_=12
	\scriptsize
	\centering
	\parbox{0.8\linewidth}{\begin{filecontents*}{experiments/lendf.csv}
Experiment	Sol	Yul	ABI	Call Stack	Trace Length	Time (s)
Lendf.Me	1519	12615	2 / 21 (+23)	3	3 / 66	2.87
\end{filecontents*}
\csvautobooktabular[separator=tab]{experiments/lendf.csv}\smallskip

	{\tiny Sol and Yul LoC exclude comments and empty lines.
	Trace lengths are measured in opponent moves / total moves.
        ABI X/Y(+Z): X out of Y non-view functions; run is unchanged by adding all Z view functions.}}\quad
      \parbox{0.15\linewidth}{\fbox{\parbox{\linewidth}{\tiny\textbf{CLI Parameters:}\\
Call Bound = 2;\\
Deploy Value = 0;\\
O-Balance = 10 ETH;\\
O-Spending = \num{1} wei;\\
No Waiting;\\
Uint256 domain = 1.}}\\\\
}
	\caption{Lendf.Me Experiment}\label{tab:lendf.me}
\end{table}

Our final real-world case study is Lendf.Me~\cite{lendfme-victim}, a decentralised lending protocol built on Ethereum.
On 19 April 2020, Lendf.Me suffered a reentrancy exploit~\cite{slowmist-lendfme-analysis} that drained nearly 
USD\,25 million worth of ETH and BTC from its pools by abusing a known callback mechanism in the ERC-777 token standard.
To analyse this vulnerability, we implemented a minimal ERC-777 token that replicates the behaviour of |transferFrom| and its callback.
\begin{lstlisting}[language=Solidity]
 contract MiniERC777 {
   ...
   function transferFrom(address from, address to, uint256 amount) external returns (bool) {
     ...
     if (isContract(from)) {      // ERC777-style call-back hook
       try ITokenSender(from).tokensToSend(msg.sender, from, to, amount, "", "") {
       } catch {}
     }
     ...
} }
\end{lstlisting}
In this attack, the |supply| function of Lendf.Me’s |MoneyMarket| contract was reentered during a call to the ERC-777 token’s |transferFrom| function, which in turn invoked |tokensToSend| on the attacker’s contract.

To detect balance manipulation we instrumented the |supply| function with an assertion, as shown below.
The instrumentation records the value of |balance.principal| before the external call and asserts that it remains unchanged afterwards.
\begin{lstlisting}[language=Solidity]
function supply(address asset, uint amount) public returns (uint) {
  ...
  uint256 prePrincipal = balance.principal;                   // Snapshot variable for analysis
  err = doTransferIn(asset, msg.sender, amount);              // External call
  if (err != Error.NO_ERROR) {
    return fail(err, FailureInfo.SUPPLY_TRANSFER_IN_FAILED);
  }
  Yult.Assert(balance.principal == prePrincipal); // Assertion for analysis
  ...
}
\end{lstlisting}
With this assertion in place and a simple |Deployer| contract that sets up a functioning Lendf.Me |MoneyMarket| instance and mints our test token, we analysed the code using \yult and successfully found an exploit trace:
\begin{lstlisting}[language=EVMTrace]
 [new opponent address: <0x..30>]
 create(object:<MiniERC777_7217> , address:<0x..aa>) -> ... ->(*@\label{ln:lendf:trace:1}@*)
 deploy(object:<Deployer_201_deployed> , address:<0x..0a>) ->(*@\label{ln:lendf:trace:2}@*)
 o-call(target:<__Yult__Toolbox_61_deployed>, sig:<mint_and_approve(uint256)>, args:<[...]>)->...->po-ret([0])->(*@\label{ln:lendf:trace:3}@*)
 o-call(target:<MoneyMarket_7043_deployed>, sig:<supply(address,uint256)>, args:<[...1...]>)->...->(*@\label{ln:lendf:trace:4}@*)
 pp-call(target:<MiniERC777_7217_deployed> , sig:<transferFrom(...)> , args:<[...]>) ->(*@\label{ln:lendf:trace:5}@*)
 po-call(target:0x...30) ->(*@\label{ln:lendf:trace:6}@*)
 o-call(target:<MoneyMarket_7043_deployed>, sig:<supply(address,uint256)>, args:<[...1...]>) -> ... ->(*@\label{ln:lendf:trace:7}@*)
 ERROR! [ASSERTION VIOLATION]
\end{lstlisting}
The trace reveals the following sequence: our |MiniERC777| token is created at address 0x..aa (ln.\ref{ln:lendf:trace:1}); the |Deployer| completes setup and the game begins (ln.\ref{ln:lendf:trace:2}); the Opponent mints and approves |MiniERC777| tokens (ln.\ref{ln:lendf:trace:3}); the Opponent calls |supply| on |MoneyMarket| (ln.\ref{ln:lendf:trace:4}); |MoneyMarket| then calls |transferFrom| on |MiniERC777| (ln.\ref{ln:lendf:trace:5}); |MiniERC777| invokes the Opponent (ln.\ref{ln:lendf:trace:6}); the Opponent reenters |supply| (ln.~\ref{ln:lendf:trace:7}), which manipulates the principal balance, and triggers the assertion.


\fi

\end{document}
\endinput